\documentstyle[epsfig,longtable]{mn}
\newif\ifAMStwofonts

\newcommand{\Halpha}{H$\alpha$}
\newcommand{\Hbeta}{H$\beta$}
\newcommand{\Hgamma}{H$\gamma$}
\newcommand{\Hdelta}{H$\delta$}

\newcommand{\oii}{\hbox{[O\,\textsc{ii}]}}
\newcommand{\oiii}{\hbox{[O\,\textsc{iii}]}}
\newcommand{\nii}{\hbox{[N\,\textsc{ii}]}}
\newcommand{\hii}{\hbox{H\,\textsc{ii}}}
\newcommand{\restcol}{$(\rmn{UV}-\rmn{B})_0$}
\newcommand{\beq}{\begin{equation}}
\newcommand{\eeq}{\end{equation}}

\ifoldfss
  \newcommand{\rmn}[1] {{\rm #1}}

  \ifCUPmtlplainloaded \else
    \NewTextAlphabet{textbfit} {cmbxti10} {}
    \NewTextAlphabet{textbfss} {cmssbx10} {}
    \NewMathAlphabet{mathbfit} {cmbxti10} {} % for math mode
    \NewMathAlphabet{mathbfss} {cmssbx10} {} %  "   "    "
  \fi
  \ifAMStwofonts
    \ifCUPmtlplainloaded \else
      \NewSymbolFont{upmath} {eurm10}
      \NewSymbolFont{AMSa} {msam10}
      \NewMathSymbol{\upi}     {0}{upmath}{19}
      \NewMathSymbol{\umu}     {0}{upmath}{16}
      \NewMathSymbol{\upartial}{0}{upmath}{40}
      \NewMathSymbol{\leqslant}{3}{AMSa}{36}
      \NewMathSymbol{\geqslant}{3}{AMSa}{3E}

      \let\leq=\leqslant \let\le=\leqslant
       \let\ge=\geqslant
    \fi
  \fi
\fi % End of OFSS

\ifnfssone
  \newmathalphabet{\mathit}
  \addtoversion{normal}{\mathit}{cmr}{m}{it}
  \addtoversion{bold}{\mathit}{cmr}{bx}{it}
  \newcommand{\rmn}[1] {\mathrm{#1}}

  \newmathalphabet{\mathbfit} % math mode version of \textbfit{..}
  \addtoversion{normal}{\mathbfit}{cmr}{bx}{it}
  \addtoversion{bold}{\mathbfit}{cmr}{bx}{it}
  \newmathalphabet{\mathbfss} % math mode version of \textbfss{..}
  \addtoversion{normal}{\mathbfss}{cmss}{bx}{n}
  \addtoversion{bold}{\mathbfss}{cmss}{bx}{n}
  \ifAMStwofonts
    \ifCUPmtlplainloaded \else
      \UseAMStwoboldmath
      \makeatletter
      \new@mathgroup\upmath@group
      \define@mathgroup\mv@normal\upmath@group{eur}{m}{n}
      \define@mathgroup\mv@bold\upmath@group{eur}{b}{n}
      \edef\UPM{\hexnumber\upmath@group}
      \new@mathgroup\amsa@group
      \define@mathgroup\mv@normal\amsa@group{msa}{m}{n}
      \define@mathgroup\mv@bold\amsa@group{msa}{m}{n}
      \edef\AMSa{\hexnumber\amsa@group}
      \makeatother
      \mathchardef\upi="0\UPM19
      \mathchardef\umu="0\UPM16
      \mathchardef\upartial="0\UPM40
      \mathchardef\leqslant="3\AMSa36
      \mathchardef\geqslant="3\AMSa3E

      \let\leq=\leqslant \let\le=\leqslant
       \let\ge=\geqslant
    \fi
  \fi
\fi % End of NFSS release 1

\ifnfsstwo
  \newcommand{\rmn}[1] {\mathrm{#1}}

  \DeclareMathAlphabet{\mathbfit}{OT1}{cmr}{bx}{it}
  \SetMathAlphabet\mathbfit{bold}{OT1}{cmr}{bx}{it}
  \DeclareMathAlphabet{\mathbfss}{OT1}{cmss}{bx}{n}
  \SetMathAlphabet\mathbfss{bold}{OT1}{cmss}{bx}{n}
  \ifAMStwofonts
    \ifCUPmtlplainloaded \else
      \DeclareSymbolFont{UPM}{U}{eur}{m}{n}
      \SetSymbolFont{UPM}{bold}{U}{eur}{b}{n}
      \DeclareSymbolFont{AMSa}{U}{msa}{m}{n}
      \DeclareMathSymbol{\upi}{0}{UPM}{"19}
      \DeclareMathSymbol{\umu}{0}{UPM}{"16}
      \DeclareMathSymbol{\upartial}{0}{UPM}{"40}
      \DeclareMathSymbol{\leqslant}{3}{AMSa}{"36}
      \DeclareMathSymbol{\geqslant}{3}{AMSa}{"3E}

      \let\leq=\leqslant \let\le=\leqslant
       \let\ge=\geqslant
    \fi
  \fi
\fi % End of NFSS release 2

\ifCUPmtlplainloaded \else
  \ifAMStwofonts \else % If no AMS fonts
    \def\upi{\pi}
    \def\umu{\mu}
    \def\upartial{\partial}
  \fi
\fi

\def\lsim{\mathrel{\rlap{\lower 3pt\hbox{$\mathchar"218$}}
     \raise 2.0pt\hbox{$\mathchar"13C$}}}
\def\gsim{\mathrel{\rlap{\lower 3pt\hbox{$\mathchar"218$}}
     \raise 2.0pt\hbox{$\mathchar"13E$}}}

\def \d{{\rm d}}

\journal{}

\title[An Ultraviolet-Selected Galaxy Redshift Survey]
{An Ultraviolet-Selected Galaxy Redshift Survey -- II: The Physical Nature of Star Formation in an Enlarged Sample}

\author[M. Sullivan et al.]{Mark~Sullivan,$^1$\thanks{E-mail: ms@ast.cam.ac.uk} Marie~A.~Treyer,$^2$ Richard~S.~Ellis,$^1$ Terry~J.~Bridges,$^3$
\newauthor 
Bruno~Milliard$^2$ \& Jos\'{e}~Donas$^2$ \\
$^1$ Institute of Astronomy, Madingley Road, Cambridge CB3 OHA, UK \\
$^2$ Laboratoire d'Astronomie Spatiale, Traverse du Siphon, 13376 Marseille, France \\
$^3$ Anglo-Australian Observatory, PO Box 296, Epping NSW 2121, Australia \\
}

\date {Accepted ---. Received ---; in original form ---.}
\begin{document}
\label{firstpage}
\maketitle

\begin{abstract}
  We present further spectroscopic observations for a sample of
  galaxies selected in the vacuum ultraviolet~(UV) at 2000~\AA\ from
  the FOCA balloon-borne imaging camera of Milliard et al. (1992).
  This work represents an extension of the initial study of Treyer et
  al.~(1998). Our enlarged catalogue contains 433 sources ($\simeq3$
  times as many as in our earlier study) across two FOCA fields. 273
  of these are galaxies, nearly all with redshifts $z\simeq 0-0.4$.
  Nebular emission line measurements are available for 216 galaxies,
  allowing us to address issues of excitation, reddening and
  metallicity.  The UV and H$\alpha$ luminosity functions strengthen
  our earlier assertions that the local volume-averaged star formation
  rate is higher than indicated from earlier surveys. Moreover,
  internally within our sample, we do not find a steep rise in the UV
  luminosity density with redshift over $0<z<0.4$.  Our data is more
  consistent with a modest evolutionary trend as suggested by recent
  redshift survey results. Investigating the emission line properties,
  we find no evidence for a significant number of AGN in our sample;
  most UV-selected sources to $z\simeq 0.4$ are intense star-forming
  galaxies. We find the UV flux indicates a consistently higher mean
  star formation rate than that implied by the H$\alpha$ luminosity
  for typical constant or declining star formation histories.
  Following Glazebrook et al.~(1999), we interpret this discrepancy in
  terms of a starburst model for our UV-luminous sources. We develop a
  simple algorithm which explores the scatter in the UV flux-H$\alpha$
  relation in the context of various burst scenarios.  Whilst we can
  explain most of our observations in this way, there remains a small
  population with extreme UV-optical colours which cannot be
  understood.

\end{abstract}


\section{Introduction}

There has been considerable progress in recent years in determining
observational constraints on the cosmic history of star formation, and
the way this relates to the far infrared background light and present
density of stars and metals (see Madau~\shortcite{m99} for a recent
summary).  Inevitably, most attention has focused on the contribution
to the global history from the most distant sources, presumably seen
at a time close to their formation. Controversial issues at the time
of writing include the interpretation of faint sub-mm sources as
young, star-forming galaxies \cite{blain}, the effect of dust on
measures derived from rest-frame ultraviolet luminosities
\cite{meurer,steidel96}, cosmic variance in the limited datasets
currently available \cite{steidel99} and uncertain non-thermal
components within the far-infrared background \cite{m99}.

At more modest redshifts ($z<1$), it might be assumed that the cosmic
star formation history is fairly well-determined. Madau et
al.'s~\shortcite{m96} original analysis in this redshift range was
based on rest frame near ultraviolet luminosities derived from the
$I$-selected Canada France Redshift Survey \cite{lilly} and local
\Halpha\ measures taken from Gallego et al.'s~\shortcite{gallego}
objective prism survey. This combination of data implied a dramatic
decline in the comoving density of star formation (by a factor of
$\simeq 10$) which is difficult to match theoretically \cite{baugh}.

The addition of further data to the low redshift component of the
cosmic star formation history has confused rather than clarified the
situation. The $b_J$-selected \textit{Autofib/LDSS} redshift survey
\cite{ellis96} satisfactorily probes the evolutionary trends from
$0.25<z<0.75$ and, whilst supporting an increase in luminosity
density over this interval, the survey illustrated the difficulty of
connecting faint survey data with similar local luminosity functions
(LFs) whose absolute normalisations remain uncertain, as well as a
fundamental difference in the luminosity dependence of the evolution
seen \cite{ellis97}. The CFRS data indicate luminosity evolution of
$\simeq 1$ mag to $z\simeq 1$ at the bright end of the galaxy LF
consistent with a decline in the star formation rate of a
well-established population. In contrast, the \textit{Autofib/LDSS}
results suggests that most of the changes in luminosity density occurred
via a rapid decline in abundance of lower luminosity (sub-$L^{\ast}$)
systems. Morphological data for both surveys from Hubble Space
Telescope (HST) \cite{brinchmann} has since shown a substantial
fraction of the rise in luminosity density arises from galaxies of
irregular morphology.

In an earlier paper in this series (Treyer et al.~1998, hereafter
Paper I), we presented the first ultraviolet~(UV)-selected constraints
on the local density of cosmic star formation. Using a flux-limited
sample of 105 spectroscopically-confirmed sources selected at 2000~\AA\ 
from a balloon-borne UV imaging camera, a local integrated luminosity
density well above optically-derived estimates was found, suggesting
claims for strong evolution in the range $0<z<1$ had been overstated.
Corrections for dust extinction would only strengthen this conclusion.

A revision of the evolutionary trends for $z<1$ is supported by a
recent re-evaluation of the field galaxy redshift survey results by
Cowie, Songaila \& Barger~\shortcite{cowie}. By selecting faint galaxies in the $U$
and $B$ bands rather than the $I$ band (c.f. CFRS), a more modest increase
with redshift in the UV luminosity density is found. Cowie et al.
propose the discrepancy with the CFRS may arise from the extrapolation
necessary in the CFRS at intermediate redshifts to determine 
2800~\AA\ luminosities from the available $I$-band magnitudes.

More generally, it is becoming increasingly apparent that different
diagnostics (UV flux, \Halpha\ luminosities, 1.4~GHz luminosities) may
lead to different star formation rates, even for the same galaxies.
Glazebrook et al.~\shortcite{glazebrook} have shown a consistent
discrepancy exists between star formation densities derived using UV
continua and nebular \Halpha\ measures, and interpreted this in terms
of both dust extinction and an erratic star formation history for the
most active sources. A similar trend is seen by Yan et al.~\shortcite{yan}.

The above developments serve to emphasize that the integrated
comoving star formation density is a poor guide to the physical
processes occurring in the various samples and, moreover, that the
evolutionary trends in the (presumed) well-studied $0<z<1$ range
remain uncertain. In this second paper in the series, we return to
the key question of the \textit{physical nature} of the star
formation observed in the local samples and particularly those
of the kind discussed in Paper I. We have extended our UV
sample and obtained uniform diagnostic spectroscopy over a wider
wavelength range so that we can compare star formation rates from
nebular and UV continuum measures.

A plan of the paper follows. In $\S$2 we discuss the enlarged
spectroscopic sample. Using the William Herschel Telescope (WHT) we
have conducted systematic spectroscopy of a further 305 sources within
Selected Area 57 (SA57) and Abell 1367, and this allows us to update
the analysis of the UV LF and SF density presented in Paper I, and
discuss the implications of possible reddening. In $\S$3 we extend our
analysis, for the first time, to include a careful discussion of the
emission line properties of our sample. A puzzling aspect revealed in
Paper I was the abnormally-strong UV fluxes and colours of a
proportion of our sources. We examine this effect in some detail and
discuss constraints on both the metallicity and AGN contamination of
our sample. In $\S$4, we interpret our various star formation
diagnostics in terms of duty-cycles exploring quantitatively the
suggestions of Glazebrook et al.~\shortcite{glazebrook} that the star
formation is erratic for a significant proportion of sources. We
discuss the implication of our results in $\S$5 and summarise our
basic conclusions in $\S$6.  Throughout this paper, all calculations
assume an $\Omega =1$, $H_0 =100~\rmn{h~km~s}^{-1}~\rmn{Mpc}^{-1}$
cosmology.

\section{The Enlarged Sample}

This paper presents the spectroscopic extension to the UV-selected
redshift survey, conducted on a sample selected using the
balloon-borne FOCA2000 camera, preliminary results of which were
presented in Paper I. A full description of the details of the FOCA
experiment can be found in Milliard et al.~\shortcite{milliard}. In
brief, the telescope is a 40cm Cassegrain mounted on a stratospheric
gondola, stabilised to within a radius of 2\arcsec\ rms. The spectral
response of the filter used on the telescope approximates a Gaussian
centred at 2015~\AA, FWHM 188~\AA. The camera was operated in two modes
-- the FOCA 1000 (f/2.56, 2.3\degr) and FOCA 1500 (f/3.85, 1.55\degr)
-- with the large field-of-view (FOV) well suited to survey work. The
limiting depth of the exposures is $\rmn{m_{UV}}=18.5$, which, for a
late-type galaxy, corresponds to $\rmn{m_B}=20-21.5$.

The extended dataset presented here is based on two FOCA fields. The
first, SA57, was partially covered in Paper I, and is centred at
$\rmn{RA}=13^{\rmn{h}}03^{\rmn{m}}53^{\rmn{s}}$, $\rmn{Dec.}=+29\degr
20\arcmin 30\arcsec$ (1950 epoch). The second field is centred on
$\rmn{RA}=11^{\rmn{h}} 42^{\rmn{m}} 46^{\rmn{s}}$,
$\rmn{Dec.}=+20\degr 10\arcmin 03\arcsec$, and contains the cluster
Abell 1367. The fields were imaged in both the FOCA 1000 and FOCA 1500
modes. The astrometric accuracy of the FOCA-1500 catalogue (around
3\arcsec\ rms, see Milliard et al~\shortcite{milliard}) is
insufficient for creating a spectroscopic target list, so the FOCA
catalogues were therefore matched with APM scans of the POSS optical
plates.  Two problems were encountered. For some UV detections, there
was more than one possible optical counterpart on the POSS plates
within the search radius used -- in these cases, the nearest optical
counterpart to the UV detection was selected.  Secondly, some of the
UV sources have no obvious counterpart on the APM plates, indicating
that either some of the detections are spurious, or that the
counterpart lies at a fainter $B$ magnitude than the limiting
magnitude of the POSS plates ($\rmn{m_B}\simeq 21$).

Paper I presented preliminary results from an optical spectroscopic
follow-up to the SA57 UV detections. After basic star/galaxy
separation, two instruments -- the Hydra instrument on the 3.5-m WIYN
telescope ($\lambda\lambda$ 3500--6600~\AA, $3.1\arcsec$ diameter fibres),
and WYFFOS on the 4.2-m William Herschel Telescope (WHT)
($\lambda\lambda$ 3500--9000~\AA, $2.7\arcsec$ diameter fibres, see
Bridges~\shortcite{bridges} for more details) -- were used to obtain
142 reliable spectra, though 14 of these came from a weather affected
exposure in which the incompleteness was very large. After further
star removal and elimination of sources with poor UV fluxes, a
complete sample of 105 galaxies with confirmed redshift remained. A
further 3 galaxies have since been found to have unreliable optical
($B$) magnitudes.

The new data sample was observed on the WHT to ensure that \Halpha\ 
emission would be visible to a redshift of $z=0.4$. The targets for
the new survey were chosen so that no identified galaxy with a
redshift from Paper I was re-observed. All the new UV sources are
taken from the deeper FOCA 1500 catalogue, which also has the
advantage of a higher imaging resolution (3\arcsec\ as opposed to
4.5\arcsec\ rms). This reduces the problem of multiple optical
counterparts for UV sources, as a smaller search radius on the optical
plates can be used, but still leaves around 9 per~cent of sources with
an uncertain identification.

Six exposures were performed of different fields within SA57, and one
was taken of Abell 1367. Each exposure is broken into several shorter
1800s exposures to help improve cosmic ray rejection, and median
spectra produced for each field. Several sources within SA57 were
observed on more than one exposure, allowing a comparison of results
between exposures. The spectra were reduced as in Paper I, but
additional flux-calibration was performed on the new spectral sample.
Details of all observing runs can be found in
Table~\ref{observinglog}.

\begin{table*}
\begin{center}
\begin{tabular}{clccccc} 
\hline \hline
Date & Field & R.A. (1950) & DEC. (1950) & Telescope/ & Exposure & Flux \\
 & & (h m s) & $(\degr\ \arcmin\ \arcsec)$& instrument & time (s) & calibration? \\
\hline
Paper I\\
28-02-96 & SA57 - 1 & 13:05:48 & 29:17:49 & WIYN/Hydra & 3x1800 & No \\
29-02-96 & SA57 - 2 & 13:05:48 & 29:17:49 & WIYN/Hydra & 3x1800 & No \\
02-04-97 & SA57 - 3 & 13:04:11 & 29:21:04 & WHT/WYFFOS & 4x1800 & No \\
 & SA57 - 4 & 13:00:59 & 29:36:28 & WHT/WYFFOS & 2x1800 & No \\
\hline
New \\
24-04-98 & SA57 - 5 & 13:04:01 & 28:59:48 & WHT/WYFFOS & 5x1800 & Yes\\
 & SA57 - 6 & 13:02:53 & 29:09:26 & WHT/WYFFOS & 4x1800 & Yes\\
25-04-98 & SA57 - 7 & 13:02:53 & 29:28:27 & WHT/WYFFOS & 3x1800 & Yes\\
 & SA57 - 8 & 13:04:02 & 29:37:53 & WHT/WYFFOS & 3x1800 & Yes\\
 & A1367 - 1 & 11:41:31 & 20:20:18 & WHT/WYFFOS & 5x1800 & Yes\\
26-04-98 & SA57 - 9 & 13:05:24 & 29:28:19 & WHT/WYFFOS & 4x1800 & Yes\\
& SA57 - 10 & 13:05:24 & 29:09:25 & WHT/WYFFOS & 5x1800 & Yes\\
\hline \hline
\end{tabular}
\caption{Details of all the observing runs that contribute to the enlarged sample. Standards were not available for the Paper I sample, hence these data is not flux-calibrated.}
\label{observinglog}
\end{center}
\end{table*}

The spectra were analysed using the \textsc{splot} facility in
\textsc{iraf} and the \textsc{figaro} package \textsc{gauss}.
Redshifts were measured by visual inspection, and the equivalent
widths (EWs) and fluxes of \oii\ (3727~\AA), \oiii\ (4959~\AA\ and
5007~\AA), \Hbeta\ (4861~\AA) and \Halpha\ (6562~\AA) determined using
both spectral analysis programs.  $1\sigma$ errors were also provided
by \textsc{splot} using an estimate of the noise in the individual
spectra. The continuum level can be fitted interactively using
polynomial fitting within the \textsc{gauss} program, and compared
with the linear fitting from the \textsc{splot} program. In most
cases, especially in the spectra with a high S/N, the two flux
measurements show an excellent agreement within the $1\sigma$ errors
provided by \textsc{splot} -- the average discrepancy is $\simeq 13$
per~cent. This provides a good reliability check on the effects of
continuum fitting on the spectra, which differ in the two routines.
Additionally, the \Halpha\ and \oii\ EWs were measured independently
by two of the authors (MS and MAT) as a check that there were no
measurement biases. The average discrepancy was $\simeq 14$ per~cent,
indicating a good agreement. Though the spectral resolution (10~\AA)
is good enough to resolve the separate \oiii\ lines, in many cases the
\Halpha\ line (6562~\AA) was blended with the nearby \nii\ lines at
6583~\AA\ and 6548~\AA, so a deblending routine was run from within
\textsc{splot} to allow determination of the fluxes of these
individual lines.

The integration error estimates are derived by error propagation
assuming a Poisson statistics model of the pixel sigmas, generated by
measuring the noise in the spectra on an individual basis. It is
assumed that the linear continuum has no errors. The \textsc{splot}
errors in the deblending routines are derived using a Monte-Carlo
simulation as follows. The model is fit to the data -- using the pixel
sigmas from the noise model -- and is used as a noise-free spectrum.
100 simulations were run, adding random Gaussian noise to this
`noise-free' spectrum using the noise model. The deviation of each new
fitted parameter to model parameter was recorded, and the error
estimate for each parameter is then the deviation containing 68.3 per
cent of the parameter estimates -- this corresponds to $1\sigma$ if
the distribution of the parameter estimates is Gaussian. This allows
calculation of the errors in cases where individual lines are blended
together.

The errors are thus random measurement errors only, i.e. they arise
from the S/N of the spectrum in question. A further source of
uncertainty will be introduced during flux calibration, as each fibre
on the spectrograph may have a slightly different throughput. Ideally,
standard stars should be observed through each fibre, but this is not
possible in practice. Note, however, that this uncertainty will only
apply to the line fluxes, and not to the EWs. Additionally, no
aperture corrections are applied at this stage (see Section~4.2 for a
discussion of this).

\begin{table*}
\begin{center}
\begin{tabular}{cccccccccc} 
\hline \hline
Field & Number & Stars & QSOs & Missing mags & Galaxies & Emission lines & \Halpha & Unidentified & $\rmn{OC}>1$\\
\hline
New SA57 & 241 & 37 & 14 & 9 & 130 & 97 & 88 & 51 & 32\\
ABELL 1367 & 64 & 5 & 4 & 3 & 51 & 38 & 37 & 1 & 3\\
Old SA57 & 128 & 8 & 5 & 13 & 92 & 81 & 34 & 10 & 14 \\
\hline
Total  & 433 & 50 & 23 & 25 & 273 & 216 & 159 & 62 & 49 \\
\hline \hline
\end{tabular}
\caption{The breakdown of spectroscopic objects in the new, old (Paper I) 
and combined sample, giving the number of each object type. \textit{Missing mags} indicates that either UV or $B$ magnitudes were not available.}
\label{spectratable}
\end{center}
\end{table*}

A summary of the new sample is given in Table~\ref{spectratable},
together with the statistics for that obtained by combining with the
data discussed in Paper I. From this enlarged sample, 48 objects have
two optical counterparts and 1 has three. Additionally, of the
galaxies with a redshift, 15 were determined to be unreliable UV
detections, and 10 have unreliable $B$-magnitude information from the
POSS plates -- these are shown as \textit{missing mags} in the table.
This leaves 234 galaxies in the spectroscopic sample, and 224 galaxies
in a restricted sample with full colour information, where there is an
unambiguous optical identification. The total area surveyed in the
enlarged sample is $1.88~\rmn{deg}^2$ in SA57 and $0.35~\rmn{deg}^2$
in Abell 1367, giving $\sim 2.2~\rmn{deg}^2$ in total.

For the new data set, 4 of the unidentified spectra suffered from
technical difficulties in extraction unrelated to the S/N ratio, so
the formal incompleteness is 48/301, or $\simeq 16$~per~cent. Of the 68
unidentified spectra in the enlarged sample, 10 suffered from
technical difficulties, so the formal incompleteness within all the
well-exposed fields -- i.e. excluding the shortened WHT exposure from
Paper I -- is therefore 52/423, or $\simeq 12$~per~cent.

In summary, therefore, the combined catalogue represents
a three-fold increase in sample size c.f. Paper I, with the
added benefit of emission line measurements for a significant
fraction of the total.

\subsection{Photometry}

The FOCA team adopted a photometric system discussed in
detail by Milliard~et al.~\shortcite{milliard} and in Paper I, which
is close to the ST system. The apparent UV magnitude to flux
conversion is given by:

\begin{equation}
\rmn{m}_{\lambda}=-2.5\log \rmn{f}_{\lambda}-21.175
\label{appmageqn}
\end{equation}

\noindent
where the flux (f$_{\lambda}$) is in erg/cm$^2$/s/\AA. The zero-point
is accurate to $\leq 0.2$~mag. Close to the limiting magnitude of this
survey however, the uncertainty in the relative photometry may reach
$\simeq0.5$~mag \cite{donas} due to non-linearities in the FOCA
camera.  Conservatively, we estimate the errors in the UV magnitudes
(m$_{\lambda}$, hereafter m$_{\rmn{uv}}$) to be 0.2 for $\rmn{m_{uv}}<17$,
and 0.5 for $\rmn{m_{uv}}>17$.

As in Paper I, the $B$-photometry was taken from the POSS database,
including saturation and isophotal loss corrections. Again, there will
be non-linearity effects near the limiting magnitude of the plates,
and also at the brighter end. The error in the $B$-photometry was
taken to be $\simeq \pm0.2$. However, the $B$ photometric scale has to
be corrected by $\rmn{m_B^{corr}}\equiv \rmn{m_B^{APM}}-0.546$ in
order to align it with the FOCA system (Paper I, Donas et
al.~\shortcite{donas})\footnote{This correction differs from that
  adopted in Paper I; we found the correction had been slightly
  underestimated.}.

\subsection{Extinction corrections}

Extinction arising along the line-of-sight to a target galaxy makes
the observed ratio of the fluxes of two emission lines differ from
their ratio as emitted in the galaxy. The extinction, $C$, can
be derived using the Balmer lines \Halpha\ and \Hbeta:

\begin{equation}
\frac{F(\rmn{H}\alpha)}{F(\rmn{H}\beta)}=D10^{-C[S(\rmn{H}\alpha)-S(\rmn{H}\beta)]}
\label{ceqn}
\end{equation}

\noindent
where $F(\rmn{H}\alpha)$ and $F(\rmn{H}\beta)$ are the measured
integrated line fluxes, and $D$ is the ratio of the fluxes as emitted
in the nebula.  Assuming case B recombination, with a density of
100~cm$^{-3}$ and a temperature of 10,000~K, the predicted ratio of
\Halpha\ to \Hbeta\ is $D=2.86$~\cite{osterbrock}. Using the standard
interstellar extinction law from Table 3 in Seaton~\shortcite{seaton},
$S(\rmn{H}\alpha)-S(\rmn{H}\beta)=-0.323$, and $C$ can be readily
determined from Eqn.~\ref{ceqn}. Any corrected emission line flux,
$F_0(\lambda)$, can then be estimated using:

\begin{equation}
F_0(\lambda)=F(\lambda)10^{C[1+(S(\lambda)-S(H_{\beta}))]}
\label{linecorrecteqn}
\end{equation}

\noindent
where the values of $S(\lambda)-S(\rmn{H}\beta)$ were taken from
Seaton~\shortcite{seaton}, with values of -0.323, -0.034, 0 and
0.255 for \Halpha, \oiii\, \Hbeta\ and \oii\ respectively. $1\sigma$
errors for the reddened fluxes have been calculated from the $1\sigma$
errors in the unreddened fluxes and in $C$ in the standard way. For
comparison with other emission line surveys, the extinction parameter
$A_V$ has also been calculated using:

\begin{equation}
A_V=E(B-V) R=\frac{CR}{1.47}~\rmn{mag}
\label{aveqn}
\end{equation}

\noindent
where the relation $E(B-V)=C/1.47$, and $R$ is the mean ratio
$R=A_V/E(B-V)$, with a value $R=3.2$, both from
Seaton~\shortcite{seaton}.

\begin{figure}
\epsfig{figure=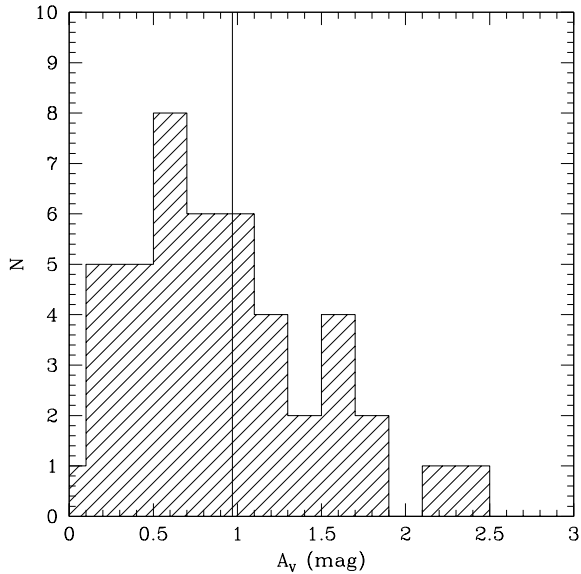,width=80mm}
\caption{The distribution of the reddening parameter $A_V$ for the galaxies 
with both \Halpha\ and \Hbeta\ in emission, after correction for 
underlying stellar absorption, The average value (0.97) is the 
solid line.}
\label{avplot}
\end{figure}

A correction must also be made for stellar absorption underlying the
\Halpha\ and \Hbeta\ lines. Though most of the galaxies in the sample
have strong emission lines, the reddening corrections are very
sensitive to the amount of absorption on the Balmer lines. Two methods
were used to analyse the contribution of stellar absorption. Firstly,
each galaxy spectrum was checked for higher order Balmer lines
(\Hgamma, \Hdelta\ etc.) and, if these lines appeared in
absorption, the EW of each was measured, and the average of these
values then applied as a correction to both the \Halpha\ and \Hbeta\ 
lines. Where the higher-order lines were not visible, or appeared in
emission, a correction of 2~\AA\ was applied, typical of such spectra
\cite{tressecfrs}. The second method was to
use the program \textsc{dipso} to fit the stellar absorption line
underneath the \Hbeta\ emission, then use this fit as the continuum
and subtract from the spectrum. The flux of the \Hbeta\ line should
then contain no absorption contribution if the fitting is done
carefully. The two methods gave similar results. The distribution of
$A_V$ after the corrections can be seen in Fig.~\ref{avplot}.

From the spectra containing both strong \Halpha\ and strong \Hbeta,
the mean value for $A_V$ without any absorption correction is 1.78,
and after correction is 0.97. These compare well with values of 1.52
for a selection of CFRS galaxies \cite{tressecfrs}, which made no
allowance for stellar absorption, and studies of individual \hii\ 
regions in local spiral galaxies ($\simeq 0.6$ \cite{oey} and $\simeq
1$ \cite{kennicutt89} ), both corrected for absorption. In spectra
where it was not possible to measure both \Halpha\ and \Hbeta, either
due to a low S/N, or, more commonly, due to the \Hbeta\ line being
badly affected by stellar absorption, a value of $A_V=0.97$ was
assumed (i.e. $C=0.45$) -- these galaxies are not shown in
Fig.~\ref{avplot}.

It is important to realise one source of possible bias this may
introduce into our corrections.  The Balmer-derived corrections
applied here require the presence of both \Halpha\ and \Hbeta\ in the
galaxy spectra. However, as the extinction increases, if the limiting
factor on determining a line flux were purely the S/N of the spectra,
the \Hbeta\ would become undetectable before \Halpha, implying the
average correction used here to be a lower limit. The presence of
significant \Hbeta\ absorption in many spectra prevents an accurate
calculation of the size of this effect.

This complication aside, the Balmer decrement remains the best way to
estimate extinction in our sample. The problem now arises of how to
convert these emission line reddening corrections to those appropriate
for our UV (and optical) magnitudes. Though the different
extinction laws (e.g. MW, SMC, LMC) are similar at optical emission
line wavelengths, and the Balmer extinction results are relatively
insensitive to the choice of extinction law used, this is not the case
in the UV. There is a wide choice of methods available in the
literature, and it is clear that for the unresolved galaxies under
study here, the reddening of the UV continuum will depend upon details
of the dust-star-gas geometry. The reddening of the stellar continuum
may be different from the obscuration of the ionised gas, as the stars
and the gas may occupy different areas within the galaxy. Indeed, it
has been shown that the continuum emission from stars is often less
obscured than line emission from the gas
\cite{fanelli,calzetti94,mas-hesse}.  From studies of the central
regions of starburst galaxies, Calzetti~\shortcite{calzetti97b}
derived the following empirical, and geometry-independent,
prescription to correct fluxes as a function of wavelength. Using the
standard form,

\begin{equation}
F_0(\lambda)=F_{obs}(\lambda)10^{0.4E(B-V)_s\mu(\lambda)}
\end{equation}

\noindent
where the colour excess of the stellar continuum, $E(B-V)_s$, is related
to that for the ionised gas, $E(B-V)_g$, and hence $C$, by:

\begin{equation}
E(B-V)_s=0.44E(B-V)_g=\frac{0.44C}{1.47}
\label{egeqn}
\end{equation}

\noindent
The function $\mu(\lambda)$ and Equation~\ref{egeqn} are empirical
relations taken from Calzetti~\shortcite{calzetti97b}, who derived
these results using a sample of star-forming galaxies. The function
$\mu(\lambda)$ has the value 9.70 at 2000~\AA, and 6.17 at $\simeq$
4100~\AA\ -- the central wavelength of the POSS $B$ filter. In terms of
magnitudes, the corrections are then:

\begin{equation} 
m_{UV}^{\rmn{corr}}=m_{UV}^{\rmn{obs}}-0.3C\mu(2000~\AA/(1+z))
\end{equation} 

\begin{equation}
m_{B}^{\rmn{corr}}=m_{B}^{\rmn{obs}}-0.3C\mu(4100~\AA/(1+z))
\end{equation}

\begin{figure}
\epsfig{figure=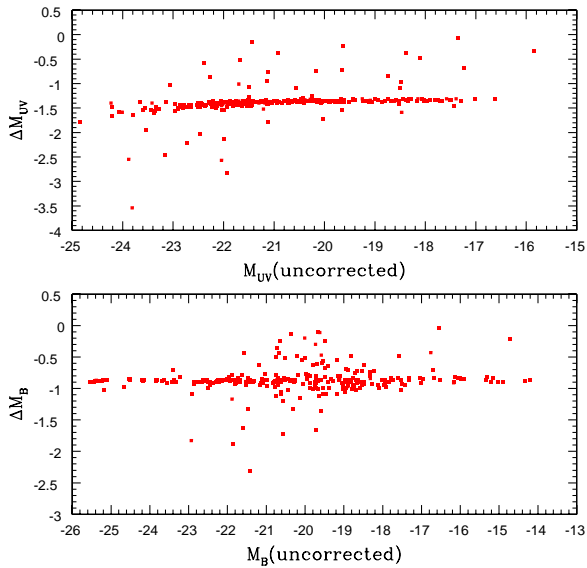,width=80mm}
\caption{Dust extinction correction as a function of uncorrected absolute
  magnitude using Calzetti (1997b)'s `recipe'. Note that the small
  scatter arises only because an average $A_V=0.97$, $C=0.45$ was
  applied for those galaxies without direct \Halpha\ / \Hbeta\ 
  extinction measurements.}
\label{calmagplot}
\end{figure}

\begin{figure*}
\epsfig{figure=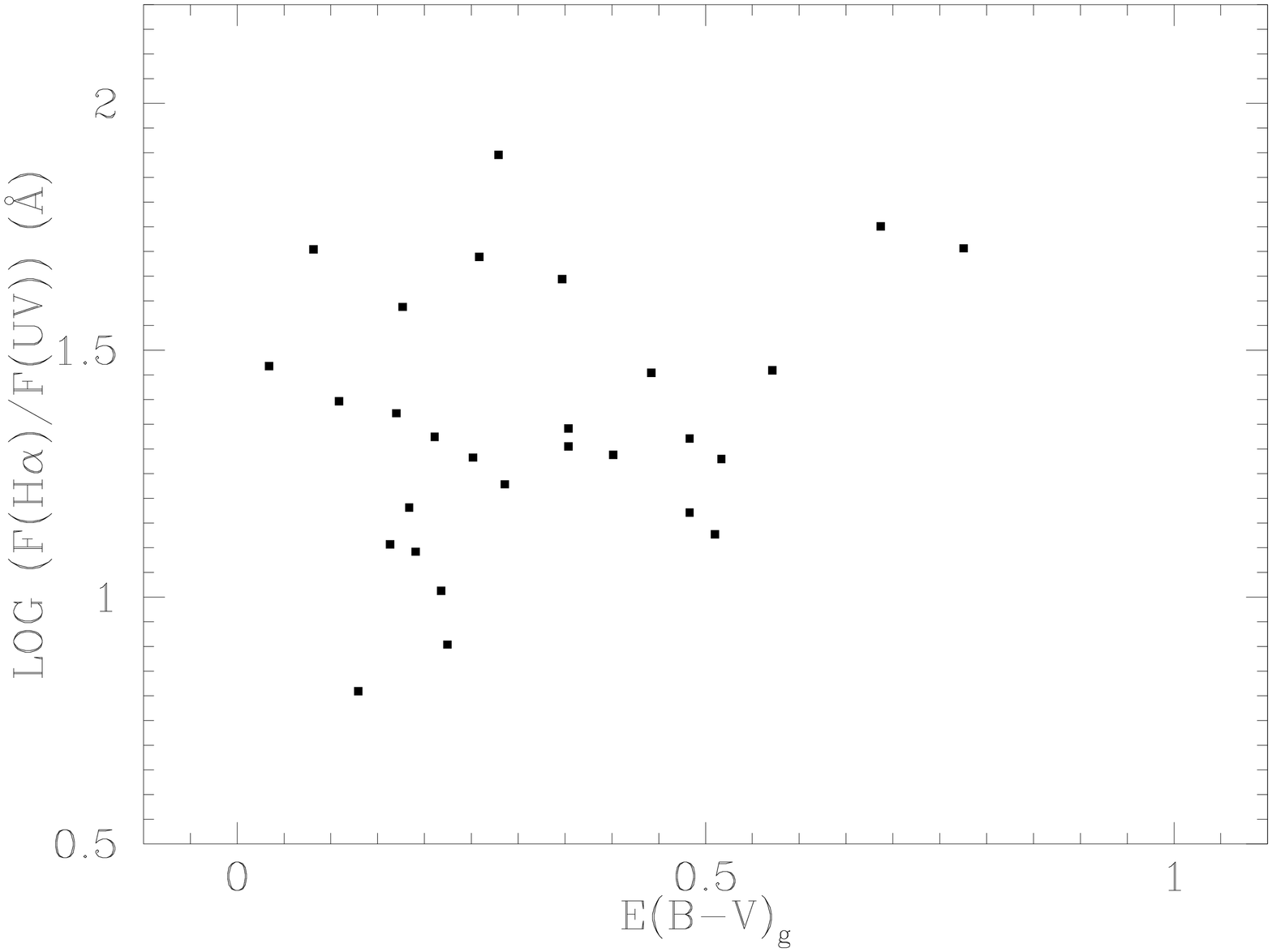,angle=-90,width=80mm}
\epsfig{figure=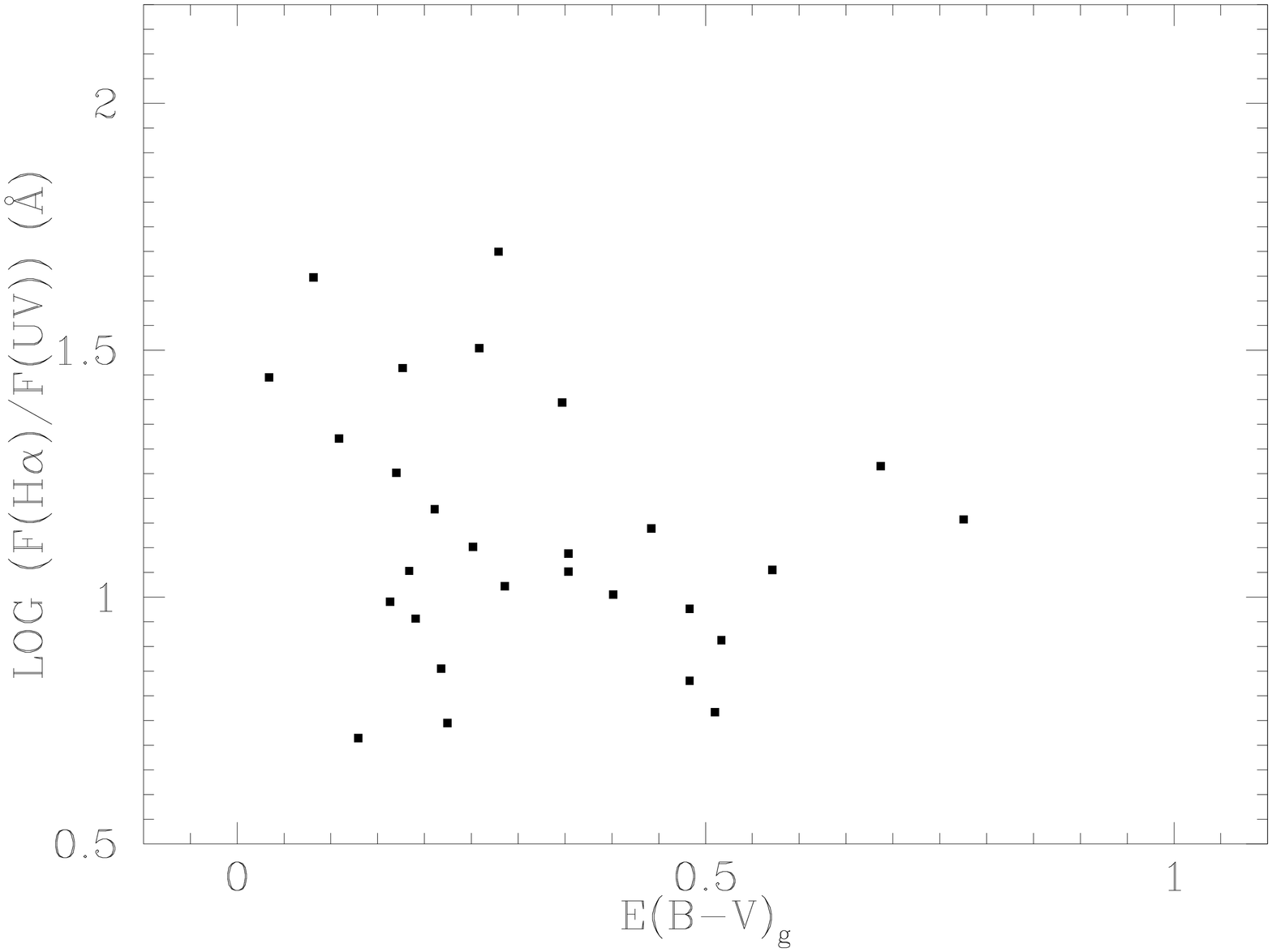,angle=-90,width=80mm}
\caption{The relationship between the ratio of \Halpha\ to UV luminosities and the ionised gas colour excess (from Section 2.2). The left plot is for the fluxes uncorrected for dust extinction (but corrected for stellar absorption), the right plot for fluxes corrected as explained in the text. Only single optical counterpart, field galaxies from SA57 are shown.}
\label{ratio_colourexcessplot}
\end{figure*}

\noindent
The effect of the reddening corrections on the absolute magnitudes is
shown in Fig.~\ref{calmagplot}, and the relation between both the
uncorrected and corrected UV and \Halpha\ fluxes and $E(B-V)_g$ are
shown in Fig.~\ref{ratio_colourexcessplot}. This is shown for the most
complete sample of our survey, the SA57 field galaxies, excluding the
Coma cluster galaxies, which may experience different dust
environments.

The 0.44 factor in Eqn.~\ref{egeqn} takes into account the fact that 
the stars and gas may occupy distinct regions with differing
amounts of dust and different dust covering factors. The \Halpha\ 
luminosity arises purely from very young, short-lived ionising stars,
which must remain close to the (dusty) regions in which they were
born. By contrast, the UV continuum at 2000~\AA\ contains a significant
contribution from older non-ionising stars, which may no longer be
associated with the regions in which they were formed, and hence will
suffer less from dust extinction.

If this simple interpretation is correct, then the reddening of the
stellar continuum and of the ionising gas should not be strongly
correlated; indeed the correlation found (left-hand plot of
Fig.~\ref{ratio_colourexcessplot}) is weaker than in previous studies
\cite{calzetti94,calzetti97a}. This may not be an entirely unexpected
result. As this survey is selected in the UV at 2000~\AA, it is likely
to be biased against those objects which are intrinsically dusty and
hence have lower measured UV fluxes. Therefore we should not be
surprised to see an absence of galaxies with a large \Halpha\ to UV
ratio.  Additionally, if the \Hbeta\ is suppressed relative to the
\Halpha\ by a significant amount, it will not be measured reliably in
the optical spectra, so galaxies with a large $E(B-V)_g$ will not be
shown in Fig.~\ref{ratio_colourexcessplot}. The trend is similar to
that noted by Meurer, Heckman \& Calzetti~\shortcite{meurer99}, who
plotted the ratio of line flux to F(1600\AA) against UV spectral
slope, $\beta$, (their Fig.~7), for a sample of local UV-selected
starbursts and a sample of 7 $U$-band `dropouts' observed by various
authors.  These are corrected for Galactic extinction only, and, for
the `dropout' galaxies, show only a weak correlation between the
line/UV ratios and extinction, similar to that found here.

For the mean correction ($C=0.45$) used above to correct the emission
lines, the Calzetti law gives corrections at 2000~\AA\ of
$A_{2000}=1.33$.  Other studies are in broad agreement with this
value, for example Buat \& Burgarella~\shortcite{buat98} derive
$A_{2000} \simeq 1.2$ using radiation transfer models to estimate
extinction. Using the parameterization of Seaton~\shortcite{seaton},
which uses a simple foreground dust screen model, we derive values of
$A_{2000}=2.70$ for the average value $C=0.45$. This last correction
introduces several complexities.  We already have uncomfortably blue
colours for a sub-sample of our galaxies; as we shall see in
Section~4, there is considerable difficulty in reproducing these using
conventional starburst models.  The UV luminosities would also become
difficult to explain (Section~4). For this reason, we adopt the
Calzetti law throughout this paper, noting that although this will
result in the smallest corrections to our UV luminosites, due to the
nature of the selection criteria for this survey, the UV continuum is
not likely to suffer from a larger degree of extinction.

\subsection{UV redshift/colour distribution}

\begin{figure*}
\epsfig{figure=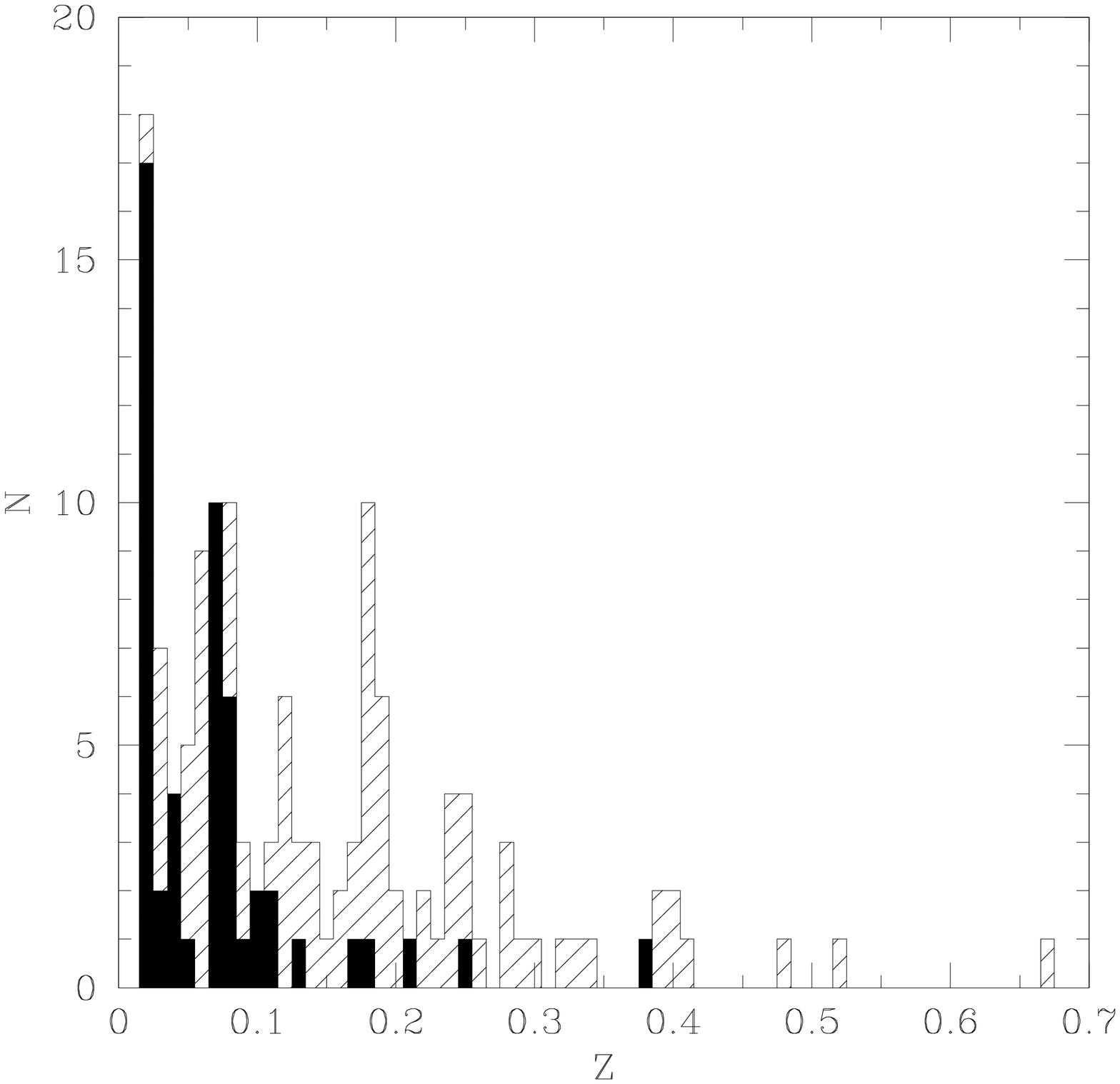,width=80mm}
\epsfig{figure=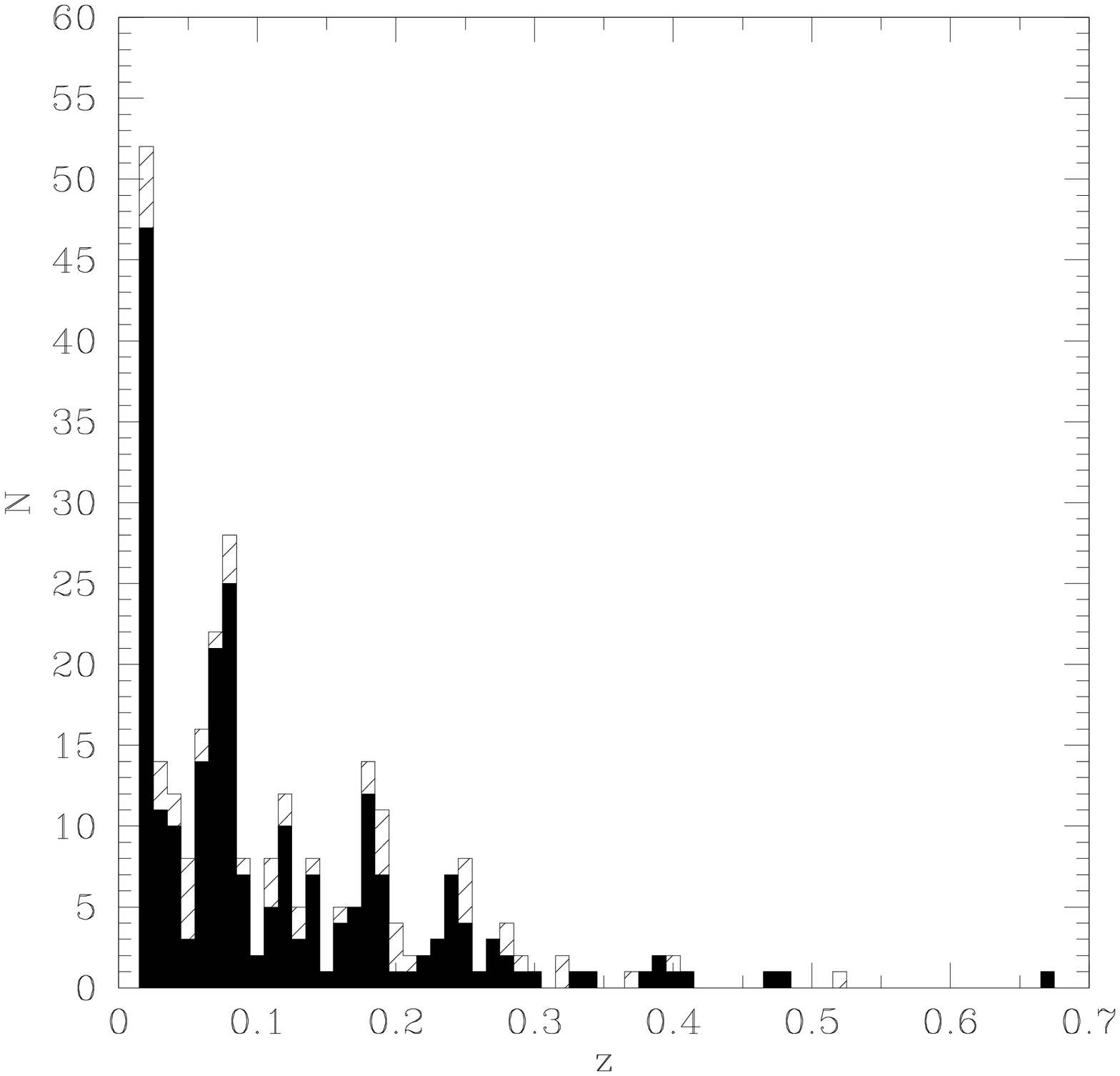,width=80mm}
\caption{The redshift distribution for the new sample (left) and the combined sample (right). 
  LEFT: The hatched area is the new SA57 data, while the superimposed
  heavy shading refers to objects from the Abell 1367 FOCA field.
  RIGHT: The hatched area is the entire sample, while the heavy
  shading refers to the restricted sample of galaxies with only one
  optical counterpart.}
\label{allzplot}
\end{figure*}

Table~7 lists the catalogue for the new observations and the old data.
The overall redshift distribution of the new sample can be seen in
Fig.~\ref{allzplot}, together with the distribution of the enlarged
sample. The distribution in the new sample has a large peak at
$z=0.02$ due to the presence of both the Coma cluster in SA57, and the
cluster Abell 1367.

Absolute magnitudes, $M_{UV}$ and $M_B$, were derived for each galaxy
as follows. The redshift was used to calculate a luminosity distance,
and the dust corrected observed colour, $(\rmn{uv}-\rmn{b})$, to
assign a spectral class and hence $k$ -correction. As in Paper I, the
spectral classes were allocated according to the (E/S0, Sa, Sb, Scd,
SB) scheme using spectral energy distributions (SEDs) from
Poggianti~\shortcite{poggiantised}. The absolute UV magnitude $M_{UV}$ of a
galaxy with dust-corrected UV magnitude $m$, redshift $z$, and
inferred type $i$, is then computed as:

\begin{equation}
M_{UV}=m_{UV} - 5\log d_L(z)-25 - k_{i}(z)
\end{equation}

\noindent
(and a similar relationship for $M_B$) where $d_L(z)$ is the
luminosity distance at redshift $z$ (we assume $\Omega=1$ and
$H_0=100~\rmn{h~km~s^{-1}~Mpc^{-1}}$).  This allows calculation of the
rest-frame colours, \restcol.

\begin{figure*}
\epsfig{figure=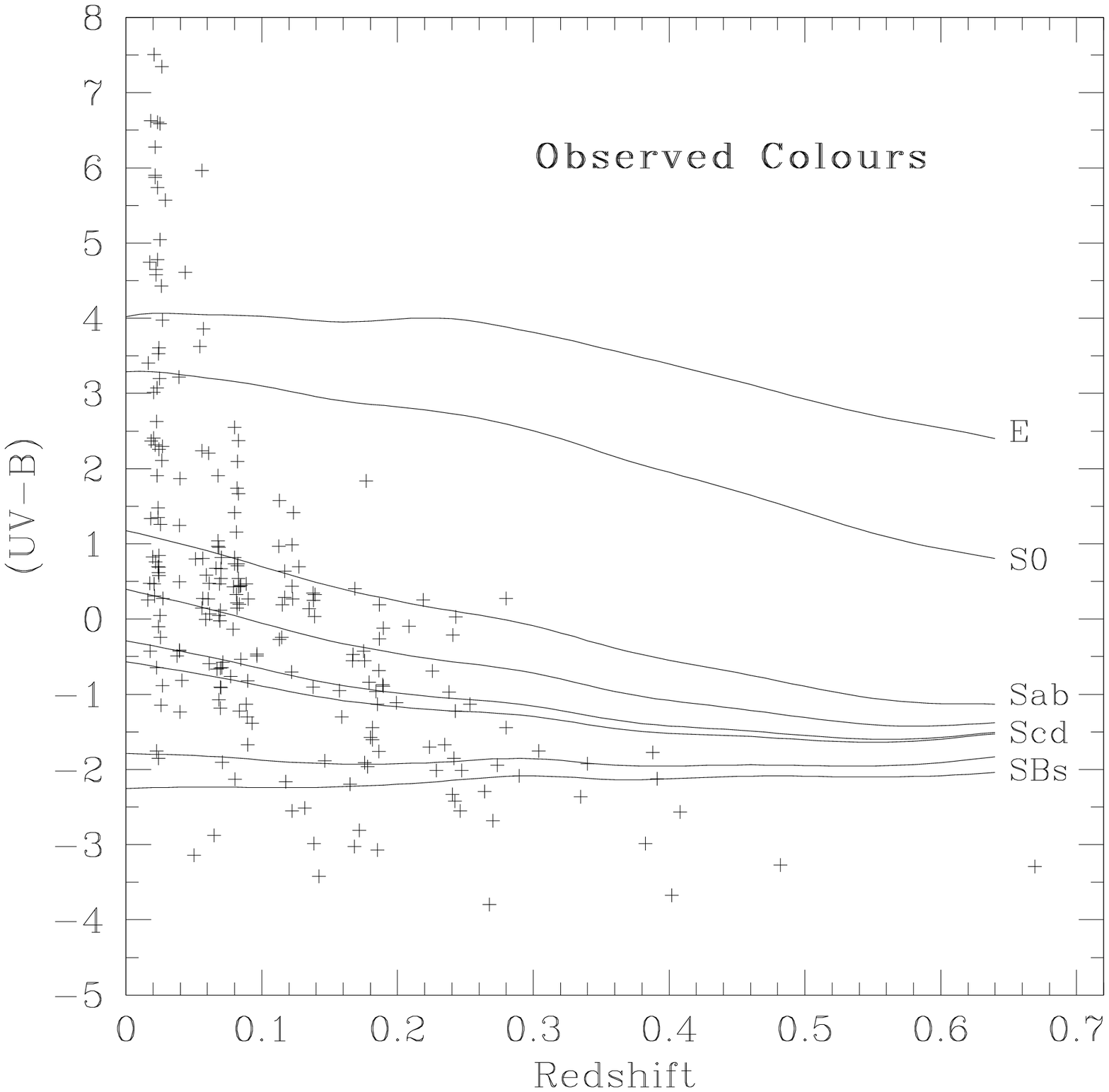,width=80mm}
\epsfig{figure=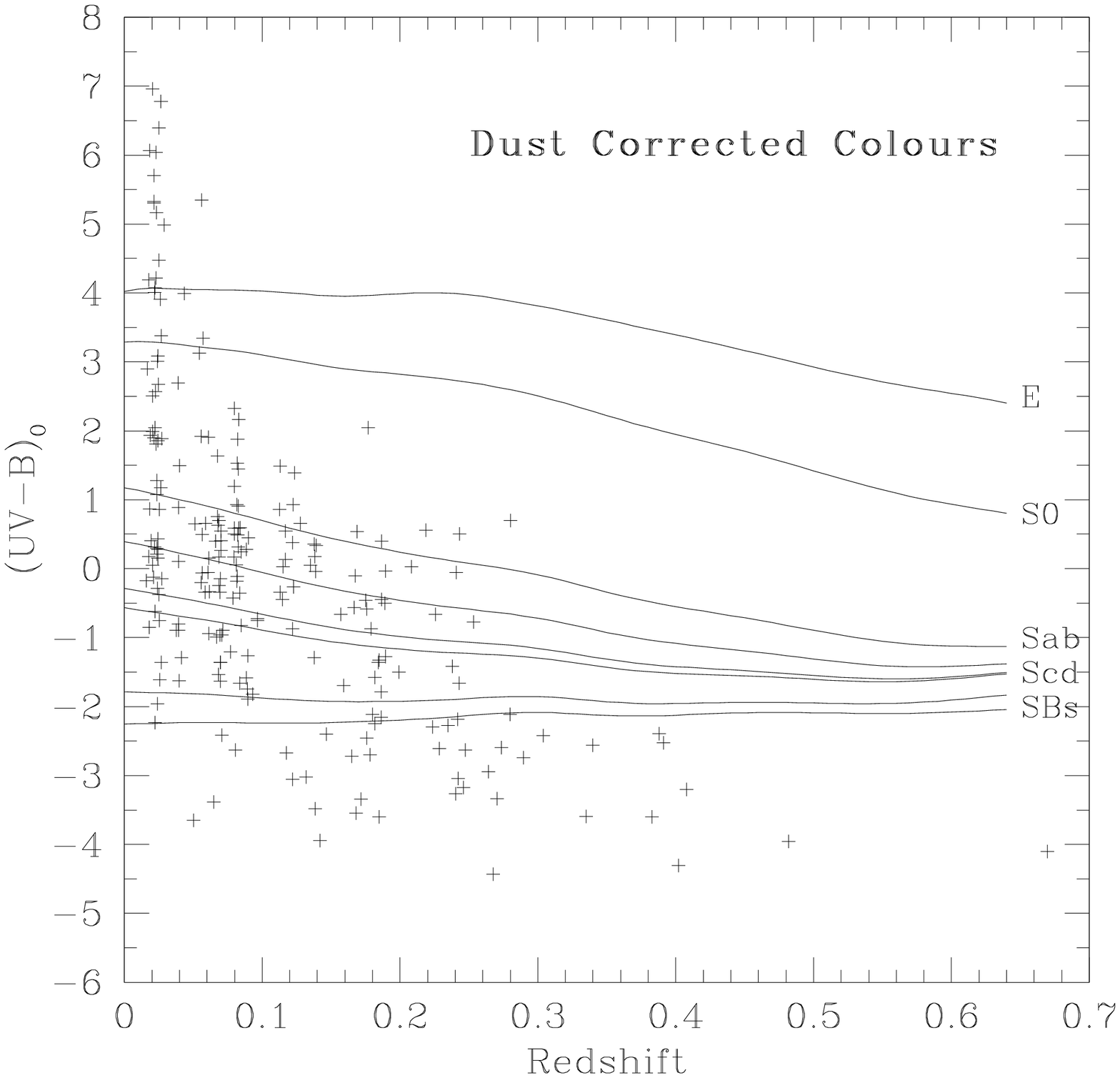,width=80mm}
\caption{The colour-redshift distribution for the full spectroscopic
sample of confirmed galaxies. The photometric system is described in
the text. The lines show the model predictions of galaxy colours as a
function of redshift for a set of SEDs from Poggianti~(1997). These are
the same models as in Paper I; see text for full details of these SED
models. Multiple counterparts are not shown. The left hand plot shows
the colours uncorrected for dust extinction, on the right the colours
have been corrected using the Calzetti law.}

\label{colzplot}
\end{figure*}

Fig.~\ref{colzplot} shows the distribution of the \restcol\ colours
with redshift, with the colours both uncorrected and corrected for
dust.  Multiple counterpart cases are not shown. Superimposed on these
distributions are various model SEDs for different galaxy types as a
function of redshift from Poggianti~\shortcite{poggiantised}.  The
bluer models (labelled SB, as in Paper I) show the colours generated
by a starburst superimposed on a passively evolving system. The redder
(upper) case, SB1, assumes a 100~Myr burst prior to observation
involving 30 per~cent of the galaxy mass. The bluer SB2 burst is a shorter
(10~Myr) but more massive (80 per~cent galaxy mass) burst.

As with previous studies using the FOCA catalogues, including Paper I,
there is a significant fraction of galaxies which have extreme
\restcol\ colours -- in this case 12 per~cent of the uncorrected
colours are bluer than the bluest burst model SED plotted; this
increases to 17 per~cent after our dust correction. The extreme
colours are typically galaxies with strong UV detections, and previous
analysis has shown that a systematic offset between the UV and optical
photometric systems could not produce effects of the size seen in
Fig.~\ref{colzplot} (Paper I). An intriguing possibility is that we do
not see all of the UV galaxies on the APM plates, which are limited to
$\rmn{b}\simeq 21$. This would suggest that we are not seeing the most
extreme objects in the colour plots, as there are no optical
counterparts to some of the UV detections. Only deeper optical images
of our studied fields can settle this issue. Possible explanations for
the objects in our sample with these extreme UV colours will be examined 
in Sections 3 and 4.

\subsection{The UV, \Halpha\ and OII luminosity functions revisited}

We now update the results of Paper I. The availability of emission line
measurements allows us to extend our luminosity functions to those
based on \Halpha\ and \oii\ luminosities as well as the UV flux. With
our enlarged sample which reaches $z\simeq 0.4$, we can also test for
the presence of evolution {\it internally within our own sample}.

We adopt the traditional $\rmn{V}_{\rmn{max}}$ method for the
luminosity function (LF) derivation (e.g. Felten~1977), corrected for
incompleteness in the number-magnitude distribution using the average
number counts of Milliard et al.~\shortcite{milliard}. The
incompleteness function $p(m)$ is defined as the ratio of the number
of galaxies with measured spectra to the total number of UV sources
per magnitude bin per square degree on the sky.

As in Paper I, we removed sources lying in the redshift range of the
intervening Coma and Abell 1367 clusters, which we conservatively take
to occupy $0.020<z<0.027$. We also discarded those with insecure
optical counterparts.  The incompleteness function was computed after
these subtractions, as the galaxy number counts of Milliard et
al.~\shortcite{milliard} are averaged over several fields and therefore
we expect the cluster contamination to be sufficiently diluted.  The
least complete magnitude bin is the faintest ($18\le m\le 18.5$), as
expected, with $p(m)=65\%$.  All other magnitude bins are over 85 per~cent
complete.  The mean incompleteness-corrected $<V/V_{max}>$ is 0.48,
i.e. the galaxy distribution can be considered uniform. The volume
$V_{max}=p(m) V(z_{max})$ is then defined as the
incompleteness-corrected comoving volume at redshift $z_{max}$, out to
which the galaxy could have been observed, i.e. satisfying $m(M,z\le
z_{max})\le 18.5$.

For the \Halpha\ and \oii\ LFs, the incompleteness function was
computed using the new data only, as emission line flux calibration
was not available for the old sample. Line widths were measured for
$\simeq 74$ per~cent of the galaxies in this new sample. However, most of the
`missing' lines probably cannot simply be attributed to a low S/N, and
we must consider them as truly being absent. For this reason, we do
not apply any correction for missing lines to the LFs. Those lines
which are unmeasured due to a low S/N, rather than being absent from
the spectrum, are, by definition, weak, and therefore ignoring them
only adds to the uncertainty at the faint end.

For \Halpha, we also account for the fact that the line
could not be observed at $z>0.4$, i.e.  $z'_{max}={\rm
min}(z_{max},0.4)$.  The \oii\ and \Halpha\ luminosities have been
corrected for extinction as described in Section 2.2. For the UV LF, we
consider both the uncorrected and extinction-corrected magnitudes
following Calzetti's prescription as described in Section 2.3.

Fig.~\ref{colzplot} shows the effect of the reddening corrections on
the colours. After correction, the \restcol\ colours are bluer and
therefore the galaxy types and $k$-corrections, as inferred from the
redshift-colour diagram, will alter slightly. $z_{max}$ will also be
slightly lower, as the extinction increases with the emission
frequency \cite{calzetti97b} and therefore with redshift.  This has
a negligible effect on the emission line LFs.

We fit each luminosity function with a Schechter~\shortcite{schechter}
function in the usual way:

\beq
\phi(L)\d L=\phi^\star \left(L \over L_\star \right)^{\alpha} \exp \left( - {L \over L_\star }\right){\d L \over L_\star}.
\eeq

\noindent
The best fit parameters -- $\phi^\star$, $\alpha$, $M_\star$ for the
UV and log $L_\star$ for the emission lines -- are listed in
Table~\ref{bestfitstable}, as well as the resulting luminosity
densities in each case. These are defined as:

\beq
{\cal L}=\int_0^{\infty} L\phi(L)\d L = \phi_\star L_\star \Gamma(\alpha +2)
\label{leqn}
\eeq

\begin{figure}
\epsfig{figure=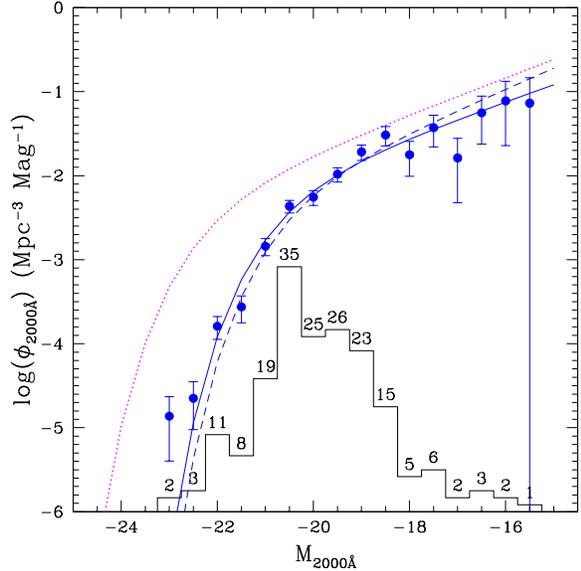,width=80mm}
\caption{The UV luminosity function derived
  from the full sample, with (dotted line) and without (solid line and
  dots) dust extinction correction. The dust correction assumes a
  Calzetti~(1997b) law as described in the text. The long-dashed line
  is the best fit derived from the old sample (Paper I). The histogram
  shows the number of galaxies contributing to each magnitude bin in
  the uncorrected case.}
\label{uvlfplot}
\end{figure}

\begin{figure}
\epsfig{figure=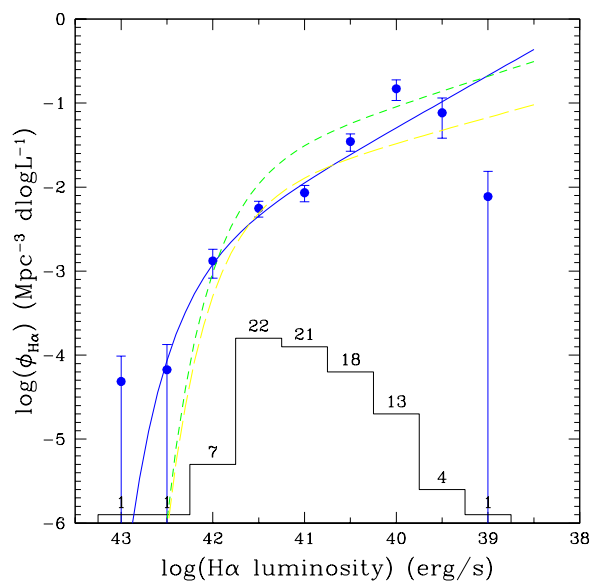,width=80mm}
\caption{The dust corrected \Halpha\ luminosity function derived from 
  the present sample. Our best fit is shown by the solid line.  The
  short-dashed line is the \Halpha\ LF derived by Tresse \&
  Maddox~(1998) in a similar redshift range, while the long-dashed
  line shows the $z\simeq 0$ estimate of Gallego et al.~(1995).}
\label{halfplot}
\end{figure}

\begin{figure}
\epsfig{figure=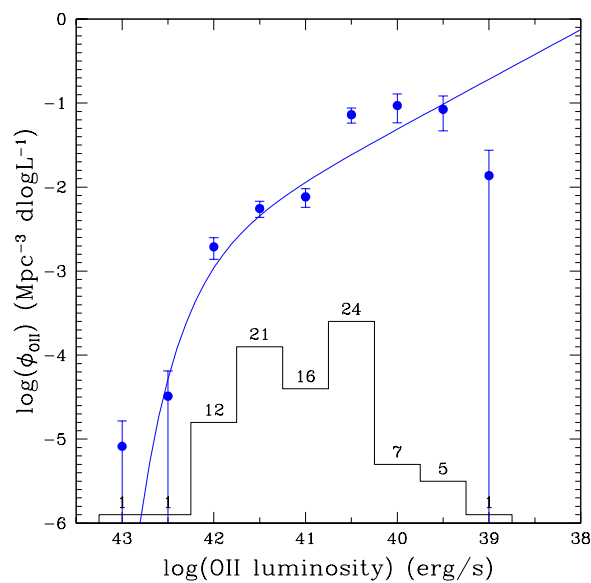,width=80mm}
\caption{The dust corrected \oii\ luminosity function derived from the present
  sample (dots). Our best fit is shown by the solid line.}
\label{o2lfplot}
\end{figure}

\noindent
The error bars are Poissonian.  The four LFs are shown in
Figs.~\ref{uvlfplot},\ref{halfplot} and \ref{o2lfplot}; we defer discussion of these
to $\S$2.5.

\begin{figure}
\epsfig{figure=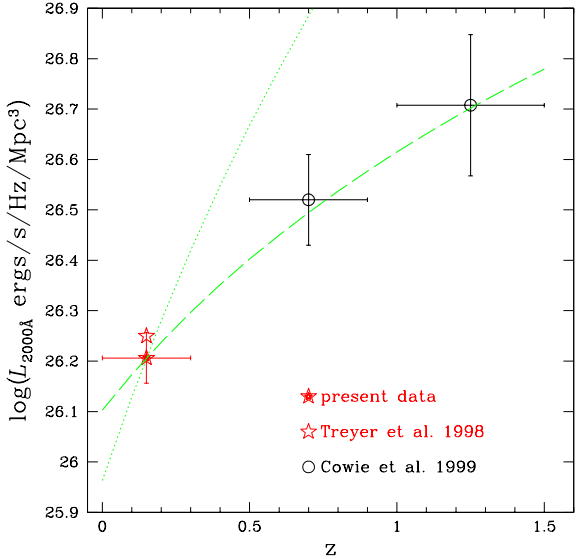,width=80mm}
\caption{The 2000~\AA\ luminosity density as a function of redshift. 
  Empty dots are from Cowie, Songaila \& Barger~(1999) extrapolated
  assuming a faint end slope of -1.5 similar to the present low
  redshift estimate. The dashed line shows the $(1+z)^{1.7}$ evolution
  law derived by Cowie et al. The dotted line is a $(1+z)^{4}$
  evolution law based on CFRS (Lilly et al.~1996). Both analytic
  trends are normalised at our new UV estimate; none of the data
  points are dust corrected.}
\label{cowiecfplot}
\end{figure}

\begin{table*}
\begin{tabular}{lrrrr}
\hline \hline
Parameter & UV uncorrected & UV dust corrected & \Halpha & \oii\\
\hline
$\alpha$ & $-1.51 \pm 0.10$ & $-1.55 \pm 0.11$ & $-1.62 \pm 0.10$  & $-1.59 \pm 0.12$\\ 
$M_\star$/log $L_\star$ (cgs) & $-20.59\pm 0.13$ & $-22.14\pm 0.20$ & $42.05 \pm 0.14$ & $41.96 \pm 0.09$\\
log $\phi_\star$ (Mpc$^{-3}$) & $-2.02 \pm 0.11$ & $-2.15 \pm 0.14$ & $-2.92 \pm 0.20$ & $-2.82 \pm 0.18$\\
log $\cal L$ (cgs Mpc$^{-3}$) & $38.08 \pm 0.05$ & $38.61 \pm 0.05$ & $39.49 \pm 0.06$ & $39.46 \pm 0.06$\\
SFR (PEGASE $f=1$) & $-1.52,-1.69,-1.74\pm 0.05$ & $-0.99,-1.16,-1.21\pm 0.05$ & $-1.56 \pm 0.06$ & $-1.62 \pm 0.05$\\
SFR (PEGASE $f=0.7$)& $-1.51,-1.68,-1.74\pm 0.05$ & $-0.98,-1.15,-1.21\pm 0.05$ & $-1.43 \pm 0.06$ & $-1.47 \pm 0.05$\\
\hline \hline
\end{tabular}
\caption{Parameters of the best fit Schechter functions for the various luminosity functions. $\cal L$ is the corresponding luminosity density integrated to infinitely faint magnitude. The cgs units are erg/s/\AA\ for the UV luminosity at 2000~\AA\ and erg/s for the \Halpha\ and \oii\ luminosities. The star-formation rates (SFRs) are derived assuming a Salpeter IMF and the PEGASE code (see section 4 for details). We consider two cases for the fraction of Lyman continuum photons reprocessed into recombination lines ($f=1$ and $f=0.7$ respectively). The three SFRs listed for the UV light densities are taken at 3 different ages of a constant SFH stellar population; the conversion factors are listed Table ~\ref{sfrtable} (first two lines).}
\label{bestfitstable}
\end{table*}

Fig.~\ref{cowiecfplot} shows the dust uncorrected 2000~\AA\ luminosity
density -- ${\cal L}$(2000~\AA) -- as a function of redshift. The high
redshift points are from Cowie, Songaila \& Barger~\shortcite{cowie},
although, unlike the authors, we do not assume a faint magnitude
cutoff.  For consistency, we integrated the Cowie et al. LFs to
infinity (Eqn.~\ref{leqn}) assuming a faint end slope of -1.5 similar
to the present low redshift estimate.  The dashed line shows the
$(1+z)^{1.7}$ luminosity evolution derived by Cowie et al., normalised
at our new UV estimate.  As thoroughly discussed by these authors,
this evolution is much less radical than the one derived from the CFRS
analysis of Lilly et al.~\shortcite{lilly96} (dotted line, similarly
normalised), implying much more star-formation has occured in recent
times than previously suspected.  In particular, the strong peak in
SFR at $z\sim 1-2$ may have been overestimated.

We looked for traces of evolution in the present sample by computing
the UV LF in two redshift bins: $[0-0.15]$ and $[0.15-0.4]$. $V_{max}$
is then defined as min$(V(0.15),V(z_{max}))$ for the low redshift
galaxies, and as $V({\rm min}(0.4,z_{max}))-V(0.15)$ for the higher
redshift bin. The mean redshifts in each bin are 0.078 and 0.22
respectively. The low and high redshift LFs overlap around $M_\star$
and both are consistent with the best fit derived for the full sample.
Therefore no statistically significant evolution can be seen in the
present data. However, the increase in light density between the mean
redshifts of the two samples expected from a $(1+z)^{1.7}$ evolution
law, as derived by Cowie et al., is only a factor of 1.2 -- within the
error bars of the present estimate. By contrast, the $(1+z)^{4}$
evolution law based on the CFRS by Lilly et al.~\shortcite{lilly96}
predicts a 60 per~cent increase in UV light density between the two
redshift bins, which is difficult to reconcile with our statistics,
assuming the Poisson fluctuations are the dominant source of
uncertainty.  Although the present data do not allow a very reliable
conclusion on this point, a weak rate of evolution for the UV light
density seems more likely.

\subsection{The low-redshift star-formation rate}

The uncorrected UV LF is in good agreement with the estimate of Paper
I, although the latter was based on a third of the present number of
redshifts. The steep faint end slope remains a significant feature, in
contrast with local optically-selected surveys. It is also apparent in
the \Halpha\ and \oii\ LFs, although the faintest data points were
excluded in both cases and the fits are relatively poor. A steep faint
end slope is also found in the 1.4~GHz LF derived from faint radio
galaxies \cite{mobasher,serjeant} confirming the preponderance of
star-forming galaxies among this population.  The shape of our
\Halpha\ LF is in poor agreement with previous low-redshift
determinations (Gallego et al.~1995, Tresse \& Maddox~1998), although
given the large uncertainties in emission line measurements, the fact
that the three estimates derive from very different selection
criteria, and that they probe different redshift ranges, the
discrepancy is probably acceptable.  Our integrated \Halpha\ 
luminosity density is $\sim 43$ per~cent lower than the Tresse \&
Maddox~\shortcite{tresself} value derived from a sample of $I$-band
selected galaxies at $z < 0.3$ from the CFRS. The mean redshift of
this sample is 0.2. Truncating the present UV-selected sample at
redshift 0.3 leaves the best fit \Halpha\ LF practically unchanged,
while slightly reducing the mean redshift to 0.12. The discrepancy
still cannot be reasonably attributed to evolution within such a short
redshift range, rather to poor statistics, differing selection effects
and possibly $k$-correction models (used in computing $V_{max}$). Our
\Halpha\ luminosity density is also $\sim 27$ per~cent higher than the
estimate of Gallego et al.~\shortcite{gallego}, which probes a more
local ($z=0$), and also probably more comparable (\Halpha\ selected),
galaxy population.  The rate of evolution resulting from the latter
discrepancy (from $z=0$ to 0.15) is actually in very good agreement
with that derived by Cowie et al.~\shortcite{cowie} from the 2000~\AA\ 
light density.

The conversion from UV, \Halpha\ or \oii\ luminosity densities into
SFRs is very model dependent (see Section 4 for a detailed
discussion).  As an illustration, we use the PEGASE stellar population
synthesis code \cite{frv} with which we were able to derive the
conversion rates for all three diagnostics self-consistently.  We
assume a Salpeter IMF with stellar masses ranging from
0.1~--~120~${\rm M}_\odot$, and consider two cases for the fraction of
Lyman continuum photons reprocessed into recombination lines ($f=1$
and $f=0.7$ respectively).  These models are described in detail in
Section 4.  The conversion factors are listed in Table~\ref{sfrtable}
(first two lines).

The \Halpha\ and \oii\ luminosity densities thus converted into SFRs
give very consistent results.  The conversion factors we use here
yield SFRs~$\sim$~30 to 40 per~cent higher than those derived from Madau
(1998)'s fiducial model based on the stellar population synthesis code
of Bruzual \& Charlot (1993) (for a similar IMF).  As our H$\alpha$
luminosity density falls between the values of Gallego et
al.~\shortcite{gallego} and of Tresse \& Maddox~\shortcite{tresself},
so does the resulting SFR for a given stellar population synthesis
model.

\begin{figure}
\epsfig{figure=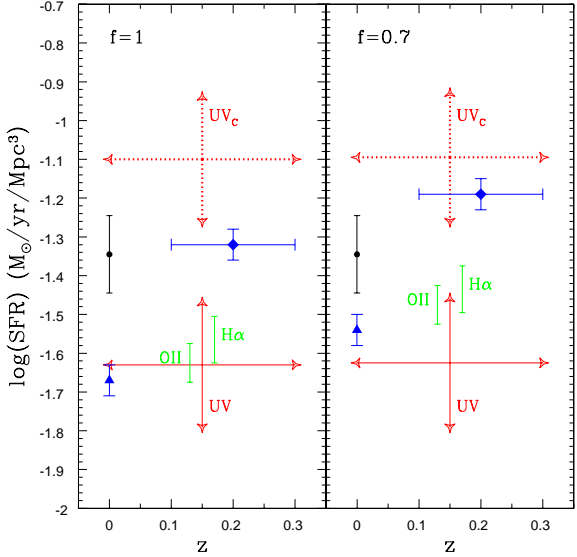,width=80mm}
\caption{The local star-formation rate derived from the 
  present and other local surveys, assuming the conversion factors
  described in the text (previous H$\alpha$ estimates are rescaled
  according to these factors). All the emission line estimates are
  dust-corrected as described in the text.  The triangle represents
  the \Halpha\ estimate of Gallego et al.~(1995) and the diamond that
  of Tresse \& Maddox~\shortcite{tresself}. The present estimates are
  indicated by error bars. The lower cross shows the range of SFRs
  estimated from the uncorrected UV continuum light (assuming the
  range of models described in the text), while the upper dotted one
  shows the dust-corrected values.  The \Halpha\ and \oii\ estimates
  are where indicated on the plot.  The $z=0$ dot shows the SFR from
  the 1.4~GHz analysis of Serjeant et al.~\shortcite{serjeant}.}
\label{sfrplot}
\end{figure}

Converting UV light into an instantaneous SFR is less straightforward
as it involves an uncertain contribution from longer-lived stars,
adding to the already large uncertainty in the dust corrections. We
consider three different ages (in a constant star-formation history) to derive the UV
conversion factors; 10, 100 and 1000~Myr (see Section 4 for details --
the conversion factors are as listed in Table~\ref{sfrtable}).  The
range of SFRs thus derived from the 2000~\AA\ light density is shown
in Fig.~\ref{sfrplot}, along with the present (dust-corrected) and
previous \Halpha\ and \oii\ estimates. The left and right panels
assume $f=1$ and $f=0.7$ respectively.  Also shown is the local SFR
estimate recently derived from 1.4~GHz data \cite{serjeant,mobasher}.

The SFR derived from the UV continuum \textit{uncorrected} for dust
extinction is in good agreement with the \textit{corrected} \Halpha\ 
and \oii\ estimates for the case $f=1$, and slightly lower than these
values for $f=0.7$.  Taking the emission line estimates at face value,
this suggests that local UV-selected galaxies are not significantly
affected by dust and that Calzetti's extinction law in the UV is
significantly overestimated \textit{for this population}. There are
many caveats however, not least of which is the model-dependency of
the conversion factors from line/UV luminosities to SFRs (see, for
example, Schaerer~\shortcite{schaerer}). The \Halpha\ extinction
corrections may be underestimated, as argued by Serjeant et
al.~\shortcite{serjeant} based on their estimate of the local SFR from
radio emission, which is dust insensitive. Our Balmer-derived dust
corrections to the \Halpha\ and \oii\ lines are certainly likely to be
lower limits, as discussed in Section~2.2. It is also possible that
the \Halpha\ and UV luminosities are measured over different effective
apertures, though we consider this unlikely (see Section~4.2 for a
discussion of this point).

The large uncertainties involved in determining the low redshift SFR
are readily apparent from the large scatter both in the data and in
the models.  These uncertainties tend to increase with redshift,
making interpretations about the SFR evolution quite unreliable at
this point. Understanding the detailed physical mechanisms of
star-formation is therefore a crucial task towards reconciling the
various SF diagnostics and finally drawing conclusions about the
nature of star formation in the nearby Universe.

\section{Emission line properties}

A significant advance over the spectra presented in Paper I is that we
now have reliable line measurements for a substantial fraction of the
UV-selected sample. For the highest S/N spectra, EWs and fluxes for up
to 5 emission lines have been measured (6 including the deblended
\nii\ line). Our analysis now proceeds in two parts. This section will
cover the emission line properties and correlations, together with
diagnostic diagrams, whilst the next section will cover comparisons of
emission lines with the UV fluxes and the subsequent star-formation
modelling.

One of the possible explanations for the abundance of extreme
\restcol\ colour objects seen in this survey is that the UV light
produced in these galaxies comes from a non-thermal source, such as an
QSO/AGN. The mean (not dust-corrected) $\rmn{UV}-\rmn{B}$ colour of the
23 such objects in our sample is -1.41, though there is large scatter,
with some as blue as $\simeq -4$. Clearly, care must be taken to
remove such objects from our `star-forming' galaxy sample. Those
galaxies with obvious AGN characteristics have been removed from the
sample; however, there remains the possibility that AGNs with strong
star-forming components have remained in the sample, giving
over-abundant UV fluxes. While the best way to assess the size of the
effect is to image the galaxies in the UV and look at the distribution
of the UV light, an indirect method of identifying AGN from starburst
galaxies is to use emission line diagnostic diagrams.

\begin{figure*}
\epsfig{figure=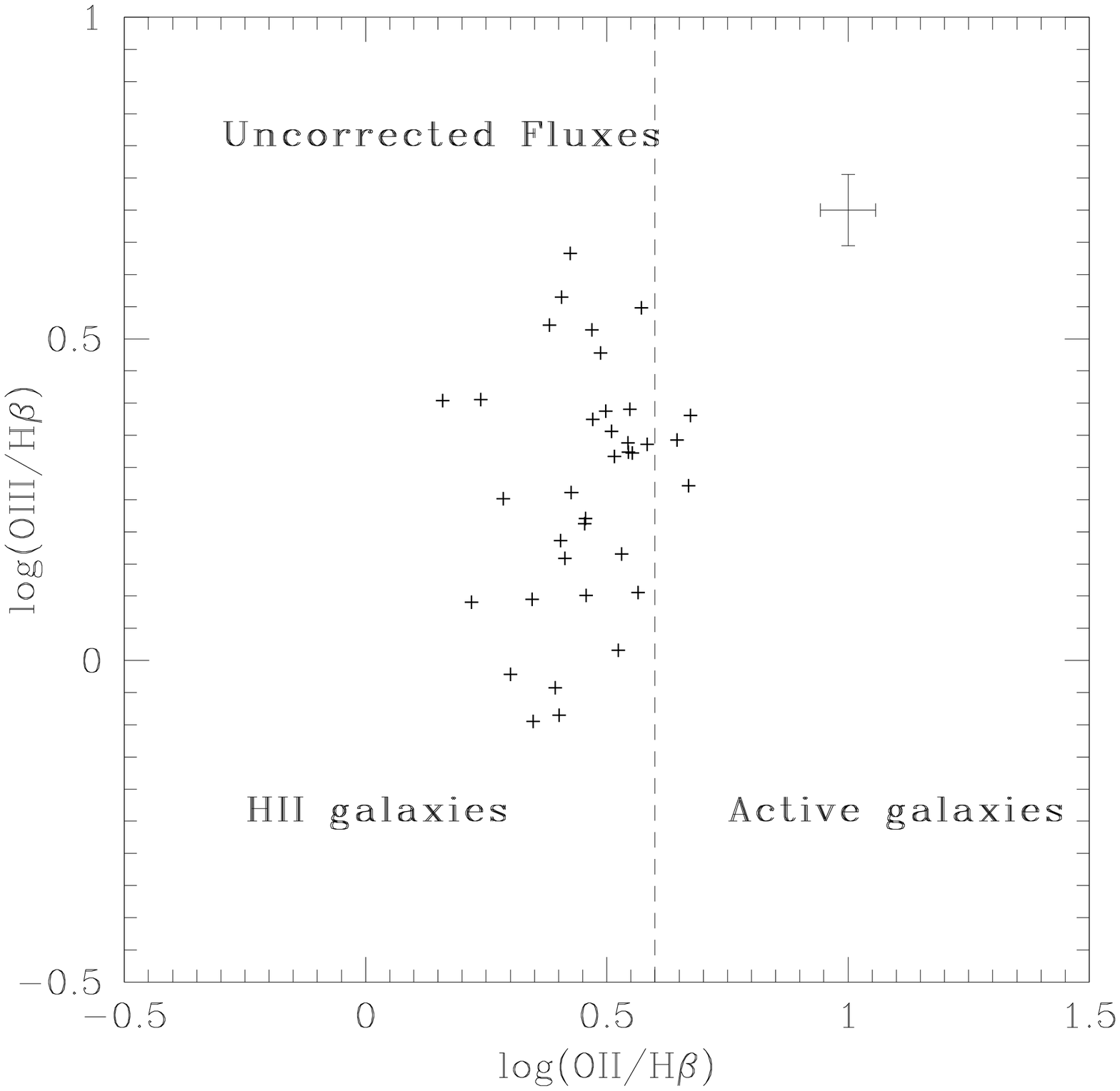,width=80mm}
\epsfig{figure=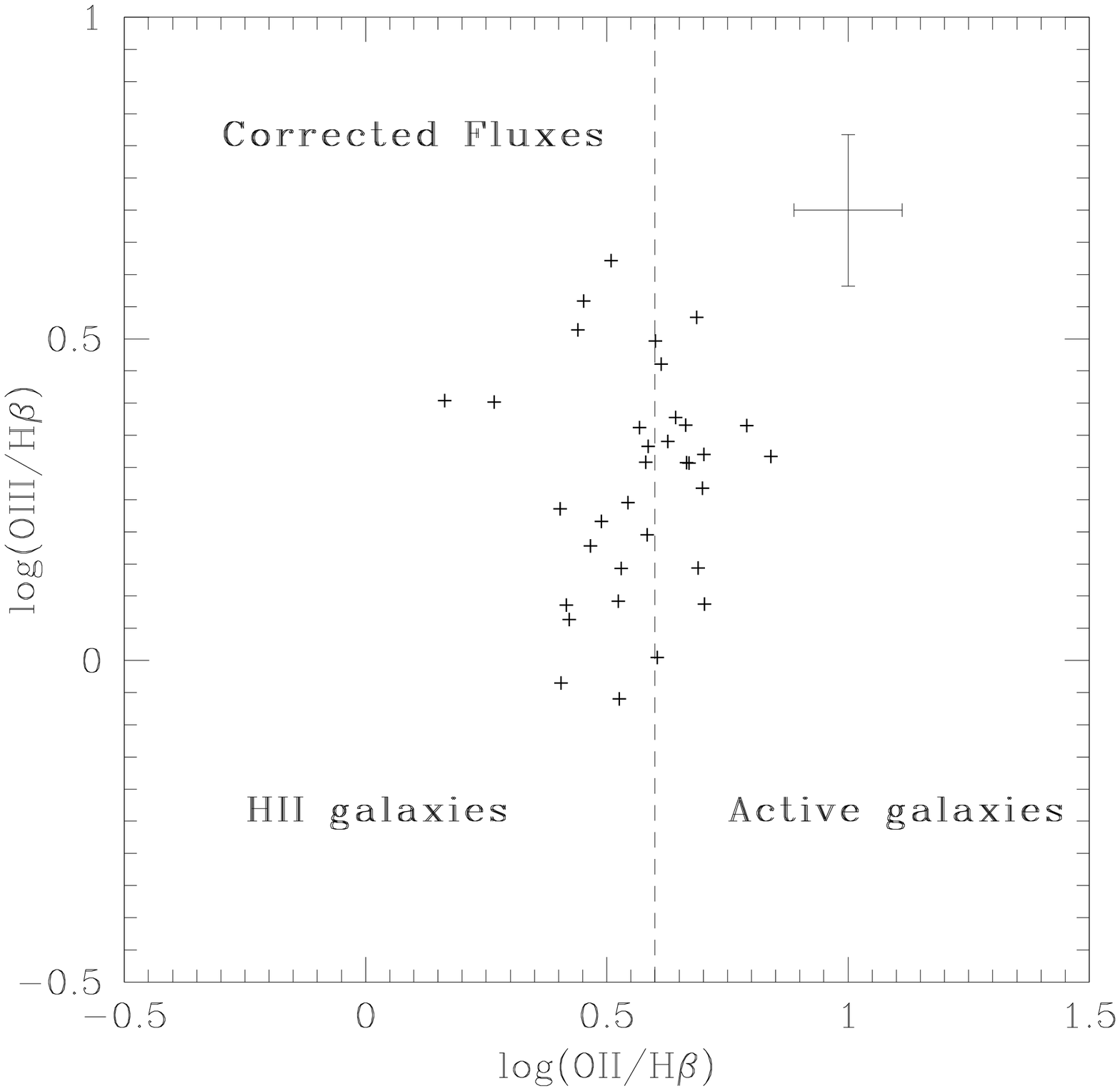,width=80mm}
\caption{The diagnostic diagrams as described in the text, for both the corrected and uncorrected fluxes. Typical $1\sigma$ error bars are shown in the top right. Note how the reddening corrections move many more galaxies away from the \hii\ region, though the error bars are larger.}
\label{diagplot}
\end{figure*}

Diagnostic diagrams can be used to separate and distinguish between
different ionisation sources in the host galaxies. Considering the lines
measured in this survey, the \oiii/\Hbeta\ versus \oii/\Hbeta\ diagram
is the most appropriate to use. Though not an ideal choice, as the
ratio of \oii/\Hbeta\ depends significantly on reddening, this diagram
allows us to look at what proportion of our sources may have
ionisation sources other than hot OB-type stars, which is vital given
the star-formation modelling attempted in Section 4.
Fig.~\ref{diagplot} shows this diagnostic diagram for both the
reddened and unreddened fluxes. The line on the diagram is taken from
Tresse et al.~\shortcite{tressecfrs}, and shows the approximate
empirical limit between \hii\ galaxies and `active' galaxies.

It is interesting to note that a significant fraction of the sources
lie to the right of the line in Fig.~\ref{diagplot}, i.e. away from
the region that is normally associated with \hii\ galaxies. The
important point however is the size of the uncertainties on
the plot, particularly in the corrected fluxes. Though not shown on
the diagram, the points to the right of the line typically have larger
errors -- up to 2 times as high -- so it is difficult to conclude
that a large fraction of the sources have a strong AGN component.
Additionally, the effect of stellar absorption on the \Hbeta\ line has
a large effect on this plot, adding to the uncertainty involved.
Although the \Hbeta\ fluxes include some allowance for the \Hbeta\
absorption due to the way in which they were measured, some individual
points may still contain significant errors associated with them.  Any
further correction will increase the measured \Hbeta\ flux, moving the
galaxy down and to the left on the diagram, back into the \hii\ galaxy
region. 

If galaxies that lie away from the \hii\ region are responsible for the
extreme \restcol\  colours seen in this survey, then we expect a
correlation between the \oii/\Hbeta\ ratio and UV-B colour; none is
found. This suggests that an explanation of the \restcol\ colours
cannot be found purely in the source of ionisation of the galaxies, at
least not with the current quality of data.

Another possible explanation for the anomalous UV colours compared
to the model predictions could be unusually low metallicities in
the galaxies concerned. To investigate this, we examined the metallicity 
using the $R_{23}$ emission line index \cite{pagel} following the
prescription from Poggianti et al.~\shortcite{poggianti99}. This 
is defined in terms of corrected fluxes as:

\begin{equation}
R_{23}=\frac{(\rmn{\oii}_{3727}+\rmn{\oiii}_{4959,5007})}{\rmn{H}_{\beta}}
\label{r23eqn}
\end{equation}

\noindent
and is calibrated using:

\begin{equation}
12+\log(\frac{\rmn{O}}{\rmn{H}})=9.265-0.33x-0.20x^2-0.21x^3-0.33x^4
\label{r23calibeqn}
\end{equation}

\noindent
where $x=\log(R_{23})$. Full details of this calibration can be found
in Zaritsky, Kennicutt \& Huchra~\shortcite{zaritsky}; in brief, the
absolute $R_{23}$ index calibration is accurate only to $\simeq
0.2~\rmn{dex}$, so this estimator is most useful for calculating
relative metallicities.

Only the SA57 galaxies -- the more complete spectroscopic sample --
were used in this analysis, creating a sub-sample of 35 galaxies that
have the complete line information required to estimate the
metallicity. An added complication is the effect of stellar absorption
on the \Hbeta\ line, as the metallicity estimates are very sensitive
to this; however, due to the emission line nature of our survey
galaxies, and they way in which the \Hbeta\ fluxes were measured, the
effect of stellar absorption should not be a large one.

\begin{figure}
\epsfig{figure=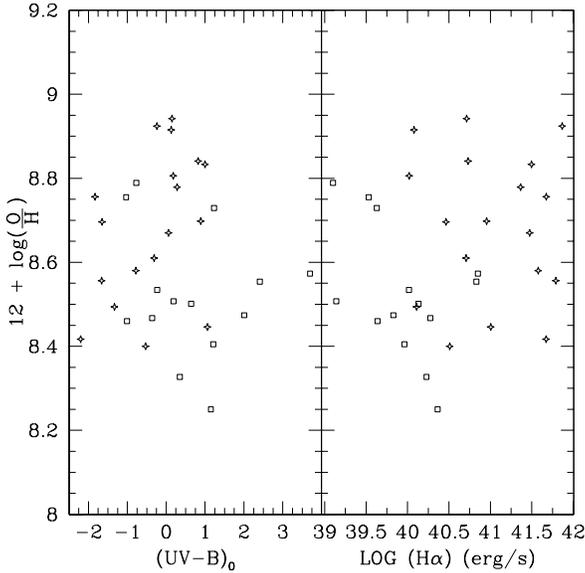,width=80mm}
\caption{Plot of the metallicity derived from the R$_{23}$ index against \restcol\ and \Halpha\ luminosity for the SA57 galaxies which have the complete line information. Coma galaxies are plotted as squares, field galaxies are stars.}
\label{colmetalplot}
\end{figure}

Fig.~\ref{colmetalplot} shows the relationship between galaxy colour,
\Halpha\ luminosity and metallicity. Those galaxies which are members
of the Coma cluster are plotted as empty squares, the field galaxies
are shown as stars. There is a slight suggestion in the left hand plot
that the bluer galaxies have a higher metallicity (a
counter-intuitive result). There is no apparent correlation with
\Halpha\ luminosity. Unfortunately, due to the significant errors on
the points (not shown) it is difficult to conclude a great deal about
metallicity being responsible for the extreme \restcol\ colours seen
in some of our galaxies.

To conclude, there is still no single convincing explanation for the
abnormal colours that we see in this survey on the basis of the
emission line information that we have at our disposal. While it is
difficult to completely rule out the possibility of significant AGN
contamination, or metallicity effects, the evidence is not strong.
Improved S/N ratios must be obtained on the relevant lines to finally
settle the issue.

\section{Star Formation Modelling}

\subsection{Discussion of spectral evolution models}

Our UV selected galaxy survey allows us to compare two different
tracers of star-formation activity which should, when converted,
produce similar SFRs.  The follow-up optical spectra have provided
\Halpha\ emission line measurements, and the FOCA experiment has
provided a measurement of the UV continuum at 2000~\AA. This UV
continuum light is dominated by short-lived, massive main-sequence
stars, with the number of these stars proportional to the SFR.
\Halpha\ emission lines are generated from re-processed ionising UV
radiation at wavelengths of less than 912~\AA. This radiation is only
produced by the most massive stars, which have short lifetimes of
$\simeq$ 20~Myr.  To convert these two tracers into actual SFRs for each
galaxy requires constructing the spectral energy distribution (SED) of
a model galaxy over time, and the following method is used.

The spectral characteristics of an instantaneous burst of
star-formation for a given IMF and set of evolutionary stellar tracks
are calculated, and, using a time-dependent SFR, then used to build
synthetic spectra over the course of a galaxy's history. This
time-varying SED is converted into time-varying UV magnitudes and
\Halpha\ luminosities (or EWs). For the UV magnitudes, the response
(or transmission) of the FOCA-2000~\AA\ filter is included. For \Halpha,
the emission line flux can be calculated from the number of ionising
Lyman continuum photons, assuming that a certain fraction of these
photons are absorbed by the hydrogen gas in the galaxy. This gas is
assumed to be optically thick to the Lyman photons (according to case
B recombination).

The number of ionising photons is assumed to be a fraction $f$ of the
number of Lyman continuum photons, and two values will be assumed here.
The first is $f=0.7$, as proposed by
DeGioia-Eastwood~\shortcite{eastwood}, who studied \hii\ regions in the
LMC. However, when the galaxy as a whole is studied, the fraction
absorbed is much higher (see Kennicutt~\shortcite{kennicuttreview} for
a discussion). Leitherer et al.~\shortcite{leitherer95} studied the
redshifted Lyman continuum in a sample of 4 starburst galaxies with the
Hopkins UV telescope, and reports that $<3$ per cent are not absorbed,
i.e. $f=1$.  These absorbed ionising photons are then re-processed into
recombination lines; for \Halpha\ we adopt a conversion of 0.45
\Halpha\ photons per ionising Lyman photon.

There are several galaxy spectral synthesis models available in the
literature (see Leitherer et al.~\shortcite{leitherer} for a recent
review). Different models predict slightly
different time-varying SEDs, and therefore different conversion
factors. For \Halpha, produced only by the most massive
stars, a constant star-formation history (SFH) will have little
time-dependence in \Halpha\ luminosity -- for the published models, the
\Halpha\ luminosity reaches a constant level after $\simeq
20-30~\rmn{Myr}$, and varies little thereafter.  However, the
conversion factor is very sensitive to the form of the IMF, as it
depends critically on the number of high-mass stars. The conversion
from UV continuum luminosity to a star-formation rate (SFR) is more
difficult, as the UV continuum at 2000~\AA\ also contains a
contribution from stars with a longer lifetime, and, unlike \Halpha,
will not settle at a constant level but will increase slowly with
time. Three different conversion factors will be used here, for ages
of 10, 100 and 1000~Myr.

There are many \Halpha\ conversion factors given in the literature
(e.g. Glazebrook et al.~\shortcite{glazebrook}, Kennicutt et
al.~\shortcite{kennicuttreview} and Madau et al.~\shortcite{m98} for
discussions). The modelling in this paper makes use of two spectral
synthesis codes: PEGASE (see Fioc and Rocca-Volmerange~\shortcite{frv} for
further details), and Starburst 99, developed by Leitherer et
al.~\shortcite{starburst}. The PEGASE code uses the evolutionary tracks
of Bressan et al.~\shortcite{bressan}, together with the stellar
spectral libraries assembled by Fioc and Rocca-Volmerange, and is
limited to solar metallicity. These libraries cover the wavelength
range of 200~\AA\ to $10\mu$m with a resolution of 10~\AA . Full details
of the tracks used in the Starburst 99 code can be found in Leitherer
et al.~\shortcite{starburst}. This code has a choice of 5
metallicities ($\rmn{Z}=0.040$, 0.020 ($=\rmn{Z}_{\odot}$), 0.008,
0.004 and 0.001).

The \Halpha\ and UV conversion values used are taken from Madau et
al.~\shortcite{m98}, Kennicutt~\shortcite{kennicuttreview} and
directly from the PEGASE and Starburst 99 spectral synthesis codes,
and tabulated in Table~\ref{sfrtable} for a SFR of
$1\rmn{M_{\odot}yr}^{-1}$ and a Salpeter~\shortcite{salpeter} IMF. The
table also lists various \oii\ conversion factors used in Section 2.4.
While the UV continuum is relatively insensitive to the fraction of
Lyman ionising photons absorbed by the nebular gas, the \Halpha\ 
fluxes are critically sensitive to this poorly known number; values of
$f=0.7$ and $f=1$ were used to produce the two PEGASE values. The SB99
UV conversion factors do not increase with decreasing metallicity for
the 10~Myr case (as would be expected) due to red supergiant (RSG)
features appearing in the population; this metallicity dependent
feature is due to the inability of most evolutionary models to predict
correctly RSG properties on this time-scale (see Leitherer et
al.~\shortcite{starburst} for a full discussion).

\begin{table*}
\begin{center}
\begin{tabular}{ccccccc} \hline \hline
Source & Assumed Salpeter IMF & L(\Halpha) & L(\oii) & \multicolumn{3}{c}{L(UV$_{2000}$)} \\
 & lower/upper mass limits & ($10^{41} \rmn{erg~s}^{-1}$) & ($10^{41} \rmn{erg~s}^{-1}$) & \multicolumn{3}{c}{($10^{39} \rmn{erg~s}^{-1} \AA^{-1}$)} \\
 & ($\rmn{M}_{\odot}$) & & & 10 Myr & 100 Myr & 1000 Myr\\
\hline
PEGASE ($f=1.0$) & 0.1/120 & 1.15 & 1.25 & 4.01 & 5.90 & 6.69 \\
PEGASE ($f=0.7$) & 0.1/120 & 0.85 & 0.88 & 3.94 & 5.84 & 6.63 \\
SB99 (Z=0.040) & 0.1/120 & 1.23 & & 3.53 & 5.01 & \\
SB99 (Z=0.020) & 0.1/120 & 1.53 & & 3.44 & 5.17 & \\
SB99 (Z=0.004) & 0.1/120 & 1.79 & & 3.55 & 5.69 & \\
SB99 (Z=0.001) & 0.1/120 & 2.01 & & 3.46 & 5.85 & \\
M98 & 0.1/125 & 1.58 & & & 6.00 &\\
K98 & 0.1/100 & 1.27 & 0.71 & & &\\
\hline \hline
\end{tabular}
\end{center}
\caption{The conversion rates used to transform \Halpha\, \oii\ and FOCA UV$_{2000}$ luminosities into SFRs (in the sense $\rmn{L}=SFR\times \rmn{conversion~factor}$). Values from Madau et al.~\shortcite{m98}, Kennicutt~\shortcite{kennicuttreview}, and the PEGASE and Starburst 99 spectral synthesis models.}
\label{sfrtable}
\end{table*}

\begin{figure*}
\epsfig{figure=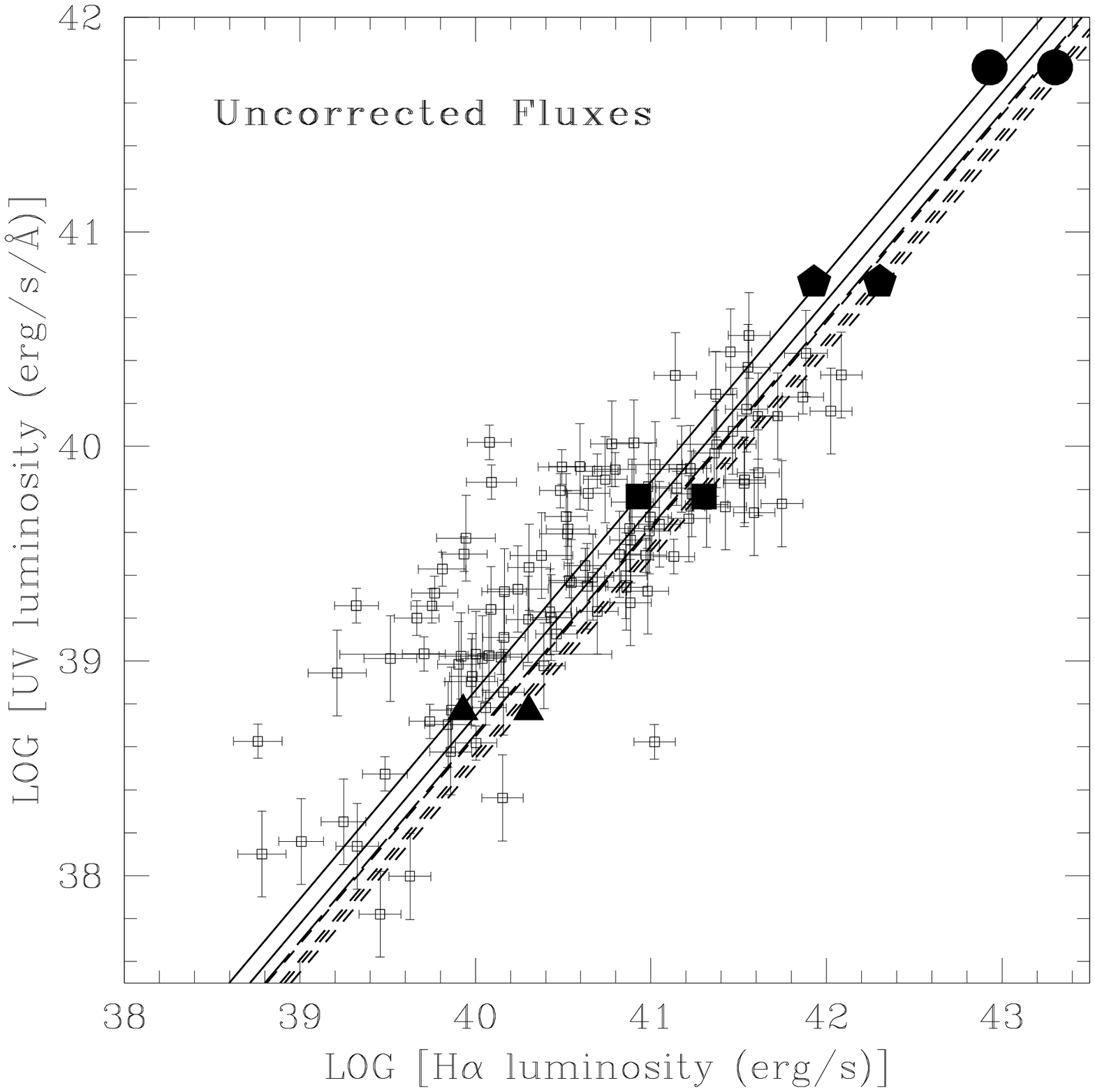,width=80mm}
\epsfig{figure=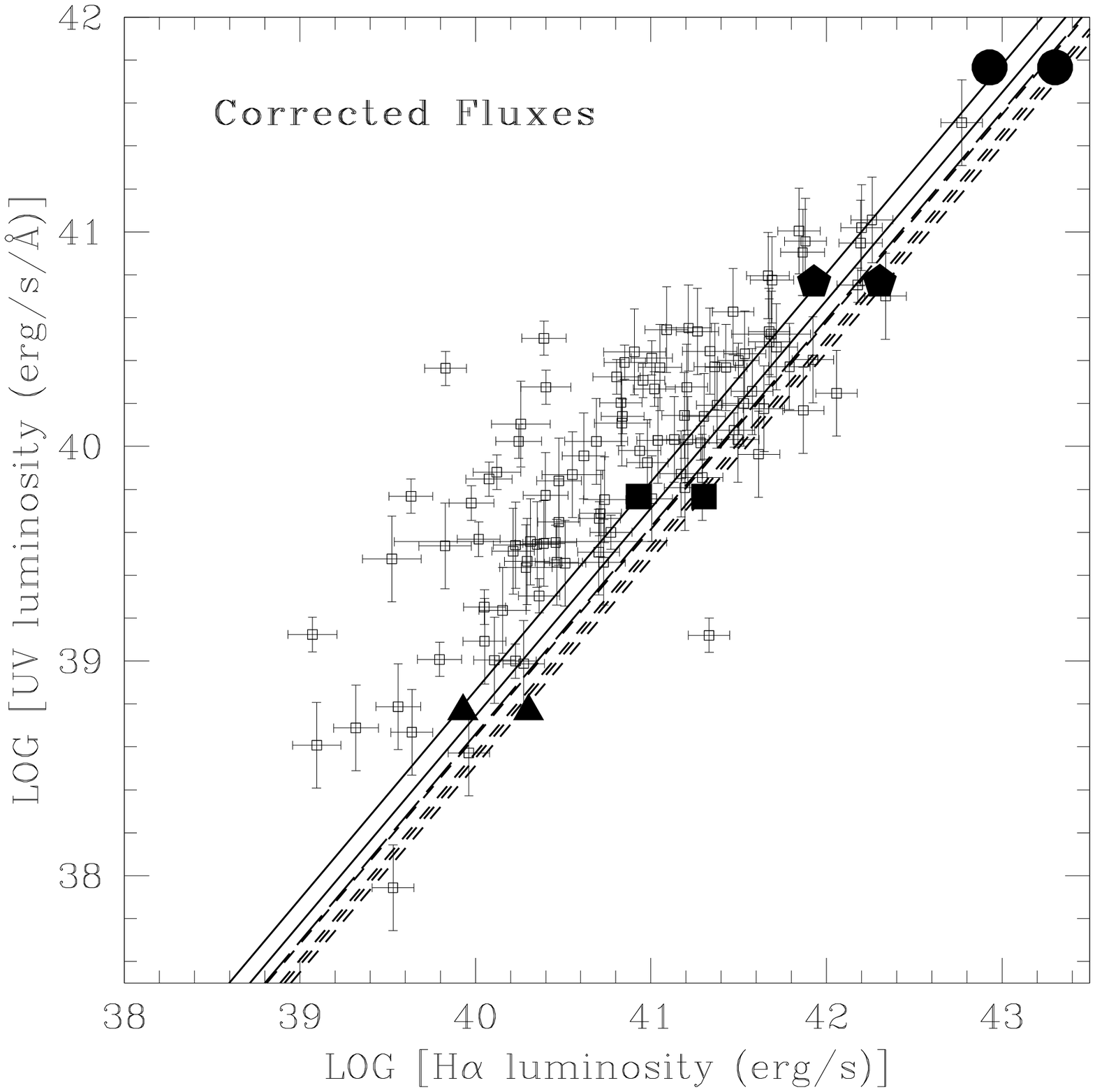,width=80mm}
\caption{The correlation between the \Halpha\ and FOCA-UV luminosities, for the most secure sample. Only field galaxies are shown. \textit{Left:} Both luminosities uncorrected, \textit{right:} both luminosities corrected. Errors are $1\sigma$. The lines show the position of galaxies with a constant SFR. The lines are (from left to right) SOLID:PEGASE $f=0.7$ and $f=1.0$, SHORT DASHED: Starburst 99, metallicities 0.04, 0.02, 0.004, 0.001, LONG DASHED: M98 \Halpha\ with PEGASE UV $f=1.0$. UV conversion factors taken at 100~Myr. The markers refer to SFRs of 0.1, 1.0, 10 and 100 M$_{\odot}~\rmn{yr}^{-1}$ respectively for the PEGASE $f=0.7$ and $\rmn{Z}=0.001$ Starburst 99 models. See text for further details.}
\label{havsuvplot}
\end{figure*}

\subsection{The star-formation diagnostic plots}

\begin{figure}
\epsfig{figure=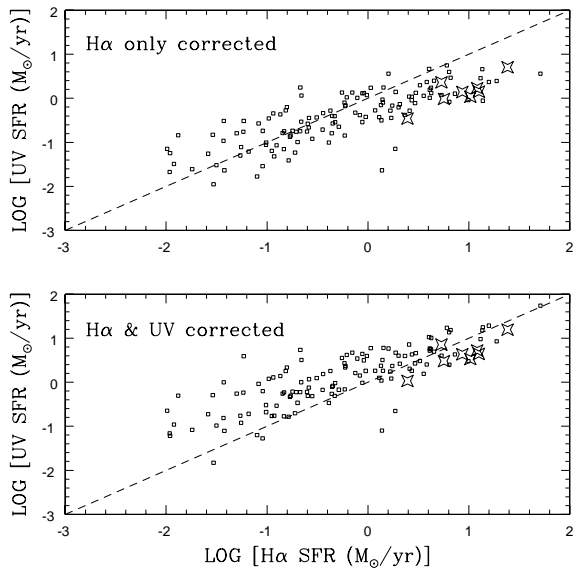,width=80mm}
\caption{A comparison of SFRs derived from our sample (open squares) and that of Glazebrook et al.~1999 (stars). SFR conversions are taken from the PEGASE program. The dashed lines shows a perfect agreement between \Halpha\ and UV SFRs. TOP: Only \Halpha\ fluxes are corrected for dust, BOTTOM: Both UV and \Halpha\ corrected. There is a good agreement between the samples, though the Glazebrook et al. sample only covers the bright end of our distribution.}
\label{kgbcompareplot}
\end{figure}

For a constant star formation history, SFRs calculated from the two
different tracers should produce the same result.
Fig.~\ref{havsuvplot} shows the correlation between the \Halpha\ and
UV luminosities from the most secure sample, for both uncorrected
(left) and dust corrected (right) data. The overlaid lines show the
conversion into SFRs using the different factors in
Table~\ref{sfrtable}. As can be seen, though there is a good
correlation over three orders of magnitude, the different range of SFH
parameters cannot reproduce the scatter around the correlation. The
other trend is that the majority of the data points lie above the
constant SFH lines, i.e. the galaxies are typically over-luminous in
the UV for a given \Halpha\ luminosity. This result is also reflected
in the integrated SFRs derived from the \Halpha\ and UV light
densities (Section~2). In this case, although Fig.~\ref{sfrplot} does
not show the \textit{uncorrected} \Halpha\ derived value against the
uncorrected UV value, it clearly shows the large discrepancy between
the \textit{corrected} \Halpha\ and UV derived SFRs. It also shows
that the corrected \Halpha\ derived SFR is only slightly above the
uncorrected UV derived SFR, (also seen in Fig.~\ref{kgbcompareplot}
for individual galaxies, a point returned to later).

The significance of the scatter can be examined by comparing with that
expected from the random measurement errors. The data points are
fitted using a least-squares technique, and the distribution of the
residual ($\log (\rmn{L}_{H\alpha})$ -- `least-squares best-fit $\log
(\rmn{L}_{H\alpha})$') compared with the distribution of the $\log
(\rmn{L}_{H\alpha})$ errors. While the distribution of the $\log
(\rmn{L}_{H\alpha})$ errors peaks strongly at $\simeq
0.05~\rmn{to}~0.1$, the distribution of the residuals is flat over a
much larger range (0 to 1). A similar pattern is seen in the $\log
(\rmn{L}_{\rmn{UV}})$ errors. An explanation for the scatter must
therefore be sought elsewhere.

A possible systematic effect which could explain the large UV flux
relative to the \Halpha\ flux concerns the relevant apertures over
which the two measurements were made. The finite (2.7\arcsec) diameter
of the WYFFOS fibres implies that some of the galaxy light may be lost
from the spectra, especially if the UV magnitudes were measured over a
larger aperture. Unfortunately due to the poor imaging resolution of
the FOCA detector, most UV sources are unresolved and therefore we do
not have any reliable size information.  However, we do not believe
this effect to be a serious one. Firstly, inspection of the POSS
images suggests that a diameter of 2.7\arcsec\ will include all of the
continuum light from most of our galaxies.  Secondly, if this were to
be a significant effect, we would expect the ratio of UV luminosity to
\Halpha\ luminosity to decrease with redshift as more of the galaxy
light is included in the WYFFOS fibres.  We examined the data for such
a trend, but found no significant trend from $z=0$ to $z=0.3$. From
this, we conclude there to be no evidence for significant aperture
mismatches \textit{internally within our sample}.

The over-luminosity in the UV and the scatter cannot be explained in
terms of simple foreground screen dust corrections, as these will
increase the discrepancy, not reduce it. Other dust geometries are
also unlikely to be the cause. To move the observed positions of the
galaxies so that they agree with the constant SFH predictions requires
large dust corrections to the \Halpha\ luminosities, but almost
negligible corrections for the UV luminosities. Though the $C=0.45$
correction derived in Section 2.2 \textit{applied solely to the
  \Halpha\ fluxes} produces a better agreement between the two SFR
tracers, to remove the scatter completely would require corrections of
up to $C=1.4$, corresponding to $\rmn{A}_{\rmn{V}}\simeq 3$ (see
Section 2.2), \textit{whilst simultaneously having no effect on the UV
  luminosities}. Such corrections are not seen from the Balmer
decrement measurements, and would require extreme dust geometries.

This trend initially appears to contradict that of Glazebrook et
al.~\shortcite{glazebrook}, who find that their \Halpha\ derived SFRs
lie above that derived from the UV continuum at 2800~\AA. As a
comparison, Fig.~\ref{kgbcompareplot} shows both our sample and that
of Glazebrook et al. converted to SFRs on an individual galaxy basis
using the appropriate PEGASE conversion factor at 2800~\AA\ (taken from
Table~3, Glazebrook et al.~\shortcite{glazebrook}). The dust
corrections applied to the \Halpha\ and the 2800~\AA\ UV are as given
in Glazebrook et al. It is clear that only the bright end of our
galaxy population is sampled by Glazebrook et al., and that in this
range there is a good agreement between the two samples. The dashed
line shows a perfect agreement between \Halpha\ and UV derived SFRs,
and it is clear that correcting just the \Halpha\ luminosities for
dust reduces the offset apparent in Fig.~\ref{havsuvplot}, indicating
that our UV corrections may be upper limits and possibly
overestimated, though the scatter is more difficult to explain.

However, both the offset and the scatter in Fig.~\ref{havsuvplot} and
the lower (dust corrected) half of Fig.~\ref{kgbcompareplot} can also
be explained by a series of starbursts superimposed on underlying
galactic SFHs.  During a starburst, a galaxy moves up and to the right
on the UV-\Halpha\ plane, increasing luminosity in both quantities. As
the burst decays, the \Halpha\ rapidly decreases due to the short
lifetimes of ionising stars, but the UV luminosity is temporarily
retained, moving the galaxy to the left.  Subsequently, the galaxy
returns to its pre-burst (quiescent) position, describing a loop on
the plot.

This scenario can also reconcile the difference between bright
galaxies, including the Glazebrook et al. sample, where the
\Halpha~SFR~$>$~UV~SFR, and the fainter sample in the lower half of
Fig.~\ref{kgbcompareplot}, where the opposite is seen. The bright
galaxies are (likely) at the peak of a particular burst, and, in a
Calzetti-like dust scenario, the UV will contain a contribution from
young massive (and hence dust obscured stars) as well as older (less
obscured) stars; hence our UV corrections are probably
\textit{underestimated} in this range and the \Halpha\ SFR will be
larger then UV derived measures. As the burst dies away, the
contribution to the UV luminosity from the obscured massive stars
decreases and the UV SFR will become larger than the (more rapidly
decaying) \Halpha\ SFR. The next section attempts to quantify the
burst parameters in this picture.

\begin{figure}
\epsfig{figure=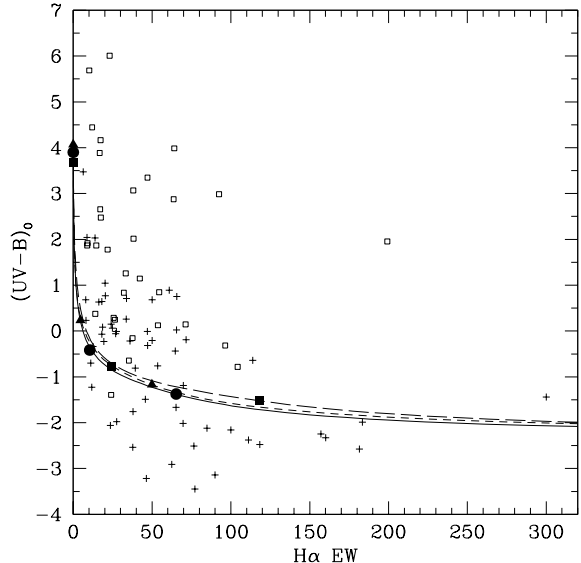,width=80mm}
\caption{Plot of \Halpha\ EW against \restcol\ colours for field and
cluster galaxies in SA57 (COMA) and Abell 1367, overlaid with various
SFHs using different IMFs. Field galaxies shown as crosses, cluster
galaxies as open squares. Colours have been dust corrected. The solid
line is the Rana-Basu IMF (also circle markers), short dash Scalo IMF
(also triangles) and long dash Salpeter IMF (squares). The markers
refer to different galaxy ages (0.5, 6 and 12~Gyr).}
\label{hacolsfhplot}
\end{figure}

At this point it is relevant to return to the vexing question of the
origin of the extreme \restcol\ colours.  Fig.~\ref{hacolsfhplot} plots
the \Halpha\ EW against \restcol\ for the galaxies in the UV
photometric system (see Section 2.2 for details).  Only those galaxies
for which errors in the \Halpha\ EWs are available with an unambiguous
optical identification are shown.  This removes some galaxies with
extreme \restcol~$\simeq -4$ colours where the UV fluxes were possibly
the sum of two galaxies, and were therefore anomalously bright when
compared to the $B$ magnitudes.  The advantage of this diagnostic plot is
that it has no complications due to uncertainties in the flux
calibration of the WHT optical spectra.

The galaxies plotted in Fig.~\ref{hacolsfhplot} represent two different
environments --  cluster members in Abell 1367 or Coma (squares), and
field galaxies (crosses). The plot shows a clear trend -- the strongest
\Halpha\ emission systems are the bluest systems. Most of the galaxies
tend to cluster at around $\rmn{H}\alpha \simeq 20-60$,
$(\rmn{UV}-\rmn{B})_{0}\simeq 0$, but there are two other areas on this
plot of interest. One consists of those galaxies with very blue
\restcol\ colours of $\simeq -2~\rmn{to}~ -4$, the other are those with
a significant \Halpha\ EW ($\simeq 50-100$) but much redder colours.

In order to distinguish between the cluster SFH and field SFH,
Fig.~\ref{hacolsfhplot} also shows various predictions according to 
the PEGASE spectral synthesis program. The histories are for exponential 
bursts of the form:

\begin{equation}
\rmn{SFR}=\tau^{-1} \exp \left( -\frac{t}{\tau} \right)
\label{secsfr}
\end{equation}

\noindent
where $\tau$ is the characteristic time of the SFH, and is equal to
1.25~Gyr. Altering this value does not change the
trajectory, only the speed at which a galaxy travels along it. The
plot shows that the cluster galaxies are solely responsible for the
group of galaxies that have redder colours and a significant \Halpha\ 
emission.

The effect of varying the IMFs on the colours is also explored. A wide
range of IMFs are available, and three are shown in
Fig.~\ref{hacolsfhplot}; these are the Scalo~\shortcite{scalo},
Salpeter~\shortcite{salpeter} and Rana-Basu~\shortcite{rb} IMFs. As in
Fig.~\ref{havsuvplot}, it is clear that these SFHs are incapable of
reproducing the scatter in the observed points of
Fig.~\ref{hacolsfhplot}, even when variations in the IMF are
considered, and certainly cannot reproduce the \restcol\ colours seen
here.

In summary, smoothly declining SFHs cannot a) reproduce the UV colours
and luminosities seen in the field galaxies, b) produce the strong
\Halpha\ emission seen in the relatively red cluster galaxies, or c)
generate the scatter seen in both
Figs.~\ref{havsuvplot}~\&~\ref{hacolsfhplot}.

\subsection{Modelling the luminosities and colours}

In this section, we aim to understand the scatter in
Fig.~\ref{havsuvplot} by examining the duty cycle of star formation in
the survey galaxies.  We adopt throughout this section the
Salpeter~\shortcite{salpeter} IMF with mass cut-offs at
$0.1~\rmn{M}_{\odot}$ and $120~\rmn{M}_{\odot}$. The modelling
techniques discussed here were also attempted using other IMFs --
those also used in Section~4.2 -- as well as varying the mass-cutoffs
used; however, no appreciable difference was obtained using these
`standard' IMFs.

In a method similar to that adopted by Glazebrook~et
al.~\shortcite{glazebrook}, who suggested that the SF in a sample of
13 CFRS $z\simeq 1$ galaxies is erratic, we will examine the effect of
superimposing a set of bursts on a smoothly declining star formation
history. We ask what range in the strength and duration of the bursts
is required to reproduce the scatter observed.

In order to check the bursting hypothesis as a solution for the scatter
in Fig.~\ref{havsuvplot} (and Fig.~\ref{hacolsfhplot}), the positions
of non-bursting galaxies should ideally be plotted as a `reality check'
on the model predictions. In the absence of UV data for a large sample
of normal galaxies, this test cannot yet be done.  Meanwhile it must be
assumed that the absolute positions of galaxies predicted by the
synthesis codes are correct, and that there is no systematic offset
when comparing model and observations.
 
To model the properties of a galaxy, a series of bursts of varying mass 
(M) and burst decay time ($\tau$) were superimposed on gradual declining
or constant SFHs. This gives three free fitting parameters: M, $\tau$, 
and the number of bursts, $N_{b}$, as well as the form of the declining
SFH. Several versions of the latter were tried; two are introduced here,
which differ in the resulting present-day colour. Their characteristics
are summarised in Table~\ref{secularsfhstable}.

The models are compared to the observed data points statistically. We
first examined a maximum likelihood method.  The likelihood, $L$, of
the observed galaxy points being drawn from a particular model is given
by (Glazebrook et al.~\shortcite{glazebrook}):

\begin{equation}
L=\prod_i \int \int  \frac{P(h,u)\exp(-\frac{(h_i-h)^2}{2\Delta h^2_i}-\frac{(u_i-u)^2}{2\Delta u^2_i})}{2\upi\Delta h_i\Delta u_i} \d u \d h
\label{maxleqn}
\end{equation}

\noindent
where the observational \Halpha\ and UV luminosities are represented
by $u_i$ and $h_i$ respectively for each galaxy $i$, $\Delta u_i$ and
$\Delta h_i$ are the observational uncertainties in these points
measured from the individual spectra ($\Delta h_i$) or based on the UV
magnitudes ($\Delta u_i$, Section 2.1), $u$ and $h$ represent the
parameterization of the PEGASE model points, and $P(h,u)$ is the
probability density of a particular model point in \Halpha\ / UV
space.

\begin{table*}
\begin{center}
\begin{tabular}{cccccc} \hline \hline
SFH & $\tau$\ (Gyr) & \restcol & \Halpha\ EW & $\rmn{L}(\rmn{H}\alpha)$ & $\rmn{L}(UV_{2000})$ \\
& & & & $10^{40}\rmn{erg~s}^{-1}$ & $10^{39}\rmn{erg~s}^{-1} \AA^{-1}$ \\
\hline
1 (Red) & 2.00 & 1.37 & 1.6 & 0.0105 & 0.149 \\
2 (Blue) & 5.00 & -0.58 & 17.1 & 0.1540 & 2.018 \\
\hline \hline
\end{tabular}
\end{center}
\caption{The two exponential SFHs on which the bursts were superimposed. Superpositions were made between galactic ages of 4 to 12~Gyr; the characteristics above are taken at 12~Gyr.}
\label{secularsfhstable}
\end{table*}

The bursts were added at random times from a galactic age of 4 to
12~Gyr, and were fitted to the data points for the period 8 to 12~Gyr.
We then calculated the likelihood of a SFH matching the observed data
points. This was repeated 100 times, and the mean likelihood obtained.

\begin{figure*}
\epsfig{figure=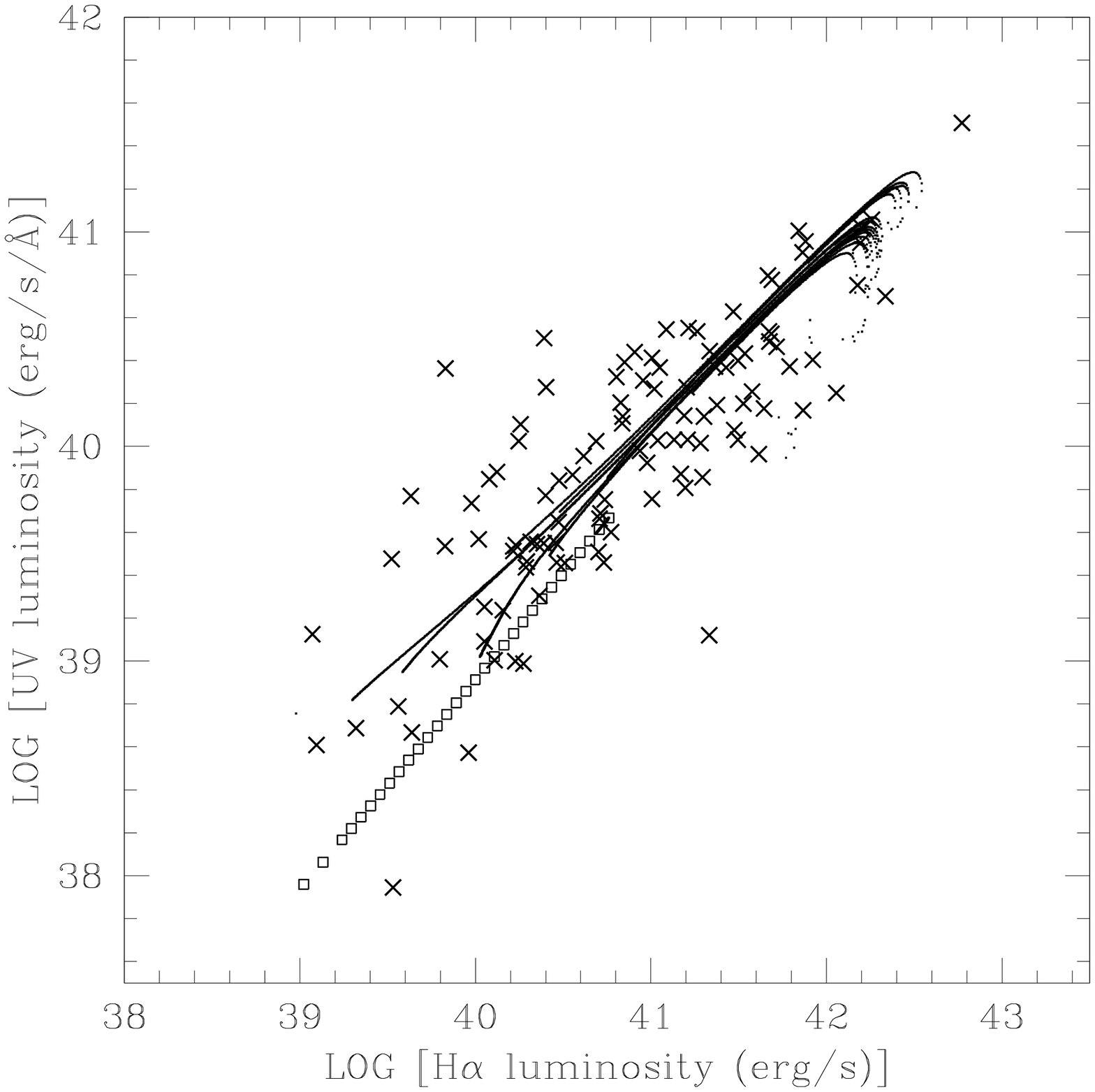,width=80mm}
\epsfig{figure=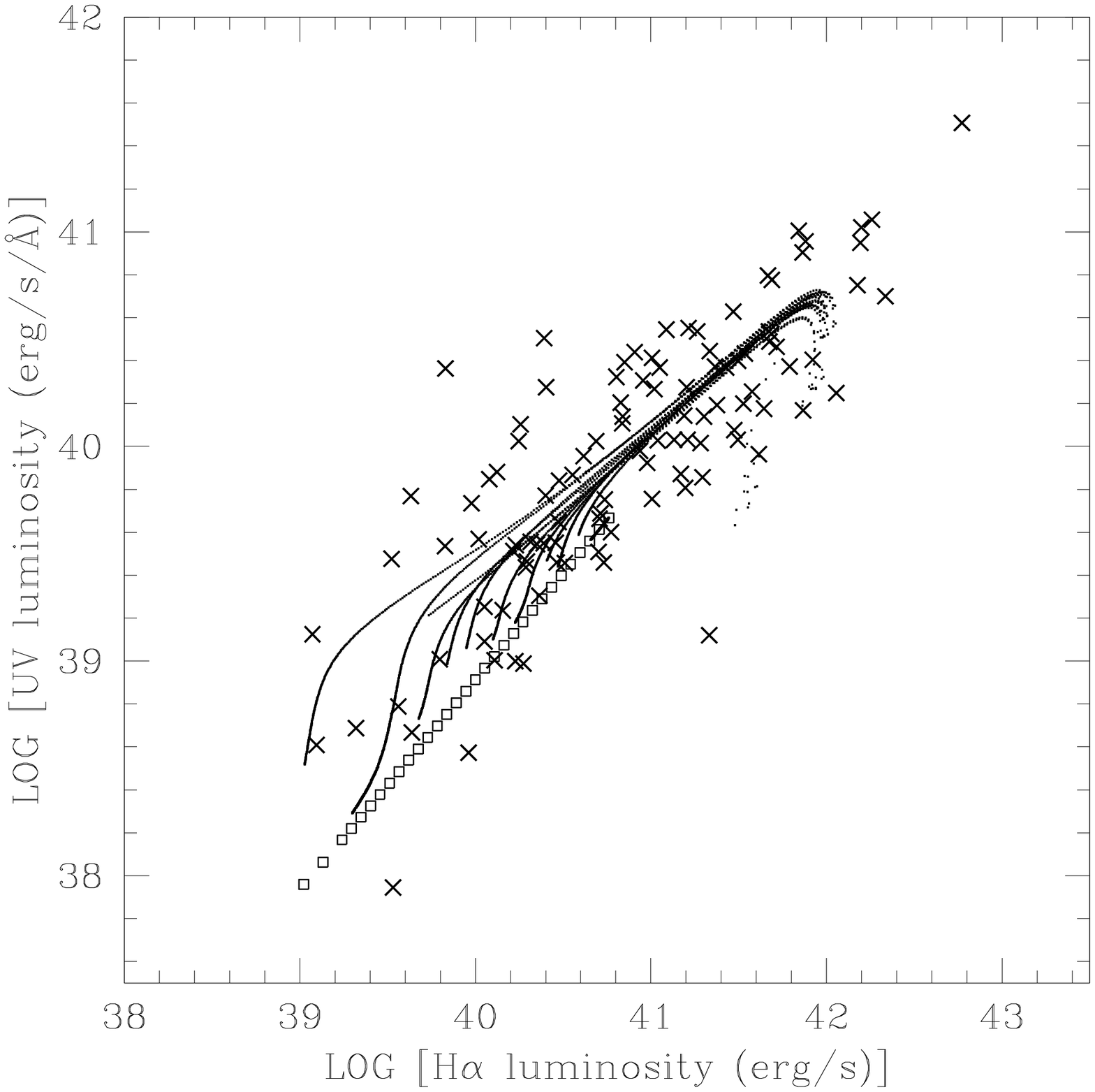,width=80mm}
\caption{Two distributions of UV-\Halpha\ measures for the sample (crosses) compared to that for model galaxies, found by superimposing a series of bursts on the SFHs described in Table~\ref{secularsfhstable} using the best-fit parameters of Table~\ref{bestfittable} (only the `redder' galaxy is shown). The open squares show the track made by a global exponentially declining SFR, forming a total mass of $10^{10}~\rmn{M}_{\odot}$, as described in the text. The lines show the paths described by a `model' galaxy during its evolution. LEFT: Maximum likelihood `best-fit'. Much of the observed scatter is not reproduced by this model. RIGHT: The second statistical test, with the emphasis on matching the scatter in the plot, rather than maximising the likelihood.}
\label{maxplot}
\end{figure*}

\begin{figure*}
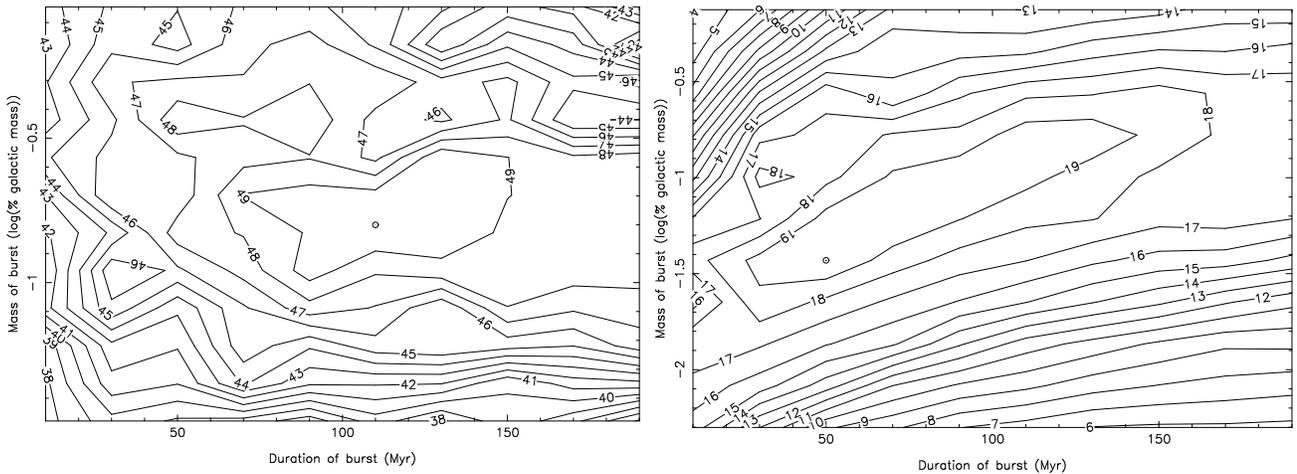

\resizebox{85mm}{!}{\rotatebox{270}{\includegraphics{figure17a.eps}}}
\resizebox{85mm}{!}{\rotatebox{270}{\includegraphics{figure17b.eps}}}
\caption{Examples of the burst mass/duration contour plots. LEFT: A maximum likelihood contour plot. The peak is at $\tau = 130$~Myr, M$\simeq 30$\%. RIGHT: The contours generated by the alternative statistical test, which reproduces more of the scatter in the points. The peak is at $\tau =50$~Myr, M$\simeq 5$\%. Both plots are for the redder SFH from Table~\ref{secularsfhstable}.}
\label{contoursfhmaxplot}
\end{figure*}

Two types of burst were considered; the first was a burst of constant
strength over its duration, the other was an exponentially declining
burst similar in form to Equation~\ref{secsfr}. Constant strength
bursts generally gave lower likelihoods as they tend to generate
stationary points in \Halpha\ / UV space; exponential bursts generate
more scatter. The `most likely' burst parameters for each galaxy type
are listed in Table~\ref{bestfittable}. A typical likelihood contour
plot is shown in Fig.~\ref{contoursfhmaxplot}. Little improvement was
obtained by varying the number of bursts; a value of 20 was used
throughout, equivalent to a burst every $\simeq 400~\rmn{Myr}$. The
best-fitting bursts are also demonstrated graphically in
Fig.~\ref{maxplot}, where the path of the model galaxy in \Halpha\ /UV
space is superimposed on the data.  As can be seen, these bursts
reproduce many of the observed data points but do not reproduce the
scatter.
 
\begin{table}
\begin{center}
\begin{tabular}{ccccc} \hline \hline
Test & SFH & Burst $\tau$ & Burst Mass & N$_{b}$ \\
& & (Myr) & (\% galaxy mass) & \\
\hline
1 & Red & 130 & 25-30 & 20 \\
1 & Blue & 50 & 10-15 & 20 \\
\hline
2 & Red & 50 & 3-7 & 20 \\
2 & Blue & 70 & 15-20 & 20 \\
\hline \hline
\end{tabular}
\end{center}
\caption{The parameters for the `best-fit' bursts for the two galaxy SFHs. The top parameters are those generated by maximum likelihood, the bottom are those from the second statistical test, which concentrates on reproducing the scatter seen in the observed points.}
\label{bestfittable}
\end{table}

A second, simpler statistical approach was adopted. The aim is to
assess the number of observed points that could be reproduced by the
maximum likelihood fits. This simply counted the number of data points
reproduced by a particular SFH. Fig.~\ref{contoursfhmaxplot} shows
an example contour plot from this method, Table~\ref{bestfittable}
lists the parameters of the `best-fitting' histories as before, and
Fig.~\ref{maxplot} shows these histories graphically.

Although the burst parameters for the two methods are similar for the
blue galaxy, they are very different for the first (redder) galaxy. The
maximum likelihood method favours longer bursts. With a short burst
the galaxy spends much of its lifetime away from the position of
the observed points; the underlying component is red and the galaxies
were selected in the UV. 

\begin{figure}
\epsfig{figure=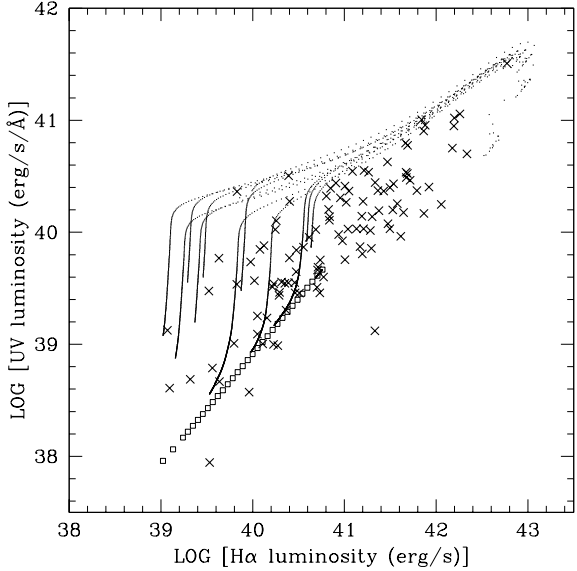,width=80mm}
\caption{As Fig.~\ref{maxplot}, but attempting to reproduce the high UV luminosity points. This shows $\tau =10$~Myr, 15\% mass bursts. The UV luminosity points can be reproduced, but even these histories are not capable of reproducing the extreme \restcol\ colours found in this survey.}
\label{examplesfhplot}
\end{figure}

It can also be seen that the histories do not reproduce the highest UV
luminosity galaxies; these are the galaxies that have the strongest
\restcol\ colours. In Section 3, we concluded that there was no
conclusive evidence that the anomalously bright \restcol\ colours
could be explained by AGN contamination or metallicity effects.  We
find that the best-fit histories from above could neither generate the
scatter nor the bluest colours seen in the data. Though the highest UV
luminosity bursts are reproducible via short, intense bursts
(Fig.~\ref{examplesfhplot}), the extreme \restcol\ colours are still not
attained.

\section{Discussion}

We have seen that the scatter in the UV-\Halpha\ plane is consistent
with a star formation history that is complex and erratic. The fundamental
observation is that most of our UV-selected galaxies show excess UV 
luminosities when compared to their \Halpha\ fluxes on the assumption 
of a simple star formation history. Moreover, dust extinction would 
serve to enhance the difference, not reduce it unless, perhaps, the
spatial distribution of dust is particularly complex. In Section 4, 
we considered more complex star formation histories invoking a regular 
pattern of bursts on top of more general exponentially-declining
histories and found that the different time-dependencies of the
stars that produce UV flux and excite nebular emission can reproduce 
the scatter observed.

It is interesting to compare this result with those (very few) surveys
for which two star formation diagnostics are available. Using
different selection criteria, Glazebrook et al.~\shortcite{glazebrook}
(13 targets colour selected from the $I$-limited CFRS) and Yan et
al.~\shortcite{yan} (slitless HST NICMOS spectroscopy of 33 targets)
both find \Halpha\ derived SFRs in excess of those from UV
measurements (at 2800~\AA) - the opposite effect to that seen here.
They attribute their discrepancy primarily to dust extinction
\cite{yan,glazebrook} but also discuss an erratic SFH
\cite{glazebrook}. The effect of dust extinction in these different
samples is very difficult to quantify; however both of these surveys
sample only the bright end of our galaxy sample, and we have shown
there is actually good agreement between these surveys and ours in
this range. The other major difference of significance is the size of
the samples available for analysis and the quality of spectral data
involved; clearly both are better for our low $z$ sample than in the
$z>1$ analyses.
 
\begin{figure}
\epsfig{figure=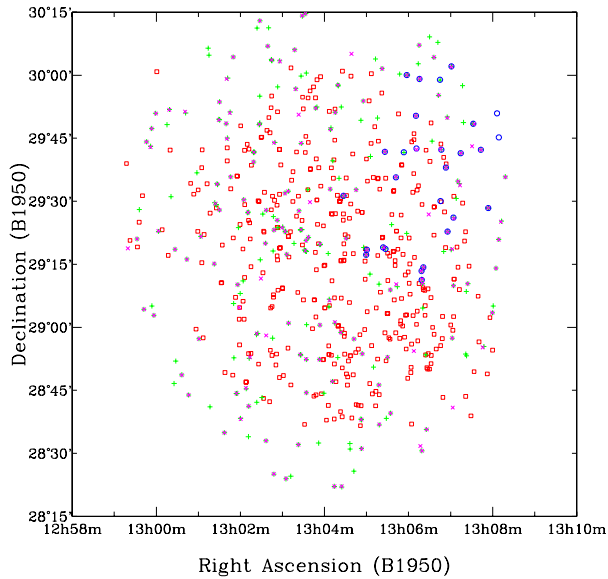,width=80mm}
\caption{The correlations between FIRST, NVSS and Windhorst et
al.~(1984) 1.4~GHz radio surveys and the FOCA 2000~\AA\ detections. FOCA
sources are shown as open squares, FIRST as square crosses, NVSS as
diagonal crosses and Windhorst as open circles. A poor correlation
between FOCA sources and radio detections is observed.}
\label{focaradioplot}
\end{figure}

The nature of our UV selected galaxies can also be assessed by
considering comparisons with 1.4~GHz radio continuum detections. This
radiation, caused by synchrotron radiation from relativistic
electrons, is thought to be generated by electrons accelerated by
supernovae from massive ($\rmn{M}>5\rmn{M}_{\odot}$) short-lived stars, and
therefore should be a further tracer of the SFR in galaxies
\cite{condon}. A literature search revealed three radio surveys which
cover some or all of SA57 -- the FIRST catalogue \cite{white}, the
NVSS \cite{condonnvss} -- both using the VLA -- and Windhorst, Heerde
\& Katgert~\shortcite{windhorst}, which uses the Westerbork Synthesis
Radio Telescope. The correlations between these surveys and the FOCA
detections are shown in Fig.~\ref{focaradioplot}.

Though the radio catalogues correlate well, there is a surprisingly
poor correlation with the FOCA sources. Only 3 FIRST sources and 1 NVSS
source have FOCA counterparts within $20\arcsec$ -- a generous search
radius -- and \Halpha\ was detected in only one of these. However,
converting 1.4~GHz luminosity to SFR for our adopted IMF
\cite{cram,serjeant}:

\begin{equation}
\rmn{SFR_{1.4}(M\ge 0.1M_{\odot})}=\frac{L_{1.4}}{7.45\times 10^{20} \rmn{~W~Hz^{-1}}}\rmn{M_{\odot}yr^{-1}}
\end{equation}

\noindent
we find that the largest SFRs in our UV-selected sample ($\simeq
15\rmn{M_{\odot}yr^{-1}}$) would give expected 1.4~GHz luminosities at
$z=0.1$ of only $\simeq 2\rmn{mJy}$, i.e. close to the detection
threshold of the FIRST survey ($\simeq 1\rmn{mJy}$). None the less, as
many sources should lie closer than $z=0.1$, we might have expected 
more 1.4~GHz detections. As previous studies \cite{serjeant} have shown that
even dust corrected \Halpha\ luminosity underestimates local SFRs when
compared with 1.4~GHz observations, this lack of correlation may
indicate that the \Halpha\ derived SFRs are marginally overestimated.
Deeper 1.4~GHz surveys are required to explore this further.

Although we have shown that the excess UV luminosities can be
reproduced by considering bursting SFHs, we have generally failed to
reproduce the full range of \restcol\ colours for our sample; in
general, colours smaller than $(\rmn{UV} -\rmn{B})_{0} \simeq -2$ are
problematic to generate, and we cannot easily explain a number of
intense UV sources which are optically faint. In Paper I we eliminated
trivial explanations for this category of sources including
photometric discrepancies between the FOCA and POSS ($B$) magnitude
systems.  It is certainly possible to generate UV-optical colours more
extreme than the Poggianti starburst SEDs. For example, using the
Starburst 99 code \cite{starburst}, $(\rmn{UV}-\rmn{B})_{0}\simeq
-2.5$~--~$-3$ can be generated (but only on a very short time-scales)
using standard IMFs and smaller ($10^4~\rmn{Myr}$) timesteps then the
Poggianti SEDs ($10^7~\rmn{Myr}$).  The number of such sources found
then can only be understood in the context of more detailed modelling
of the SF duty cycle.  On the whole, however, it is becoming apparent
that we do not understand the SEDs of these systems well.

The burst duty cycle produces results consistent with that of earlier
workers. The original analysis of Glazebrook et
al.~\shortcite{glazebrook} adopted slightly longer and more massive
bursts. Marlowe, Meurer \& Heckman~\shortcite{marlowe} examined the 
nature of starbursts in dwarf galaxies and found similar burst parameters. 
They find such bursts typically account for only a few per cent of the 
stellar mass. Theoretical studies \cite{babul} suggest that the burst 
durations may be shorter than those found here, as well as being more
intense in terms of stellar mass. A comprehensive study of bursts in
irregular galaxies has also been undertaken by Mas-Hesse \&
Kunth~\shortcite{mas-hesse}. They also find that star-formation
episodes are essentially short, with a mean age of around 4~Myr. Though
this is much shorter than the `best-fit' duty cycles found in this
survey, it does correlate well with the bursts that we require in order
to reproduce the full range of UV luminosities ($\simeq 10~\rmn{Myr}$).

There remain many avenues for further investigation into the nature of
star-formation in local samples of field galaxies. There is an urgent
need for resolved images so that the morphological details and physical
location of the star forming regions can be determined. Ultimately,
only such images can rule out significant AGN components in the most
extreme UV sources in our sample. Secondly, more accurate UV photometry
is needed over more bandpasses to check for dust and SED differences.
Finally, we comment in our analysis of the importance of verifying and
constraining the various stellar synthesis models in the UV against a
control sample of quiescent objects with well-behaved star formation
histories. The absence of such a body of data simply highlights our
surprising ignorance of the UV properties of normal galaxies.

\section{Conclusions}

We sought, in this paper, to illustrate very simply how little
is known about the star formation properties of the bulk population
of field galaxies with $z<0.4$. For a well-defined population of 
UV-selected sources, for which detailed spectroscopic analyses
have been carried out, we have found:

\begin{enumerate}
  
\item{} A volume-density of star formation well in excess of previous
  estimates both when using the UV and \Halpha\ luminosity functions.
  The UV luminosity function has a remarkably steep faint end slope
  making optical redshift surveys prone to underestimating the
  luminosity density at modest redshift.

\item{} No evidence, internally within our sample, for a strong
  increase in UV luminosity density with redshift, such as would be
  expected on the basis of earlier work based on the CFRS (Lilly et
  al. 1996).
  
\item{} A significant fraction of UV sources have UV-optical colours
  at the extreme limit of those reproducible in starburst models, and
  some are beyond this limit. From detailed emission line studies, we
  find no evidence for AGN contamination or anomalous metallicities in
  these sources.

\item{} The star formation rates derived from UV flux and
  extinction-corrected \Halpha\ line measurements are not consistent
  in the framework of model galaxies with smoothly-declining star
  formation histories. We find the UV luminosities are consistently
  higher than expected and dust effects would only exacerbate this
  discrepancy.
  
\item{} We also find a significant scatter in the UV-\Halpha\ plane
  and can reproduce this (and the offset discussed above) in terms of
  a duty cycle of starbursts superimposed upon longer term histories.
  We discuss ways of physically constraining such a model and produce
  illustrative examples where 5--20\% of the galactic mass is involved
  in bursts with decay time $\tau \simeq 50~\rmn{Myr}$ and a frequency
  of one every $\simeq 400~\rmn{Myr}$.

\end{enumerate}

\section*{Acknowledgements}

We thank the anonymous referee for his detailed comments which
improved this manuscript. We also wish to thank Chris Blake, Veronique
Buat, Lawrence Cram, Gerhardt Meurer, Bianca Poggianti, and especially
Max Pettini for their many helpful discussions in preparing this
paper. We are also grateful for the assistance provided by the La
Palma support staff in securing the optical spectra. The WIYN
Observatory is a joint facility of the University of
Wisconsin-Madison, Indiana University, Yale University, and the
National Optical Astronomy Observatories. The William Herschel
Telescope is operated on the island of La Palma by the Isaac Newton
Group in the Spanish Observatorio del Roque de los Muchachos of the
Instituto de Astrofisica de Canarias.

\onecolumn
\begin{center}
\begin{longtable}{cccccccccccl}
\caption{The updated UV-z catalogue. `OC' indicates the number of optical counterparts on the POSS plates within 10$\arcsec$. The position (1950) is that of the closest optical counterpart. `T' is the galaxy type, classified on the basis of UV-B colour; $\rmn{T}=1 - 8$ refers to types E, S0, Sa, Sb, Sc, Sd, SB1 \& SB2. All colours are rest-frame and dust corrected. \Halpha\ and \Hbeta\ give rest-frame EWs.}\\
\hline \hline
OC & RA & DEC & uv & z & O\textsc{ii} & \Halpha & \Hbeta & T & $\rmn{M}_{UV}$ & $(\rmn{UV-B})_0$ & Comments\\
\hline
\endfirsthead
\hline \hline
\endlastfoot
\caption[]{(continued)}\\
\hline \hline
OC & RA & DEC & uv & z & O\textsc{ii} & \Halpha & \Hbeta & T & $\rmn{M}_{UV}$ & $(\rmn{UV-B})_0$ & Comments\\
\hline
\endhead
\hline \hline
\endfoot
\multicolumn{12}{l}{\textbf{SA 57}}\\
\hline
1 & 13:06:32.16 & +29:48:38.2 & 16.55 & 0.0228 & 16.2 & 21.8 & 5.0 & 3 & -18.10 & 1.63 & OII,wOIII,s\Hbeta,s\Halpha\\
1 & 13:06:46.29 & +29:43:37.9 & 16.89 & 0.1130 & 25.3 & 36.0 & 5.5 & 6 & -21.70 & -0.52 & OII,\Hbeta,\Halpha,HK\\
1 & 13:06:50.23 & +29:40:26.7 & 17.20 & 0.2430 & 13.7 & 18.2 & 2.1 & 4 & -23.07 & 0.33 & OII,HK,\Halpha \\
1 & 13:06:07.71 & +29:46:05.6 & 18.21 & 0.2800 & 0.0 & 0.0 & 0.0 & 4 & -22.86 & 0.52 & OII \\
1 & 13:06:30.66 & +29:39:30.4 & 17.95 & 0.2285 & 64.1 & 118.4 & 0.0 & 8 & -22.50 & -2.79 & OII,\Hbeta,OIII,\Halpha\\
1 & 13:06:07.74 & +29:44:40.2 & 17.74 & 0.0241 & 96.5 & 96.4 & 22.0 & 5 & -17.23 & -0.46 & OII,\Hbeta,OIII,\Halpha,\Hgamma\\
1 & 13:05:07.10 & +30:00:23.6 & 17.58 & 0.4020 & 0.0 & 0.0 & 0.0 & 8 & -24.01 & -4.48 & OII,OIII,\Halpha\\
1 & 13:05:21.88 & +29:55:46.3 & 18.34 & 0.0663 & 0.0 & 0.0 & 0.0 & 4 & -19.52 & 0.22 & OII\\
1 & 13:07:21.56 & +29:22:45.6 & 17.87 & 0.1420 & 0.0 & 0.0 & 0.0 & 8 & -21.54 & -4.12 & HK\\
1 & 13:06:51.43 & +29:27:27.7 & 18.18 & 0.4820 & 0.0 & 0.0 & 0.0 & 8 & -23.79 & -4.13 & HK\\
1 & 13:06:07.65 & +29:36:35.6 & 18.29 & 0.1227 & 24.7 & 71.6 & 7.0 & 5 & -22.04 & -0.44 & OII,HK,\Halpha\\
2 & 13:06:36.80 & +29:28:41.0 & 18.33 & 0.2110 & 0.0 & 0.0 & 0.0 & 8 & -21.94 & -3.85 & HK\\
1 & 13:04:25.82 & +30:00:58.3 & 17.41 & 0.1465 & 0.0 & 0.0 & 0.0 & 8 & -22.06 & -2.58 & OII\\
2 & 13:07:05.26 & +29:11:28.8 & 16.94 & 0.0627 & 71.7 & 73.0 & 17.6 & 6 & -20.16 & -0.58 & OII,\Hbeta,OIII,\Halpha\\
1 & 13:05:00.02 & +29:40:04.3 & 16.09 & 0.0176 & 37.0 & 71.3 & 9.9 & 5 & -18.49 & 0.00 & OII,HK,\Hbeta,OIII,\Halpha\\
1 & 13:04:15.30 & +29:51:58.9 & 18.49 & 0.1752 & 6.7 & 12.5 & 0.0 & 6 & -21.45 & -0.64 & HK, \Halpha \\
2 & 13:05:16.43 & +29:35:22.5 & 17.74 & 0.3203 & 49.2 & 405.6 & 0.0 & 8 & -23.44 & -4.46 & OII,OIII,\Halpha \\
1 & 13:03:49.57 & +29:58:56.5 & 17.89 & 0.3348 & 20.0 & 77.2 & 10.4 & 8 & -25.20 & -3.77 & \Halpha,OII,HK \\
1 & 13:07:04.39 & +29:02:33.2 & 17.31 & 0.1350 & 31.9 & 0.0 & 0.0 & 5 & -22.06 & -0.13 & OII\\
1 & 13:05:01.46 & +29:34:48.4 & 18.38 & 0.1717 & 28.2 & 46.4 & 0.0 & 8 & -21.43 & -3.52 & OIII,\Halpha\\
1 & 13:05:49.35 & +29:21:16.6 & 16.94 & 0.1170 & 0.0 & 0.0 & 0.0 & 4 & -22.14 & 0.37 & HK;noisy;\\
1 & 13:03:39.69 & +29:56:17.2 & 17.95 & 0.1790 & 0.0 & 0.0 & 0.0 & 6 & -22.04 & -1.05 & OII\\
1 & 13:04:14.72 & +29:44:32.4 & 17.67 & 0.0846 & 4.4 & 11.1 & 0.0 & 6 & -20.71 & -1.00 & HK, \Halpha\\
1 & 13:06:48.52 & +29:02:35.2 & 17.33 & 0.0211 & 119.4 & 37.6 & 10.3 & 5 & -18.02 & -0.30 & OII,\Hbeta,OIII,\Halpha\\
1 & 13:06:23.47 & +29:08:47.4 & 15.98 & 0.0609 & 56.6 & 13.9 & 0.0 & 3 & -21.71 & 1.73 & OII,HK,wOII,\Halpha\\
1 & 13:06:21.20 & +29:08:58.6 & 17.23 & 0.2640 & 0.0 & 0.0 & 0.0 & 8 & -23.56 & -3.12 & noisy;HK\\
2 & 13:05:38.28 & +29:18:00.1 & 18.12 & 0.1441 & 16.1 & 15.1 & 0.0 & 7 & -21.34 & -1.55 & OII,HK,\Halpha \\
2 & 13:05:23.08 & +29:21:27.4 & 17.31 & 0.1131 & 95.9 & 0.0 & 31.2 & 8 & -21.62 & -3.67 & OII,\Hbeta,OIII,\Halpha,\Hgamma\\
2 & 13:05:22.71 & +29:21:19.6 & 17.31 & 0.1260 & 0.0 & 0.0 & 0.0 & 8 & -21.84 & -4.33 & HK\\
1 & 13:03:29.88 & +29:51:59.4 & 16.78 & 0.0558 & 8.0 & 8.8 & 0.0 & 3 & -20.72 & 1.74 & HK,OIII,\Halpha\\
2 & 13:04:13.54 & +29:39:51.1 & 18.29 & 0.0892 & 35.6 & 54.7 & 9.5 & 8 & -20.15 & -2.37 & OII,\Hbeta,OIII,\Halpha\\
2 & 13:04:53.91 & +29:27:21.2 & 17.89 & 0.5213 & 61.3 & 0.0 & 99.6 & 8 & -24.22 & -4.09 & OII,OIII,\Hbeta\\
1 & 13:05:40.80 & +29:14:43.1 & 16.47 & 0.0800 & 0.0 & 33.7 & 7.2 & 4 & -21.80 & 0.41 & OII,OIII,\Halpha,\Hbeta\\
1 & 13:04:09.89 & +29:37:58.8 & 16.04 & 0.0246 & 0.0 & 17.1 & 3.4 & 2 & -19.77 & 2.50 & wOII,\Hbeta,OIII,\Halpha\\
1 & 13:04:37.54 & +29:29:21.4 & 17.85 & 0.1673 & 48.0 & 65.6 & 11.9 & 5 & -20.92 & -0.28 & OII,\Hbeta,OIII,\Halpha\\
1 & 13:03:37.14 & +29:44:00.2 & 18.03 & 0.0511 & 30.1 & 20.4 & 3.5 & 4 & -18.73 & 0.47 & OII,HK,wOIII,\Hbeta,\Halpha\\
1 & 13:04:23.45 & +29:30:37.4 & 18.07 & 0.0697 & 0.0 & 11.8 & 0.0 & 7 & -19.87 & -1.54 & wOII,\Halpha\\
1 & 13:03:48.44 & +29:40:05.1 & 17.24 & 0.1817 & 86.2 & 300.1 & 22.9 & 7 & -21.43 & -1.75 & OII,\Hbeta,OIII,\Halpha\\
1 & 13:06:11.39 & +29:00:41.8 & 16.84 & 0.0558 & 30.6 & 18.1 & 3.7 & 5 & -20.64 & -0.38 & OII,HK,\Halpha,\Hbeta,OIII\\
2 & 13:04:22.12 & +29:29:59.3 & 18.22 & 0.1821 & 10.2 & 0.0 & 0.0 & 6 & -21.81 & -1.03 & HK,OII\\
1 & 13:03:52.02 & +29:38:00.4 & 17.07 & 0.0244 & 78.4 & 26.4 & 0.0 & 4 & -18.61 & 0.10 & OII,OIII,\Halpha\\
1 & 13:05:55.86 & +29:02:53.4 & 17.24 & 0.0394 & 44.1 & 53.6 & 8.2 & 6 & -19.65 & -1.07 & OII,\Hbeta,OIII,\Halpha,HK\\
1 & 13:03:32.85 & +29:41:38.0 & 18.39 & 0.1837 & 0.0 & 0.0 & 0.0 & 7 & -21.60 & -1.54 & OII,HK\\
1 & 13:02:53.21 & +29:51:18.0 & 16.26 & 0.0236 & 45.0 & 33.3 & 7.4 & 3 & -18.38 & 1.10 & OII,\Hbeta,OIII,\Halpha\\
2 & 13:03:40.30 & +29:37:37.9 & 18.06 & 0.2859 & 129.8 & 0.0 & 0.0 & 8 & -22.90 & -4.20 & OII,OIII,w\Hbeta \\
1 & 13:02:43.94 & +29:52:15.3 & 18.25 & 0.1850 & 42.2 & 69.9 & 11.7 & 7 & -21.12 & -1.50 & OII,\Hbeta,OIII,\Halpha,HK\\
1 & 13:05:48.77 & +29:01:33.6 & 17.89 & 0.1176 & 39.0 & 37.7 & 9.4 & 8 & -21.12 & -2.85 & OII,OIII in sky\\
1 & 13:04:33.69 & +29:21:34.7 & 18.00 & 0.1817 & 13.2 & 84.85 & 3.1 & 8 & -22.72 & -2.42 & OII,HK,\Halpha \\
1 & 13:06:27.44 & +28:50:05.1 & 17.08 & 0.1650 & 0.0 & 0.0 & 0.0 & 8 & -22.65 & -2.90 & OII \\
1 & 13:05:19.98 & +29:08:21.6 & 18.54 & 0.3913 & 21.0 & 111.2 & 8.4 & 8 & -22.27 & -2.70 & OII,HK,w\Halpha \\
2 & 13:06:04.39 & +28:55:59.2 & 18.12 & 0.1970 & 0.0 & 0.0 & 0.0 & 7 & -22.02 & -1.51 & z=.197 \\
1 & 13:02:40.62 & +29:49:55.6 & 17.61 & 0.4080 & 0.0 & 0.0 & 0.0 & 8 & -24.01 & -3.38 & HK \\
2 & 13:05:29.52 & +29:03:20.2 & 17.82 & 0.2471 & 0.0 & 0.0 & 0.0 & 8 & -22.81 & -3.13 & HK,Mgb,wOII \\
2 & 13:05:30.34 & +29:03:16.3 & 17.82 & 0.2475 & 45.3 & 114.3 & 16.7 & 8 & -23.88 & -3.31 & OII,\Hbeta,OIII,\Halpha \\
1 & 13:04:29.07 & +29:17:29.9 & 18.36 & 0.2424 & 30.5 & 62.5 & 0.0 & 8 & -22.22 & -3.22 & OII,OIII,\Halpha \\
1 & 13:02:38.07 & +29:46:52.4 & 18.35 & 0.1861 & 15.5 & 65.2 & 3.4 & 7 & -23.81 & -1.96 & OII, HK, \Halpha \\
1 & 13:06:03.45 & +28:49:13.9 & 17.47 & 0.1151 & 41.8 & 24.0 & 0.0 & 5 & -21.56 & -0.15 & OII,abn,w\Halpha \\
1 & 13:05:32.05 & +28:57:31.5 & 18.40 & 0.0566 & 27.1 & 24.9 & 2.4 & 5 & -19.11 & -0.24 & OII,HK,OIII,\Halpha \\
1 & 13:04:23.62 & +29:15:48.9 & 16.63 & 0.0616 & 36.2 & 50.1 & 9.4 & 6 & -21.21 & -0.51 & sOII,\Hbeta,OIII,\Halpha\\
1 & 13:05:26.55 & +28:58:41.3 & 16.90 & 0.0702 & 9.6 & 8.0 & 0.0 & 4 & -21.09 & 0.37 & OII,abn,w\Halpha \\
2 & 13:02:57.71 & +29:38:39.1 & 18.35 & 0.1858 & 0.0 & 0.0 & 0.0 & 8 & -21.64 & -2.73 & OII,HK,\Halpha: OII\\
2 & 13:02:30.70 & +29:45:16.1 & 17.95 & 0.0480 & 0.0 & 0.0 & 0.0 & 8 & -19.17 & -2.49 & HK \\
1 & 13:03:28.24 & +29:25:25.9 & 17.33 & 0.2416 & 17.6 & 23.5 & 3.8 & 8 & -22.38 & -2.36 & OII,OIII,\Halpha,abn \\
2 & 13:05:28.03 & +28:52:01.4 & 17.55 & 0.1860 & 25.0 & 0.0 & 0.0 & 8 & -22.44 & -2.52 & OII,HK,\Halpha \\
1 & 13:02:00.66 & +29:47:57.0 & 18.06 & 0.2235 & 85.2 & 100.0 & 15.6 & 8 & -22.34 & -2.47 & OII,\Hbeta,OIII,\Halpha \\
2 & 13:04:00.52 & +29:14:47.2 & 17.18 & 0.2849 & 23.4 & 180.1 & 9.8 & 8 & -23.65 & -4.62 & OII,\Hbeta,OIII,\Halpha \\
1 & 13:04:10.95 & +29:11:04.1 & 18.07 & 0.0613 & 47.2 & 39.4 & 0.0 & 6 & -19.61 & -1.12 & OII,OIII,\Halpha \\
2 & 13:02:09.94 & +29:43:33.8 & 17.18 & 0.0230 & 0.0 & 0.0 & 0.0 & 8 & -18.35 & -2.44 & \Halpha,OIII,wOII,\Hbeta \\
1 & 13:02:12.45 & +29:42:24.8 & 18.05 & 0.0650 & 0.0 & 0.0 & 0.0 & 8 & -19.72 & -3.56 & HK \\
1 & 13:04:50.67 & +28:58:14.8 & 18.14 & 0.2474 & 51.4 & 76.6 & 0.0 & 8 & -22.49 & -2.81 & OII,\Hbeta,OIII,\Halpha \\
1 & 13:02:06.14 & +29:42:07.9 & 18.37 & 0.2403 & 76.0 & 89.9 & 8.7 & 8 & -23.16 & -3.44 & OII,\Hbeta,OIII,\Halpha \\
2 & 13:02:48.95 & +29:30:12.8 & 17.38 & 0.0230 & 0.0 & 0.0 & 0.0 & 6 & -18.16 & -0.85 & OIII,\Halpha,wOII,\Hbeta \\
1 & 13:02:15.80 & +29:38:34.0 & 16.87 & 0.3040 & 0.0 & 0.0 & 0.0 & 8 & -24.22 & -2.60 & HK \\
1 & 13:04:48.41 & +28:54:49.3 & 17.95 & 0.0397 & 56.3 & 45.8 & 8.1 & 7 & -18.51 & -1.80 & OII,OIII,\Halpha,abs,\Hbeta \\
2 & 13:03:09.04 & +29:21:37.1 & 18.06 & 0.1270 & 0.0 & 0.0 & 0.0 & 7 & -21.13 & -1.71 & wS/N,abn \\
1 & 13:04:07.33 & +29:05:10.8 & 17.80 & 0.0799 & 20.2 & 16.2 & 0.0 & 4 & -20.47 & 0.33 & OII,HK,\Halpha \\
2 & 13:04:19.63 & +29:00:19.0 & 17.51 & 0.4010 & 0.0 & 0.0 & 0.0 & 8 & -24.07 & -4.62 & HK,OII \\
2 & 13:04:19.78 & +29:00:26.9 & 17.51 & 0.1126 & 29.1 & 48.4 & 7.0 & 6 & -21.53 & -0.69 & OII,HK,\Hbeta,\Halpha \\
1 & 13:04:28.46 & +28:57:50.8 & 17.92 & 0.3880 & 28.9 & 157.1 & 10.8 & 8 & -23.59 & -2.57 & OII,\Hbeta,OIII,\Halpha \\
1 & 13:01:41.45 & +29:42:43.6 & 17.91 & 0.3398 & 67.0 & 0.0 & 24.1 & 8 & -23.36 & -2.74 & OII, OIII \\
1 & 13:04:00.52 & +29:04:38.0 & 17.49 & 0.0930 & 0.0 & 0.0 & 0.0 & 7 & -21.05 & -2.00 & HK \\
1 & 13:02:26.87 & +29:29:58.1 & 17.25 & 0.1389 & 0.0 & 18.6 & 0.0 & 5 & -22.18 & -0.22 & OII,\Halpha,HK,vwOII \\
2 & 13:05:04.61 & +28:46:37.0 & 18.05 & 0.1151 & 35.3 & 36.6 & 0.0 & 7 & -20.94 & -1.45 & OII,HK,\Halpha \\
2 & 13:04:37.52 & +28:53:13.7 & 17.45 & 0.0480 & 0.0 & 0.0 & 0.0 & 8 & -19.67 & -4.76 & HK \\
1 & 13:03:09.40 & +29:16:44.0 & 16.59 & 0.0176 & 5.1 & 17.4 & 0.0 & 1 & -18.52 & 4.01 & HK,abs,\Halpha,sS/N \\
2 & 13:01:55.86 & +29:35:55.8 & 18.32 & 0.0492 & 0.0 & 0.0 & 0.0 & 8 & -18.85 & -3.82 & OII \\
2 & 13:01:55.26 & +29:36:03.0 & 18.32 & 0.1861 & 0.0 & 0.0 & 0.0 & 8 & -21.67 & -2.58 & OII,HK,\Hbeta,\Halpha \\
1 & 13:01:41.32 & +29:39:36.8 & 18.18 & 0.1667 & 49.0 & 64.8 & 11.9 & 6 & -21.49 & -0.74 & OII,\Hbeta,OIII,\Halpha\\
2 & 13:02:56.12 & +29:18:53.6 & 18.11 & 0.0178 & 35.2 & 100.6 & 15.7 & 6 & -15.85 & -1.14 & OII,OIII,\Halpha,\Hbeta \\
1 & 13:03:21.92 & +29:08:18.0 & 15.41 & 0.0268 & 37.0 & 42.2 & 6.4 & 3 & -21.93 & 1.00 & OII,OIII,\Halpha,\Hbeta \\
1 & 13:04:25.49 & +28:50:20.3 & 17.39 & 0.0828 & 0.0 & 0.0 & 0.0 & 4 & -20.95 & 0.14 & HKabs,\Halpha \\
1 & 13:04:09.89 & +28:53:20.3 & 17.84 & 0.6695 & 264.6 & 0.0 & 49.9 & 8 & -24.85 & -4.28 & OII,\Hbeta,OIII\\
1 & 13:02:11.57 & +29:25:28.0 & 18.14 & 0.0896 & 66.6 & 38.0 & 9.6 & 7 & -19.66 & -2.07 & OII,\Hbeta,OIII,\Halpha\\
1 & 13:02:13.53 & +29:24:51.9 & 17.47 & 0.0195 & 17.3 & 14.0 & 0.0 & 4 & -17.72 & 0.23 & OII,OIII,\Halpha\\
1 & 13:04:32.93 & +28:46:14.5 & 17.26 & 0.0828 & 7.3 & 0.0 & 0.0 & 3 & -21.10 & 1.27 & HK,broad \Halpha,wOII \\
1 & 13:02:14.19 & +29:21:01.9 & 17.90 & 0.1799 & 125.9 & 183.4 & 28.2 & 8 & -22.02 & -2.29 & vsOII,\Hbeta,OIII,\Halpha \\
1 & 13:04:16.43 & +28:47:04.0 & 17.94 & 0.0270 & 5.9 & 8.9 & 0.0 & 3 & -17.97 & 1.71 & HK,\Halpha \\
1 & 13:04:51.57 & +28:36:35.6 & 16.29 & 0.0186 & 34.3 & 9.1 & 0.0 & 3 & -18.80 & 1.76 & abs,\Halpha,wOII,wOIII \\
1 & 13:01:05.01 & +29:38:00.2 & 18.16 & 0.0500 & 0.0 & 0.0 & 0.0 & 8 & -19.05 & -3.83 & HK\\
1 & 13:02:47.46 & +29:07:09.6 & 18.02 & 0.1781 & 90.1 & 181.5 & 20.9 & 8 & -22.47 & -2.88 & OII,\Hbeta,OIII,\Halpha \\
1 & 13:02:15.16 & +29:14:25.5 & 16.48 & 0.0254 & 31.1 & 32.1 & 4.4 & 4 & -19.25 & 0.68 & OII,OIII,\Halpha,\Hbeta \\
1 & 13:04:25.24 & +28:38:42.3 & 18.38 & 0.1863 & 35.2 & 69.7 & 6.8 & 8 & -21.14 & -2.33 & OII,OIII,\Halpha,\Hbeta \\
1 & 13:02:45.39 & +29:04:56.9 & 17.06 & 0.1861 & 23.3 & 47.2 & 5.1 & 6 & -23.53 & -0.62 & OII,\Hbeta,OIII,\Halpha \\
1 & 13:00:49.01 & +29:34:15.4 & 17.11 & 0.0835 & 18.3 & 19.4 & 2.0 & 6 & -22.00 & -0.53 & OII,\Hbeta,OIII,\Halpha \\
1 & 13:00:53.53 & +29:31:42.1 & 17.58 & 0.0817 & 20.9 & 27.0 & 4.1 & 5 & -21.12 & -0.36 & OII,\Halpha,HK \\
1 & 13:03:48.16 & +28:43:55.0 & 15.85 & 0.0676 & 92.8 & 60.9 & 11.5 & 4 & -22.05 & 0.58 & OII,\Hbeta,OIII,\Halpha \\
1 & 13:03:45.03 & +28:43:35.2 & 16.06 & 0.0679 & 38.4 & 65.7 & 10.3 & 4 & -22.03 & 0.45 & OII,OIII,\Halpha,\Hbeta \\
1 & 13:04:02.91 & +28:37:54.6 & 17.02 & 0.0393 & 7.6 & 8.1 & 0.0 & 5 & -19.69 & -0.07 & HK,abs,\Halpha,wOIII \\
1 & 13:02:51.82 & +28:53:38.2 & 16.97 & 0.0218 & 36.6 & 25.7 & 4.0 & 4 & -18.52 & 0.14 & OII,\Hbeta,OIII,\Halpha\\
1 & 13:00:59.29 & +29:22:49.2 & 17.55 & 0.0825 & 22.6 & 20.3 & 3.6 & 4 & -19.64 & 0.73 & OII,HK,OIII,\Halpha \\
1 & 13:02:42.78 & +28:54:32.0 & 17.75 & 0.0184 & 49.8 & 54.5 & 7.8 & 4 & -17.43 & 0.69 & OII,OIII,\Halpha,\Hbeta \\
1 & 13:01:53.07 & +29:07:07.0 & 17.52 & 0.0254 & 74.1 & 104.3 & 22.8 & 6 & -18.47 & -0.93 & OII,OIII,\Halpha,\Hbeta \\
1 & 13:03:01.48 & +28:48:32.0 & 18.33 & 0.1170 & 30.0 & 33.6 & 0.0 & 5 & -20.73 & -0.04 & OII,\Halpha,HK \\
1 & 13:01:59.06 & +29:04:42.8 & 14.25 & 0.0267 & 40.3 & 47.1 & 6.9 & 2 & -21.99 & 3.20 & OII,OIII,\Halpha,\Hbeta \\
2 & 13:01:33.44 & +29:10:37.6 & 18.03 & 0.1978 & 11.8 & 23.6 & 0.0 & 6 & -22.19 & -0.81 & wOII, HK, \Halpha \\
2 & 13:01:32.86 & +29:10:34.5 & 18.03 & 0.0760 & 0.0 & 0.0 & 0.0 & 8 & -20.07 & -3.90 & \Halpha\ only \\
1 & 13:02:05.41 & +29:01:54.8 & 18.20 & 0.0268 & 26.9 & 24.1 & 4.1 & 7 & -17.67 & -1.54 & OII,OIII,\Halpha \\
2 & 13:02:55.68 & +28:48:03.4 & 18.25 & 0.1645 & 25.0 & 0.0 & 0.0 & 8 & -21.47 & -2.72 & OII,vw\Hbeta,OIII,\Halpha \\
1 & 13:03:31.77 & +28:36:53.0 & 18.09 & 0.1759 & 0.0 & 160.1 & 0.0 & 8 & -21.78 & -2.63 & OIIwOIII, \Halpha\\
2 & 13:01:58.51 & +29:01:55.0 & 17.48 & 0.0184 & 0.0 & 79.7 & 21.5 & 7 & -17.51 & -1.44 & wOII,\Hbeta,OIII,\Halpha \\
1 & 13:01:05.21 & +29:15:22.6 & 17.80 & 0.0830 & 0.0 & 0.0 & 0.0 & 3 & -20.57 & 1.99 & HK,\Halpha \\
1 & 13:02:25.99 & +28:51:14.9 & 18.16 & 0.2532 & 78.0 & 113.8 & 16.9 & 6 & -21.67 & -0.95 & OII,\Hbeta,OIII,\Halpha \\
1 & 13:02:15.28 & +28:53:47.1 & 15.48 & 0.0570 & 4.1 & 6.3 & 0.0 & 2 & -22.28 & 3.17 & HK,vwOII,\Halpha \\
1 & 13:02:11.95 & +28:53:43.2 & 15.95 & 0.0224 & 27.0 & 38.3 & 4.2 & 3 & -20.02 & 1.87 & OII,abs,OIII,\Halpha,\Hbeta \\
1 & 13:01:58.98 & +28:55:49.6 & 18.46 & 0.2798 & 0.0 & 27.6 & 0.0 & 8 & -22.46 & -2.29 & HK,abs \\
1 & 13:02:41.80 & +28:42:17.0 & 18.73 & 0.2187 & 14.9 & 50.1 & 0.0 & 4 & -21.74 & 0.38 & OII,abs,\Halpha,\Hbeta, OIII \\
1 & 13:02:33.29 & +28:43:05.3 & 17.60 & 0.0696 & 28.5 & 27.4 & 5.8 & 5 & -20.24 & -0.32 & OII,OIII,\Halpha,\Hbeta \\
2 & 13:02:08.38 & +28:46:55.4 & 18.53 & 0.0370 & 0.0 & 0.0 & 0.0 & 8 & -18.03 & -3.68 & z:,HK\\
1 & 13:01:49.26 & +28:48:37.7 & 16.00 & 0.0273 & 59.6 & 47.0 & 17.9 & 5 & -19.92 & -0.32 & OII,\Hbeta,OIII,\Halpha \\
1 & 13:01:07.18 & +28:57:30.8 & 16.83 & 0.0224 & 15.1 & 35.3 & 5.1 & 6 & -17.36 & -0.80 & wOII,\Hbeta,OIII,\Halpha\\
\hline
\multicolumn{12}{l}{\textbf{Abell 1367}}\\
\hline
1 & 11:41:52.98 & +20:47:39.8 & 17.48 & 0.0965 & 0.0 & 0.0 & 0.0 & 6 & -21.18 & -0.93 & OII,HK, OIII \\
1 & 11:42:30.43 & +20:42:57.5 & 16.49 & 0.0231 & 0.0 & 23.1 & 0.0 & 1 & -19.24 & 5.86 & \Halpha,HK,\Hbeta abn \\
1 & 11:41:28.99 & +20:45:35.4 & 18.19 & 0.0669 & 26.4 & 21.2 & 0.0 & 6 & -19.68 & -1.17 & OII,HK,\Hbeta,OII,\Halpha \\
1 & 11:40:59.21 & +20:46:04.6 & 16.91 & 0.1687 & 30.9 & 95.67 & 11.5 & 4 & -22.96 & 0.36 & OII,\Hbeta,OIII,\Halpha \\
1 & 11:42:53.69 & +20:40:01.9 & 17.91 & 0.0821 & 14.8 & 17.1 & 0.0 & 4 & -20.41 & 0.31 & wOII,HK,OIII,\Halpha \\
1 & 11:42:58.57 & +20:39:27.8 & 17.48 & 0.1129 & 10.6 & 33.1 & 3.1 & 3 & -21.55 & 1.31 & OII,HK,\Hbeta,\Halpha \\
1 & 11:42:02.60 & +20:42:14.0 & 16.53 & 0.0391 & 27.6 & 10.5 & 0.0 & 2 & -20.34 & 2.52 & wOII,HK,\Halpha \\
2 & 11:42:57.11 & +20:36:32.9 & 17.26 & 0.0480 & 0.0 & 0.0 & 0.0 & 3 & -19.91 & 1.56 & HK,z: \\
1 & 11:41:26.28 & +20:39:30.9 & 16.76 & 0.0399 & 77.1 & 82.9 & 14.5 & 3 & -20.00 & 1.32 & OII,\Hbeta,OIII,\Halpha \\
1 & 11:41:12.18 & +20:38:27.7 & 17.30 & 0.0222 & 0.0 & 0.0 & 0.0 & 1 & -18.34 & 3.90 & HK,\Halpha+b in abn\\
1 & 11:40:54.03 & +20:38:39.9 & 18.22 & 0.2084 & 7.2 & 49.0 & 0.0 & 5 & -22.12 & -0.15 & OII,HK,\Halpha \\
1 & 11:43:21.82 & +20:27:28.1 & 16.93 & 0.0815 & 7.1 & 16.6 & 1.0 & 4 & -21.38 & 0.75 & wOII,HK,s\Halpha \\
1 & 11:40:40.45 & +20:34:40.5 & 17.22 & 0.1770 & 0.0 & 0.0 & 0.0 & 3 & -22.79 & 1.87 & HK,z: \\
1 & 11:41:02.34 & +20:32:20.4 & 17.02 & 0.0413 & 36.7 & 37.6 & 0.0 & 7 & -19.79 & -1.47 & OII,HK,OIII,\Halpha \\
1 & 11:41:13.29 & +20:31:34.6 & 16.51 & 0.0205 & 89.2 & 199.2 & 40.4 & 3 & -18.79 & 1.81 & OII,\Hgamma,\Hbeta,OIII,\Halpha \\
1 & 11:40:22.05 & +20:34:08.2 & 17.45 & 0.0700 & 0.0 & 0.0 & 0.0 & 4 & -20.53 & 0.09 & \Hbeta,\Halpha \\
1 & 11:42:16.91 & +20:27:56.5 & 18.38 & 0.0243 & 0.0 & 0.0 & 0.0 & 4 & -17.29 & 0.26 & HK,\Hgamma,\Halpha,\Hbeta e+a\\
1 & 11:41:41.45 & +20:29:54.4 & 15.91 & 0.0216 & 0.0 & 10.2 & 0.0 & 1 & -19.67 & 5.53 & \Halpha,HK,\Hbeta(abn) \\
1 & 11:43:04.16 & +20:22:49.6 & 18.52 & 0.3826 & 22.4 & 49.4 & 10.4 & 8 & -22.96 & -3.78 & OII,\Hbeta,OIII,\Halpha\\
1 & 11:41:22.60 & +20:27:44.9 & 15.63 & 0.0220 & 16.2 & 64.1 & 5.3 & 1 & -19.99 & 3.83 & OII,HK,\Hbeta,OIII,\Halpha \\
1 & 11:42:11.59 & +20:24:09.7 & 15.39 & 0.0214 & 13.9 & 0.0 & 0.0 & 1 & -20.16 & 5.13 & \Hbeta abn,HK,wOII,\Halpha \\
2 & 11:40:24.84 & +20:29:35.5 & 18.07 & 0.0706 & 65.3 & 24.6 & 0.0 & 8 & -19.87 & -2.25 & OII,OIII,\Halpha \\
1 & 11:41:56.61 & +20:23:02.9 & 18.08 & 0.0240 & 26.9 & 92.5 & 14.6 & 2 & -17.68 & 2.83 & OII,\Hbeta,OIII,s\Halpha\\
1 & 11:40:48.41 & +20:25:49.3 & 17.45 & 0.0707 & 53.9 & 73.6 & 13.6 & 6 & -20.54 & -1.14 & OII,\Hbeta,OIII,\Hgamma,\Halpha \\
1 & 11:42:07.45 & +20:21:20.2 & 17.80 & 0.0885 & 15.0 & 15.8 & 0.0 & 7 & -20.64 & -1.76 & OII,HK,OIII,\Halpha \\
1 & 11:41:23.40 & +20:21:16.6 & 14.97 & 0.0250 & 5.1 & 12.0 & 0.0 & 1 & -20.94 & 4.30 & wOII,HK,\Halpha \\
1 & 11:41:01.40 & +20:21:04.4 & 17.48 & 0.0684 & 44.4 & 61.8 & 13.0 & 7 & -20.42 & -1.71 & OII,\Hgamma,\Hbeta,OIII,\Halpha\\
1 & 11:41:08.38 & +20:19:19.9 & 18.15 & 0.0692 & 7.4 & 15.5 & 0.0 & 6 & -19.79 & -0.52 & OII,HK,\Halpha \\
1 & 11:41:18.49 & +20:16:31.8 & 17.36 & 0.0966 & 4.2 & 3.5 & 0.0 & 6 & -21.30 & -0.90 & OIII,HK,\Halpha \\
1 & 11:40:09.47 & +20:18:35.4 & 15.65 & 0.0289 & 0.0 & 0.0 & 0.0 & 1 & -20.60 & 4.81 & HK,\Halpha,\Hbeta abs \\
1 & 11:39:39.87 & +20:19:34.2 & 16.26 & 0.0204 & 16.7 & 17.5 & 1.3 & 2 & -19.13 & 2.33 & OII,HK,OIII,\Halpha \\
1 & 11:42:40.08 & +20:09:19.8 & 17.62 & 0.0814 & 0.0 & 0.0 & 0.0 & 5 & -20.68 & -0.12 & abn:HK,\Hbeta,w\Halpha \\
1 & 11:42:19.65 & +20:09:28.8 & 17.63 & 0.0818 & 11.4 & 22.1 & 23.7 & 5 & -20.68 & -0.29 & OII,HK,OIII,\Halpha \\
1 & 11:41:51.04 & +20:10:44.6 & 18.48 & 0.2460 & 0.0 & 0.0 & 0.0 & 8 & -22.14 & -3.35 & HK,z: \\
1 & 11:40:20.70 & +20:14:37.2 & 15.02 & 0.0244 & 28.8 & 38.1 & 4.6 & 2 & -20.78 & 2.91 & OII,HK,\Hbeta,OIII,\Halpha \\
1 & 11:40:03.61 & +20:14:47.5 & 17.01 & 0.0245 & 98.4 & 53.7 & 8.7 & 5 & -18.67 & -0.02 & OII,\Hbeta,OIII,\Halpha \\
1 & 11:41:20.91 & +20:10:20.7 & 16.87 & 0.0214 & 0.0 & 0.0 & 0.0 & 1 & -18.68 & 5.15 & abn:HK,G,\Hbeta,others \\
1 & 11:39:38.39 & +20:15:09.9 & 15.53 & 0.0261 & 21.2 & 16.8 & 0.0 & 2 & -20.42 & 3.73 & OII,wHK,OIII,\Halpha \\
1 & 11:40:18.41 & +20:13:05.5 & 18.11 & 0.1318 & 44.8 & 121.6 & 11.3 & 8 & -21.14 & -3.20 & OII,\Hbeta,OIII,\Halpha \\
2 & 11:42:25.90 & +20:05:34.0 & 18.34 & 0.0812 & 0.0 & 14.0 & 0.0 & 7 & -19.92 & -1.38 & OII,HK,wOIII,\Halpha \\
1 & 11:40:52.95 & +20:10:16.2 & 17.02 & 0.1126 & 6.1 & 10.4 & 0.0 & 4 & -21.98 & 0.68 & OII,HK,\Hbeta,\Halpha \\
1 & 11:43:10.71 & +20:03:06.0 & 17.24 & 0.0183 & 0.0 & 0.0 & 0.0 & 1 & -17.96 & 5.89 & abn:HK,\Hbeta,\Halpha,others \\
1 & 11:42:13.44 & +20:05:08.6 & 17.09 & 0.0677 & 7.0 & 13.3 & 0.0 & 3 & -20.84 & 1.46 & wOII,HK,\Halpha \\
1 & 11:40:52.30 & +20:08:05.2 & 18.12 & 0.0713 & 15.0 & 9.9 & 0.0 & 6 & -19.89 & -1.07 & OII,HK,\Halpha \\
1 & 11:42:27.23 & +20:02:00.1 & 16.87 & 0.0688 & 17.8 & 33.6 & 4.5 & 5 & -21.06 & -0.01 & OII,HK,\Hbeta,OIII,\Halpha \\
1 & 11:41:26.36 & +20:03:43.3 & 15.04 & 0.0165 & 79.0 & 63.7 & 10.9 & 2 & -19.87 & 2.72 & OII,\Hbeta,OIII,vs\Halpha \\
1 & 11:40:09.70 & +20:07:20.9 & 18.05 & 0.0708 & 34.1 & 54.2 & 4.8 & 8 & -19.90 & -2.59 & OII,\Hbeta,OIII,\Halpha \\
1 & 11:39:42.36 & +20:06:54.9 & 16.79 & 0.0216 & 9.1 & 14.6 & 0.0 & 3 & -18.63 & 1.72 & wOII,HK,\Halpha \\
1 & 11:42:10.07 & +19:58:29.9 & 16.50 & 0.0821 & 0.0 & 8.0 & 0.0 & 3 & -21.85 & 1.70 & HK,\Halpha \\
1 & 11:40:55.61 & +19:54:19.7 & 15.51 & 0.0437 & 0.0 & 0.0 & 0.0 & 1 & -21.72 & 3.82 & abn:HK,G,\Hbeta,others \\
1 & 11:40:26.50 & +19:55:38.8 & 16.56 & 0.0233 & 0.0 & 0.0 & 0.0 & 1 & -19.19 & 4.99 & abn:HK,\Halpha,\Hbeta,Mgb \\
\hline
\multicolumn{12}{l}{\textbf{Paper I}}\\
\hline
2 & 13:03:58.95 & 28:52:21.8 & 18.02 & 0.2531 & 12.0 & 9.0 & 0.0 & 8 & -22.64 & -3.25 & OII \\
1 & 13:04:24.07 & 29:06:57.9 & 17.33 & 0.0160 & 0.0 & 7.0 & 0.0 & 5 & -17.39 & -0.35 & Balmer, \Halpha \\
1 & 13:04:44.85 & 28:54:00.4 & 16.41 & 0.0393 & 25.0 & 26.0 & 0.0 & 4 & -20.28 & 0.71 & OII, OIII, \Halpha \\
1 & 13:06:00.96 & 29:10:29.5 & 18.35 & 0.2702 & 49.0 & -9.0 & 0.0 & 8 & -22.46 & -3.51 & OII, OIII \\
2 & 13:05:55.61 & 29:12:27.8 & 17.38 & 0.1959 & 41.0 & 32.0 & 0.0 & 8 & -22.69 & -3.79 & OII, OIII, \Halpha: \\
1 & 13:05:59.62 & 29:13:10.1 & 18.08 & 0.1757 & 8.0 & 13.0 & 0.0 & 6 & -21.83 & -0.76 & OII, \Halpha \\
1 & 13:06:01.98 & 29:15:06.0 & 17.42 & 0.0256 & 32.0 & 86.0 & 0.0 & 7 & -18.32 & -1.79 & 0II, \Halpha \\
1 & 13:05:32.99 & 29:16:56.8 & 18.17 & 0.1377 & 5.0 & 4.0 & 0.0 & 7 & -21.16 & -1.47 & OII, HK \\
1 & 13:04:29.55 & 29:22:03.1 & 17.86 & 0.0224 & 6.0 & 48.0 & 0.0 & 8 & -17.58 & -2.41 & OII, OIII:, \Halpha \\
2 & 13:05:38.86 & 29:32:23.6 & 18.22 & 0.0493 & 6.0 & 1.0 & 0.0 & 8 & -18.93 & -3.59 & OII, OIII, \Halpha: \\
1 & 13:05:19.19 & 29:32:52.8 & 18.15 & 0.0848 & 5.0 & 6.0 & 0.0 & 4 & -20.21 & 0.07 & OII, HK, \Halpha \\
2 & 13:05:27.53 & 29:35:30.0 & 18.28 & 0.2525 & 15.0 & 12.0 & 0.0 & 8 & -22.37 & -2.84 & OII, Balmer, \Halpha \\
1 & 13:04:35.77 & 29:25:56.9 & 17.44 & 0.0607 & 27.0 & 29.0 & 0.0 & 5 & -20.19 & -0.23 & OII, OIII, \Halpha \\
1 & 13:05:10.07 & 29:35:40.6 & 18.04 & 0.1275 & 0.0 & 0.0 & 0.0 & 4 & -21.19 & 0.48 & Balmer only! \\
1 & 13:04:27.19 & 29:26:58.7 & 17.65 & 0.1382 & 3.0 & 7.0 & 0.0 & 5 & -21.74 & 0.00 & OII, HK, \Halpha \\
1 & 13:04:31.61 & 29:31:20.9 & 17.79 & 0.1379 & 9.0 & 35.0 & 0.0 & 4 & -21.61 & 0.18 & OII, HK, \Halpha \\
1 & 13:04:40.52 & 29:41:16.2 & 17.44 & 0.2677 & 16.0 & 0.0 & 0.0 & 8 & -23.35 & -4.61 & OII \\
1 & 13:04:30.97 & 29:35:33.5 & 17.63 & 0.2256 & 13.0 & 25.0 & 0.0 & 6 & -22.86 & -0.84 & OII, HK \\
2 & 13:04:09.94 & 29:27:03.8 & 16.65 & 0.0312 & 0.0 & 13.0 & 0.0 & 2 & -19.67 & 2.50 & HK, \Halpha \\
1 & 13:03:54.86 & 29:33:35.6 & 18.20 & 0.2736 & 54.0 & 70.0 & 0.0 & 8 & -22.64 & -2.77 & OII, \Hbeta, OIII, \Halpha \\
1 & 13:04:16.07 & 29:27:03.2 & 18.19 & 0.1850 & 32.0 & 0.0 & 0.0 & 8 & -21.76 & -3.78 & OII \\
1 & 13:03:20.22 & 29:41:43.9 & 17.91 & 0.0894 & 50.0 & 49.0 & 0.0 & 7 & -20.52 & -1.92 & OII, OIII, \Halpha \\
1 & 13:03:36.41 & 29:32:45.2 & 18.13 & 0.0266 & 0.0 & 0.0 & 0.0 & 1 & -17.90 & 6.60 & Balmer only! \\
1 & 13:03:15.26 & 29:37:02.5 & 16.85 & 0.0238 & 11.0 & 6.0 & 0.0 & 5 & -18.74 & 0.03 & OII, HK \\
1 & 13:03:11.64 & 29:22:44.9 & 17.98 & 0.0819 & 0.0 & 13.0 & 0.0 & 3 & -20.33 & 1.35 & HK, \Halpha \\
1 & 13:03:13.39 & 29:35:00.4 & 18.05 & 0.0897 & 28.0 & 33.0 & 0.0 & 7 & -20.39 & -1.44 & OII, OIII, \Halpha \\
1 & 13:03:23.02 & 29:31:13.0 & 18.23 & 0.2897 & 125.0 & 250.0 & 0.0 & 8 & -22.73 & -2.92 & OII, OIII, \Halpha \\
1 & 13:03:15.09 & 29:29:57.6 & 17.78 & 0.1390 & 2.0 & 4.0 & 0.0 & 4 & -21.63 & 0.16 & OII, HK, \Halpha \\
1 & 13:03:50.38 & 29:24:30.5 & 16.57 & 0.0235 & 0.0 & 4.0 & 0.0 & 4 & -18.99 & 0.12 & OIII, \Halpha \\
1 & 13:02:34.52 & 29:32:08.1 & 17.49 & 0.1682 & -9.0 & -9.0 & 0.0 & 8 & -22.25 & -3.72 & OII, \Halpha (poor ex) \\
2 & 13:02:57.34 & 29:18:58.1 & 18.32 & 0.0176 & 0.0 & 28.0 & 0.0 & 7 & -16.61 & -1.30 & \Halpha, \Hbeta, OIII, SII \\
2 & 13:02:52.59 & 29:16:59.9 & 18.49 & 0.0332 & 0.0 & 0.0 & 0.0 & 8 & -17.80 & -3.60 & HK, abs \\
1 & 13:02:32.35 & 29:12:59.0 & 18.09 & 0.2427 & 11.0 & 30.0 & 0.0 & 7 & -22.51 & -1.84 & OII, Balmer, \Halpha \\
2 & 13:02:53.91 & 29:08:50.9 & 18.39 & 0.3229 & 10.0 & -9.0 & 0.0 & 6 & -22.90 & -0.89 & OII, OIII, \Halpha \\
1 & 13:00:19.80 & 29:42:12.0 & 17.23 & 0.0900 & 8.0 & 27.0 & 0.0 & 4 & -21.26 & 0.27 & OII, H \\
1 & 12:59:17.64 & 29:38:59.6 & 16.12 & 0.0590 & 48.0 & 77.0 & 0.0 & 4 & -21.46 & 0.48 & OII, OIII, \Halpha \\
1 & 12:59:40.35 & 29:31:19.2 & 17.59 & 0.0250 & 0.0 & 0.0 & 0.0 & 1 & -18.29 & 6.22 & HK, abn \\
2 & 12:59:22.62 & 29:20:41.6 & 16.16 & 0.0620 & 21.0 & 20.0 & 0.0 & 5 & -21.52 & -0.03 & OII, \Halpha \\
1 & 12:59:33.16 & 29:19:06.8 & 17.44 & 0.0240 & 90.0 & 49.0 & 0.0 & 8 & -18.15 & -2.14 & OII, \Halpha \\
3 & 13:00:17.07 & 29:17:06.6 & 17.52 & 0.0380 & 17.0 & 18.0 & 0.0 & 6 & -19.09 & -1.08 & OII, OIII, \Halpha \\
1 & 13:02:06.03 & 29:11:51.5 & 17.32 & 0.0830 & 0.0 & 65.0 & 0.0 & 4 & -21.00 & 0.41 & OIII, \Halpha \\
1 & 13:02:03.66 & 29:18:39.3 & 18.06 & 0.0840 & 0.0 & 49.0 & 0.0 & 4 & -20.28 & 0.42 & HK, \Halpha \\
1 & 13:01:59.75 & 29:24:29.1 & 17.89 & 0.1890 & 12.0 & 21.0 & 0.0 & 6 & -22.19 & -0.68 & OII, HK, \Halpha \\
1 & 13:00:00.88 & 30:00:51.0 & 17.67 & 0.1570 & 10.0 & 13.0 & 0.0 & 6 & -21.99 & -0.84 & OII, \Halpha \\
1 & 13:06:01.14 & 29:21:14.7 & 18.53 & 0.2407 & 0.0 & 0.0 & 0.0 & 5 & -22.12 & -0.23 & HK,Hd \\
1 & 13:04:31.58 & 28:48:55.0 & 17.68 & 0.0206 & 0.0 & 0.0 & 0.0 & 1 & -17.76 & 6.78 & HK,abs \\
1 & 13:05:01.02 & 29:03:43.2 & 17.75 & 0.1146 & 3.0 & 0.0 & 0.0 & 6 & -21.24 & -0.62 & OII, HK \\
1 & 13:06:27.79 & 29:09:09.3 & 16.77 & 0.0800 & 4.0 & 0.0 & 0.0 & 3 & -21.49 & 2.15 & OII,HK \\
1 & 13:06:08.41 & 29:16:38.7 & 18.19 & 0.0249 & 16.0 & 0.0 & 0.0 & 6 & -17.50 & -0.55 & OII, OIII \\
1 & 13:07:29.10 & 28:38:54.5 & 15.23 & 0.0231 & 0.0 & 0.0 & 0.0 & 1 & -20.47 & 4.04 & HK, abs \\
1 & 13:06:13.24 & 28:56:47.1 & 18.20 & 0.1222 & 15.0 & 0.0 & 0.0 & 4 & -20.94 & 0.20 & OII,HK,abn \\
1 & 13:05:13.62 & 28:44:45.9 & 18.15 & 0.0686 & 22.0 & 0.0 & 0.0 & 5 & -19.75 & -0.42 & OII,\Hbeta,OIII \\
2 & 13:06:29.96 & 28:50:03.5 & 18.27 & 0.1849 & 6.0 & 0.0 & 0.0 & 7 & -21.70 & -1.72 & OII,HK,abn \\
1 & 13:06:26.18 & 28:50:23.5 & 17.22 & 0.0886 & 41.0 & 0.0 & 0.0 & 4 & -21.24 & 0.10 & OII,\Hbeta,OIII \\
1 & 13:05:50.11 & 29:06:51.2 & 16.36 & 0.0544 & 0.0 & 0.0 & 0.0 & 2 & -21.26 & 2.95 & HK, abn \\
1 & 13:05:08.45 & 28:44:10.1 & 17.32 & 0.0685 & 11.0 & 0.0 & 0.0 & 4 & -20.58 & 0.52 & OII, abn \\
2 & 13:03:57.96 & 28:52:18.3 & 18.04 & 0.1865 & 47.0 & 0.0 & 0.0 & 8 & -21.92 & -3.03 & OII, OIII \\
1 & 13:05:13.02 & 28:41:19.4 & 18.51 & 0.0770 & 42.0 & 0.0 & 0.0 & 7 & -19.61 & -1.38 & OII, \Hbeta, OIII \\
1 & 13:03:47.89 & 29:02:28.4 & 17.49 & 0.0836 & 42.0 & 0.0 & 0.0 & 7 & -20.80 & -1.84 & OII, OIII \\
1 & 13:07:03.27 & 28:59:17.8 & 17.99 & 0.0376 & 23.0 & 0.0 & 0.0 & 6 & -18.59 & -1.07 & OII, \Hbeta \\
1 & 13:06:04.40 & 28:36:59.0 & 17.53 & 0.0237 & 25.0 & 0.0 & 0.0 & 3 & -18.06 & 0.90 & OII, \Hbeta, OIII \\
1 & 13:07:40.56 & 28:56:33.1 & 18.21 & 0.1220 & 11.0 & 0.0 & 0.0 & 6 & -20.91 & -1.05 & OII, HK, abs \\
1 & 13:06:20.66 & 29:09:25.9 & 16.93 & 0.0612 & 35.0 & 0.0 & 0.0 & 5 & -20.72 & -0.02 & OII,\Hbeta,OIII \\
1 & 13:06:14.98 & 29:10:25.8 & 16.59 & 0.0395 & 28.0 & 0.0 & 0.0 & 6 & -20.10 & -0.98 & OII,\Hbeta,OIII \\
1 & 13:04:52.46 & 28:41:20.3 & 17.71 & 0.0698 & 17.0 & 0.0 & 0.0 & 7 & -20.20 & -1.54 & OII,\Hbeta,OIII \\
2 & 13:04:45.12 & 28:41:37.3 & 18.04 & 0.0346 & 19.0 & 0.0 & 0.0 & 6 & -18.36 & -0.58 & OII \\
1 & 13:03:55.90 & 28:44:11.1 & 17.40 & 0.2379 & 30.0 & 0.0 & 0.0 & 7 & -23.15 & -1.59 & OII,Balmer,\Hbeta,OIII \\
1 & 13:05:45.10 & 29:18:35.4 & 16.59 & 0.0587 & 21.0 & 0.0 & 0.0 & 6 & -20.97 & -0.52 & OII,HK,\Hbeta,OIII \\
1 & 13:04:06.93 & 28:53:34.4 & 17.84 & 0.1866 & 5.0 & 0.0 & 0.0 & 4 & -22.23 & 0.22 & OII,HK,\Hbeta,OIII \\
1 & 13:05:54.39 & 28:57:38.7 & 16.57 & 0.0804 & 13.0 & 0.0 & 0.0 & 8 & -21.62 & -2.81 & OII,OIII \\
1 & 13:06:03.73 & 28:50:27.4 & 16.41 & 0.0700 & 25.0 & 0.0 & 0.0 & 4 & -21.54 & 0.23 & OII,Balmer,\Hbeta,OIII \\
1 & 13:06:49.06 & 28:43:19.2 & 17.91 & 0.0789 & 26.0 & 0.0 & 0.0 & 6 & -20.29 & -0.60 & OII, OIII \\
1 & 13:06:23.77 & 29:05:09.9 & 17.19 & 0.1384 & 31.0 & 0.0 & 0.0 & 8 & -22.13 & -3.66 & OII,\Hbeta,OIII \\
1 & 13:07:26.48 & 29:14:32.2 & 16.79 & 0.1222 & 12.0 & 0.0 & 0.0 & 8 & -22.27 & -3.23 & OII \\
1 & 13:04:48.14 & 28:41:26.5 & 17.95 & 0.0179 & 30.0 & 0.0 & 0.0 & 6 & -17.02 & -1.03 & OII, OIII \\
1 & 13:04:33.35 & 28:48:11.3 & 18.23 & 0.1589 & 0.0 & 0.0 & 0.0 & 7 & -21.41 & -1.87 & HK \\
1 & 13:05:52.13 & 29:17:22.0 & 16.82 & 0.0206 & 25.0 & 0.0 & 0.0 & 5 & -18.45 & -0.13 & OII,\Hbeta,OIII \\
1 & 13:05:41.78 & 28:58:41.9 & 16.51 & 0.0800 & 4.0 & 0.0 & 0.0 & 3 & -21.75 & 1.02 & OII,HK,abn \\
1 & 13:05:16.20 & 29:17:05.5 & 18.04 & 0.1235 & 4.0 & 0.0 & 0.0 & 3 & -21.15 & 1.21 & OII,HK,abn \\
1 & 13:06:32.30 & 28:51:22.4 & 18.27 & 0.0696 & 33.0 & 0.0 & 0.0 & 7 & -19.63 & -1.81 & OII,\Hbeta,OIII \\
2 & 13:08:00.21 & 28:54:33.6 & 17.61 & 0.1223 & 11.0 & 0.0 & 0.0 & 6 & -21.52 & -0.79 & OII,HK \\
1 & 13:07:39.47 & 29:07:24.7 & 18.06 & 0.2348 & 14.0 & 0.0 & 0.0 & 8 & -22.42 & -2.45 & OII,HK \\
1 & 13:04:05.73 & 28:41:24.0 & 17.89 & 0.0695 & 23.0 & 0.0 & 0.0 & 6 & -20.03 & -1.14 & OII,HK \\
1 & 13:05:58.96 & 28:47:52.5 & 18.25 & 0.1993 & 8.0 & 0.0 & 0.0 & 7 & -21.89 & -1.68 & OII,HK \\
1 & 13:05:40.90 & 29:15:52.3 & 16.20 & 0.0242 & 23.0 & 0.0 & 0.0 & 3 & -19.44 & 1.68 & OII,\Hbeta,OIII \\
1 & 13:04:57.97 & 29:15:45.5 & 17.81 & 0.1894 & 26.0 & 0.0 & 0.0 & 7 & -22.21 & -1.45 & OII,OIII \\
1 & 13:06:34.50 & 28:59:59.0 & 17.60 & 0.0560 & 0.0 & 0.0 & 0.0 & 1 & -20.21 & 5.17 & HK, abn \\
1 & 13:06:28.46 & 28:53:45.1 & 16.16 & 0.0227 & 17.0 & 0.0 & 0.0 & 2 & -19.44 & 2.39 & OII,Balmer,OIII \\
1 & 13:06:24.43 & 28:53:25.3 & 16.66 & 0.1893 & 79.0 & 0.0 & 0.0 & 5 & -23.42 & -0.21 & OII,\Hbeta,OIII \\
1 & 13:04:25.09 & 28:59:31.8 & 18.09 & 0.0565 & 5.0 & 0.0 & 0.0 & 4 & -19.39 & 0.32 & OII,HK \\
1 & 13:06:32.05 & 29:07:04.6 & 17.66 & 0.0794 & 4.0 & 0.0 & 0.0 & 5 & -20.55 & -0.01 & OII,HK, abn \\
1 & 13:04:14.22 & 28:51:15.7 & 17.26 & 0.1225 & 0.0 & 0.0 & 0.0 & 4 & -21.88 & 0.75 & HK,abn \\
\hline
\multicolumn{12}{l}{\textbf{QSO / AGN}}\\
\hline
1 & 13:05:45.71 & +29:50:48.6 & 18.23 & 1.1760 &   &   &   &   &   &   & QSO\\
2 & 13:05:42.92 & +29:22:52.9 & 17.49 & 0.4020 &   &   &   &   &   &   & QSO\\
1 & 13:05:30.13 & +29:17:38.2 & 18.44 & 0.6750 &   &   &   &   &   &   & QSO\\
1 & 13:04:27.11 & +29:25:29.8 & 16.98 & 1.3370 &   &   &   &   &   &   & QSO\\
2 & 13:04:27.70 & +29:15:50.9 & 16.76 & 0.1836 &   &   &   &   &   &   & AGN\\
1 & 13:05:15.88 & +28:52:26.6 & 17.06 & 0.5789 &   &   &   &   &   &   & QSO\\
2 & 13:03:45.69 & +29:06:30.7 & 16.94 & 0.0790 &   &   &   &   &   &   & QSO\\
2 & 13:02:56.66 & +29:18:51.3 & 18.11 & 0.0760 &   &   &   &   &   &   & QSO\\
1 & 13:04:41.25 & +28:48:41.1 & 19.08 & 1.5628 &   &   &   &   &   &   & QSO:CIV,CIII\\
1 & 13:02:06.39 & +29:29:13.4 & 17.42 & 1.0160 &   &   &   &   &   &   & QSO\\
2 & 13:01:05.74 & +29:42:15.0 & 18.58 & 1.7590 &   &   &   &   &   &   & QSO:CIV,CIII\\
1 & 13:04:14.14 & +28:38:10.2 & 18.36 & 1.3625 &   &   &   &   &   &   & CIV,CIII,MgI QSO\\
1 & 13:03:07.94 & +28:55:03.4 & 16.61 & 0.1840 &   &   &   &   &   &   & OII,Balmer,AGN\\
1 & 13:02:45.26 & +28:53:19.3 & 17.99 & 1.6700 &   &   &   &   &   &   & QSO:CIV,CIII\\
1 & 11:39:52.35 & +20:26:34.2 & 18.03 & 1.4150 &   &   &   &   &   &   & QSO:CIV,CIII\\
1 & 11:41:07.70 & +20:22:36.4 & 17.21 & 1.0580 &   &   &   &   &   &   & QSO\\
1 & 11:42:32.69 & +19:55:58.8 & 17.68 & 1.0320 &   &   &   &   &   &   & QSO\\
1 & 11:40:20.54 & +19:56:54.1 & 18.03 & 1.1610 &   &   &   &   &   &   & QSO\\
1 & 13:04:12.79 & 29:35:29.7 & 18.30 & 1.0195 &   &   &   &   &   &   & QSO:CIII,MgII\\
1 & 13:06:36.24 & 29:21:56.4 & 17.80 & 0.7460 &   &   &   &   &   &   & QSO:MgII\\
1 & 13:07:21.82 & 28:43:32.7 & 16.41 & 0.7370 &   &   &   &   &   &   & QSO:MgII\\
1 & 13:04:41.44 & 29:17:32.8 & 17.69 & 1.5866 &   &   &   &   &   &   & QSO:CIV,CIII\\
1 & 13:04:23.17 & 28:39:55.6 & 17.17 & 0.9186 &   &   &   &   &   &   & QSO:MgII\\
\hline
\multicolumn{12}{l}{\textbf{Stars}}\\
\hline
2 & 13:06:25.11 & +29:42:09.2 & 16.17 & * &   &   &   &   &   &   & STAR \\
1 & 13:07:04.67 & +29:31:10.5 & 17.30 & * &   &   &   &   &   &   & STAR \\
1 & 13:06:46.99 & +29:29:54.3 & 18.21 & * &   &   &   &   &   &   & STAR \\
1 & 13:07:20.12 & +29:20:31.2 & 15.59 & * &   &   &   &   &   &   & STAR \\
1 & 13:06:19.31 & +29:29:14.2 & 18.32 & * &   &   &   &   &   &   & STAR \\
2 & 13:05:17.21 & +29:44:18.9 & 0.00 & * &   &   &   &   &   &   & STAR \\
1 & 13:03:54.71 & +29:59:38.9 & 16.83 & * &   &   &   &   &   &   & STAR \\
1 & 13:04:07.05 & +29:51:24.9 & 18.61 & * &   &   &   &   &   &   & STAR \\
1 & 13:05:36.36 & +29:26:50.8 & 18.09 & * &   &   &   &   &   &   & STAR \\
1 & 13:03:44.60 & +29:57:14.7 & 17.39 & * &   &   &   &   &   &   & STAR \\
1 & 13:03:47.43 & +29:54:37.7 & 18.44 & * &   &   &   &   &   &   & STAR \\
1 & 13:03:12.29 & +30:01:45.1 & 17.48 & * &   &   &   &   &   &   & STAR \\
1 & 13:06:23.62 & +29:02:51.5 & 16.57 & * &   &   &   &   &   &   & STAR \\
1 & 13:02:44.74 & +29:55:03.3 & 17.34 & * &   &   &   &   &   &   & STAR \\
1 & 13:05:39.76 & +29:06:32.0 & 15.76 & * &   &   &   &   &   &   & STAR \\
1 & 13:04:11.26 & +29:26:54.5 & 16.68 & * &   &   &   &   &   &   & STAR \\
1 & 13:04:27.83 & +29:19:52.0 & 17.20 & * &   &   &   &   &   &   & STAR \\
1 & 13:05:47.53 & +28:57:23.2 & 18.17 & * &   &   &   &   &   &   & STAR \\
1 & 13:02:42.36 & +29:46:19.1 & 17.96 & * &   &   &   &   &   &   & STAR \\
1 & 13:03:58.77 & +29:24:08.3 & 17.96 & * &   &   &   &   &   &   & STAR \\
1 & 13:05:33.64 & +28:57:10.0 & 17.21 & * &   &   &   &   &   &   & STAR \\
1 & 13:03:20.26 & +29:26:26.6 & 18.14 & * &   &   &   &   &   &   & STAR \\
2 & 13:03:59.68 & +29:14:38.8 & 17.18 & * &   &   &   &   &   &   & STAR \\
1 & 13:05:07.25 & +28:55:37.9 & 18.11 & * &   &   &   &   &   &   & STAR \\
1 & 13:02:52.40 & +29:29:02.4 & 17.79 & * &   &   &   &   &   &   & STAR \\
1 & 13:03:48.62 & +29:13:43.6 & 16.41 & * &   &   &   &   &   &   & STAR \\
1 & 13:02:44.35 & +29:27:59.6 & 18.11 & * &   &   &   &   &   &   & STAR \\
2 & 13:02:25.48 & +29:30:07.2 & 17.68 & * &   &   &   &   &   &   & STAR \\
1 & 13:02:53.26 & +29:16:02.8 & 17.02 & * &   &   &   &   &   &   & STAR \\
1 & 13:01:12.62 & +29:34:10.7 & 17.86 & * &   &   &   &   &   &   & STAR \\
1 & 13:01:50.15 & +29:21:17.3 & 18.03 & * &   &   &   &   &   &   & STAR \\
1 & 13:00:59.77 & +29:32:42.4 & 17.36 & * &   &   &   &   &   &   & STAR \\
1 & 13:01:16.97 & +29:27:10.6 & 15.87 & * &   &   &   &   &   &   & STAR \\
1 & 13:01:08.23 & +29:22:04.7 & 17.20 & * &   &   &   &   &   &   & STAR \\
1 & 13:03:12.23 & +28:46:17.7 & 15.76 & * &   &   &   &   &   &   & STAR \\
1 & 13:02:50.69 & +28:51:02.8 & 17.95 & * &   &   &   &   &   &   & STAR \\
1 & 13:01:38.04 & +28:49:14.5 & 17.38 & * &   &   &   &   &   &   & STAR \\
1 & 13:00:56.66 & +28:59:33.2 & 16.76 & * &   &   &   &   &   &   & STAR \\
1 & 11:43:10.44 & +20:29:09.0 & 18.39 & * &   &   &   &   &   &   & STAR \\
1 & 11:40:11.78 & +20:22:29.7 & 16.97 & * &   &   &   &   &   &   & STAR \\
1 & 11:42:09.41 & +19:56:06.9 & 18.02 & * &   &   &   &   &   &   & STAR \\
1 & 11:41:19.47 & +19:55:56.6 & 18.22 & * &   &   &   &   &   &   & STAR \\
1 & 13:04:58.22 & 29:11:27.1 & 18.45 & * &   &   &   &   &   &   & STAR \\
1 & 13:03:47.20 & 29:17:50.1 & 8.34 & * &   &   &   &   &   &   & STAR \\
1 & 13:00:08.49 & 29:23:14.1 & 17.50 & * &   &   &   &   &   &   & STAR \\
1 & 13:01:34.41 & 29:19:34.1 & 16.85 & * &   &   &   &   &   &   & STAR \\
1 & 13:07:33.80 & 28:49:27.4 & 15.48 & * &   &   &   &   &   &   & STAR \\
1 & 13:07:56.30 & 29:02:17.9 & 18.47 & * &   &   &   &   &   &   & STAR \\
2 & 13:05:13.95 & 29:04:36.3 & 15.29 & * &   &   &   &   &   &   & STAR \\
2 & 13:07:52.45 & 28:48:54.9 & 17.56 & * &   &   &   &   &   &   & STAR \\

\end{longtable}
\end{center}
\twocolumn

\label{lastpage}

\end{document}